\documentclass[12pt]{CSM_template}

\usepackage{lineno}
\usepackage[draft]{hyperref}
\modulolinenumbers[5]
\usepackage[numbers]{natbib}
\usepackage{url}
\usepackage{float}
\usepackage{amsmath}
\usepackage{amsfonts}
\usepackage{array}
\usepackage{dsfont}
\usepackage{amssymb}
\usepackage{amsthm}
\usepackage{bbold}
\usepackage{fullpage}
\usepackage{mathtools}
\usepackage{enumerate}
\usepackage{mathrsfs}
\usepackage{subfigure}
\usepackage{txfonts}
\usepackage{color}
\usepackage{multicol}
\usepackage{tcolorbox}
\tcbuselibrary{skins,breakable}
\usepackage{tikz}
\usetikzlibrary{shadings,shadows}

\newcommand{\oprocendsymbol}{\hbox{$\bullet$}}
\newcommand{\oprocend}{\relax\ifmmode\else\unskip\hfill\fi\oprocendsymbol}

\newcommand{\longthmtitle}[1]{\mbox{}\textup{\textbf{(#1):}}}

\newcommand*{\SetSuchThat}[1][]{} 
\newcommand*{\MvertSets}{%
    \renewcommand*\SetSuchThat[1][]{%
        \mathclose{}%
        \nonscript\;##1\vert\penalty\relpenalty\nonscript\;%
        \mathopen{}%
    }%
}
\MvertSets 
\DeclarePairedDelimiterX \Set [2] {\lbrace}{\rbrace}
    {\,#1\SetSuchThat[\delimsize]#2\,}

\allowdisplaybreaks

\newcommand{\R}{\varmathbb{R}}

\newcommand{\Rplus}{\varmathbb{R}_{> 0}}
\newcommand{\Rpluseq}{\varmathbb{R}_{\geq 0}}

\newcommand{\F}{\mathbf{F}}

\newcommand{\norm}[1]{\Vert #1 \Vert}
\newcommand{\x}{\mathbf{x}}
\newcommand{\m}{\mathbf{m}}
\newcommand{\W}{\mathbf{W}}

\renewcommand{\c}{\mathbf{c}}
\newcommand{\B}{\mathbf{B}}
\newcommand{\K}{\mathbf{K}}
\newcommand{\eye}{\mathbf{I}}
\newcommand{\zero}{\mathbf{0}}
\renewcommand{\u}{\mathbf{u}}
\newcommand{\w}{\mathbf{w}}
\renewcommand{\v}{\mathbf{v}}
\renewcommand{\d}{\mathbf{d}}
\newcommand{\pre}{\text{pre}}
\newcommand{\post}{\text{post}}
\newcommand{\range}{\mathrm{range}}
\newcommand{\all}{\mathrm{all}}
\newcommand{\supp}{\text{supp}}

\newtheorem{theorem}{Theorem}
\newtheorem{definition}[theorem]{Definition}

\newtheorem{remark}[theorem]{Remark}

\newtheorem{corollary}[theorem]{Corollary}

\newcounter{thebox}

\newenvironment{myblock}[1]{%
\refstepcounter{thebox}%
\bigskip
\tcolorbox[beamer, no shadow,%
noparskip,breakable,
colback=black!60!green!5,colframe=black!60!green!5,%
title={\smallskip \color{black}Box~\arabic{thebox}. #1 \medskip \hrule}]\vspace*{-10pt}}%
{\endtcolorbox}

\graphicspath{{epsfiles/}}

\begin{document}



  \title{Linear-Threshold Network Models for Describing and Analyzing
    Brain Dynamics}

\author{Michael McCreesh \quad Erfan Nozari \quad Jorge Cort\'es \\ Corresponding Author: M. McCreesh (mmccreesh@ucsd.edu)}

\date{\today}
\maketitle

  
%


\begin{abstract}


Over the past two decades, an increasing array of control-theoretic
methods have been used to study the brain as a complex dynamical
system and better understand its structure-function relationship.
This article provides an overview on one such family of methods, based
on the linear-threshold rate (LTR) dynamics, which arises when
modeling the spiking activity of neuronal populations and their impact
on each other.  LTR dynamics exhibit a wide range of behaviors based
on network topologies and inputs, including mono- and multi-stability,
limit cycles, and chaos, allowing it to be used to model many complex
brain processes involving fast and slow inhibition, multiple time and
spatial scales, different types of neural behavior, and higher-order
interactions.  Here we investigate how the versatility of LTR dynamics
paired with concepts and tools from systems and control can provide
a computational theory for explaining the dynamical mechanisms
enabling different brain processes. Specifically, we illustrate
stability and stabilization properties of LTR dynamics and how they
are related to goal-driven selective attention, multistability and its
relationship with declarative memory, and bifurcations and
oscillations and their role in modeling seizure dynamics in
epilepsy. We conclude with a discussion on additional properties of
LTR dynamics and an outlook on other brain processess that for which
they might be play a similar role.

\end{abstract}





\section{Introduction}\label{sec:introduction} 

The brain is one of the most complex dynamical systems known to
exist. It is composed of billions of interconnected neurons and
involved in countless cognitive and physical functions. It has been
the holy grail of neuroscience for centuries to explain, at various
levels of analysis, how these functions arise from the brain's complex
and ever-changing structure. Computational modeling has long been a
pillar in this quest for multi-level understanding.
Furthermore, the complexity that arises from the interplay between
neuronal dynamics, multiple temporal and spatial scales, and the
intricate interconnection and rich topological structure observed in
the brain makes the use of system-theoretic techniques a natural
choice to analyze and understand such computational models. This
approach has led to a growing number of recent works at the
intersection of control theory and neuroscience, see
e.g.,~\citep{SJS:12,SC-MYL-JJC-MBW-JDK-KS-PLP-ENB:13,SG-FP-MC-QKT-ABY-AEK-JDM-JMV-MBM-STG-DSB:15,SVS-PS:18,MSM-NJC:20,EN-JC:21-tacI,EN-JC:21-tacII,PS-EN-JZK-HJ-DZ-CB-FP-GJP-DSB:20,GD-TOL-JD-AF-RF:15,MEB:21,MEB:22}
for a small sample.

A wide variety of processes and phenomena in the brain have been
studied through the use of dynamical systems models. Visual
processing~\citep{CAC:15,AK-WT:11}, voluntary
movement~\citep{MK-KF-RS:87,KVS-MTK-MS-MMC:11}, and pathological
behavior due to disorders such as Parkinson's or
epilepsy~\citep{CIC-JBB-MSJ:00,EJM-SJvA-JWK-PAR:17} are only a few
examples. Indeed, the wide range of functions and behaviors exhibited
by the brain exceed that of any single model.
Nevertheless, generalizable, multipurpose models are increasing sought
whose dynamics are rich enough to describe multiple dynamical
behaviors using a single model structure.
In this work we investigate one such model consistent with empirical
descriptions of neural physiology, the linear-threshold rate dynamics,
and illustrate its richness by using it to describe multiple dynamical
behaviors observed in the brain.
Specifically, we consider the dynamical brain behaviors of goal-driven
selective attention (GDSA), declarative memory, and epileptic seizure
activity (see Box~\ref{box:neural_activities_summary} for an overview
of these applications). Each of these applications can be tackled from
a dynamical systems perspective, a choice made due to their dependence
on both the structure of the brain and the dynamics occurring within
it. For the purpose of fully exploring the rich behavior of the
linear-threshold model, the key point is that each of these
applications can be associated with distinct dynamical properties,
illustrating the versatility of the linear-threshold model.

While these three applications are distinct functions of the brain,
there does exist overlap between them. For example, GDSA is a key part
of working memory in order to select relevant information from
internally stored
representations~\citep{AG-ACN:12,RQ-JK-ES-NF-RT-ERB-LGC:19}. Meanwhile,
epilepsy and memory are tied in that during epileptic seizures the
ability of the brain to both encode and retrieve memories is
inhibited~\citep{GLH:15,YH-ET:15,CRB-AZZ:09}. As such, while
individual models of each dynamical behavior are important, the
ability to use a single model that can show all three behaviors is
desirable. In this work, through an examination of a dynamical systems
approach to GDSA, epileptic seizures, and declarative
memory we provide a review of the extensive properties of the
linear-threshold dynamics, covering topics such as stability,
stabilizability, bifurcations, and oscillations.

\begin{myblock}{Examples of Neural Activities Through the Lens of
    Dynamical Systems}\label{box:neural_activities_summary}
  In this article we will pay particular attention goal-driven
  selective attention (GDSA), declarative memory, and epileptic
  seizure activity.
  \begin{itemize}
  \item GDSA is the process in which the brain actively determines
    which subset of the currently incoming sensory information is
    relevant for the current task in order to process it, while
    simultaneously suppressing irrelevant information. These two
    components are referred to as \emph{selective recruitment} and
    \emph{selective inhibition}, respectively.
    Examples of GDSA include selective vision in busy visual fields,
    selective listening in a noisy environment, and selective
    smell. While GDSA has been widely studied in the neuroscience
    community, see
    e.g.,~\citep{JS:90,AMT:69,ECC:53,RD-JD:95,LI-CK:01,BCM:93,MAP-GMD-SK:04},
    the recent work~\citep{EN-JC:21-tacI,EN-JC:21-tacII,MM-JC:24-tcns}
    approaches it from a dynamical systems perspective.
  \item Epilepsy is a disease of the brain characterized by recurrent
    seizures. While seizures can take different forms and have varying
    symptoms, they all include abnormal brain activity, typically
    either excessive or highly synchronous oscillatory
    behavior~\citep{KT-KTH-AK-LE-AB-LE-ZJ-GN-AS-DF-IU-LW:18}. Seizures have
    been studied extensively through the dynamical systems perspective
    with a variety of
    models~\citep{HGEM-TLE-BK-JFN-CAS-RGE-RRG-GMM-CJM-AKT-JDC-SAvG-WvD:15,AC:22,VKJ-WCS-PPQ-AII-CB:14} and
    approaches~\citep{JT-FW-PC-OF:11,DS-SVS:14,DE-DS-SVS:15,ETW-MV-ZH-RM-VRR-SC:22,FLdS-WB-SNK-JP-PS-DNV:03},
    including the linear-threshold
    model~\citep{FC-AA-FP-JC:21-csl,AA-FC-FP-JC:22-ojcsys}.  Dynamical
    systems approaches to seizures have been effective due to the
    ability to relate seizures with bifurcations in the network,
    capturing the sudden switch from healthy to seizure behavior.

  \item
    Research of memory models dates back to the late $19^{\text{th}}$
    century~\citep{WJ:90}, and are
    extensive~\citep{GM:56,RCA-RMS:68,GRL-EFL:76} with general
    divisions between short-term and long-term memory being constant
    across the literature. By the $1990$'s, models included further
    divisions, with working memory being key for short-term memory and
    long term memory divided into declarative and non-declarative
    memory~\citep{DLS-ET:94}. In this paper we consider declarative
    memory, which is the component of long-term memory that includes
    the ability to consciously store and retrieve personal information
    (episodic memory) and general knowledge (semantic
    memory)~\citep{AB:20}.  In the $1980$'s both short and long-term
    memory models started to be considered using a dynamical systems
    approach~\citep{JJH:82,GEH-TJS:86}, with memories being defined by
    properties of the dynamics. Since these early studies the modeling
    of memory has been extensively studied with a systems theory
    approach~\citep{SG-DH-HS:08,RP-HJ:11,SA-AK-PEK:99,RHRH-HSS-JJS:03,IT-EK-HS:87},
    motivated by the ability to relate the structure of the network
    with the observed activity patterns through the models.
  \end{itemize}
\end{myblock}

%

\subsection*{Outline}
The paper is organized as follows. In
Section~\ref{sec:brain_modeling}, we discuss methods of modeling the
brain, focusing on firing rate models, leading to the derivation of
the linear-threshold model. We also discuss the role of feedback and
feedforward control employed in the context of the brain. In
Section~\ref{sec:single-region}, we explore properties of the LTR
dynamics in single networks through the modeling of selective
attention, epilepsy, and memory.  We exploit the description of the
dynamics as a piecewise-affine state-dependent switched system to
leverage system-theoretic tools in unveiling the relationship between
the network interconnection structure and its dynamical behavior.  In
Section~\ref{sec:multi-region}, we examine the construction of
interconnected brain networks and study their resulting properties,
with a view on the application to selective attention and
epilepsy. For selective attention, we consider hierarchical and
star-connected interconnected topologies and illustrate the role of
feedback/feedforward control in ensuring recruitment of task-relevant
regions and inhibition of task-irrelevant regions.  For epilepsy, we
consider interconnected excitatory-inhibitory pairs and describe
conditions for oscillations to emerge and spread throughout the
resulting network of networks.  We finish in
Section~\ref{sec:discussion-conclusions} with a discussion of
additional properties of the dynamics and an outline of additional
applications that could be studied using the linear-threshold model.

\section{Linear Threshold Rate Models}\label{sec:brain_modeling}

Computational modeling of the brain is particularly challenging due,
in part, to the different scales of information in which one can
phrase and approach the problem. At the smallest
level, or the ``microscale'', the brain is composed of billions of
neurons whose dynamics can be measured at the individual level through
their electrical activity. At the opposite end of the spectrum, the
``macroscale'', the brain can be divided into large regions, each
composed of tens of millions of neurons with ``activity'' patterns
recorded using modalities such as functional magnetic
resonance imaging (fMRI) or electroencephalography (EEG). Various
levels also fall in-between the two extremes, often referred to as
the ``mesoscale''. Each scale is the host to different network
structures, elemental components, and connectivity patterns. This
heterogeneity of spatial scales alone makes it infeasible to study the
brain and all its functions and emerging phenomena using a single
computational model.

Both microscale models, illustrating the voltage dynamics of
individual neurons, and macroscale models, showing aggregate
connectivity between regions, have been the subject of much research
using computational models. Details on microscale models can be found
in~\cite{ALH-AFH:52,GBE-DHT:10,EMI:07,PD-LFA:01} and references
therein. For macroscale models, we direct the reader
to~\cite{EN-MAB-JS-LC-EJC-XH-ASM-GJP-DSB:24,MB:17} and references
within.  Interestingly, while the microscale dynamics at the neuronal
level must be nonlinear, at the macroscale this is not
necessarily the case. Despite the frequent assumption that accurate
brain models must be
nonlinear~\citep{XL-DC-LM-TMM-HB:11,KES-LK-LMH-JD-HEMdO-MB-KJF:08}, the
recent comparison~\cite{EN-MAB-JS-LC-EJC-XH-ASM-GJP-DSB:24} of a large
variety of linear and nonlinear macroscopic models did not find any
advantage in the latter.

In this work we are interested in discussing computational brain
models at the mesoscale, describing the interaction between
populations of neurons each having similar function and statistical
properties. Two main types of mesoscale models are local field
potential models (LFPs) and firing rate models. LFP models are based
upon measuring the electric potential in the extracellular space
around neuron populations, while firing rate models measure the
average firing rate of all the neurons within a
population~\citep{PD-LFA:01}. Both LFPs and firing rate models have
been used extensively for studying brain
function~\citep{GTE-CK-NKL-SP:13,SK-AB-AF-JR:19,TS-AVC:19} and, in
their simplest forms, can be transformed into each other through an
affine transformation~\citep{JF-KPH:96}. In this work, we study firing
rate models, in particular the linear-threshold rate model, and its
application to multiple brain functions.


In addition to the spatial scale of information in the brain, there
exist vast temporal differences between distributed neural
processes. Each region/circuit in the brain operates on its own
timescale, that might be different to the timescales of other
regions and circuits.
As such, when considering a model at any scale of spatial information,
it is necessary to encode a timescale into the model dynamics of each
region, and examine the impact in the dynamical behavior of their
relative differences.

\begin{figure}[htb]
\centering
  \includegraphics[scale=0.25]{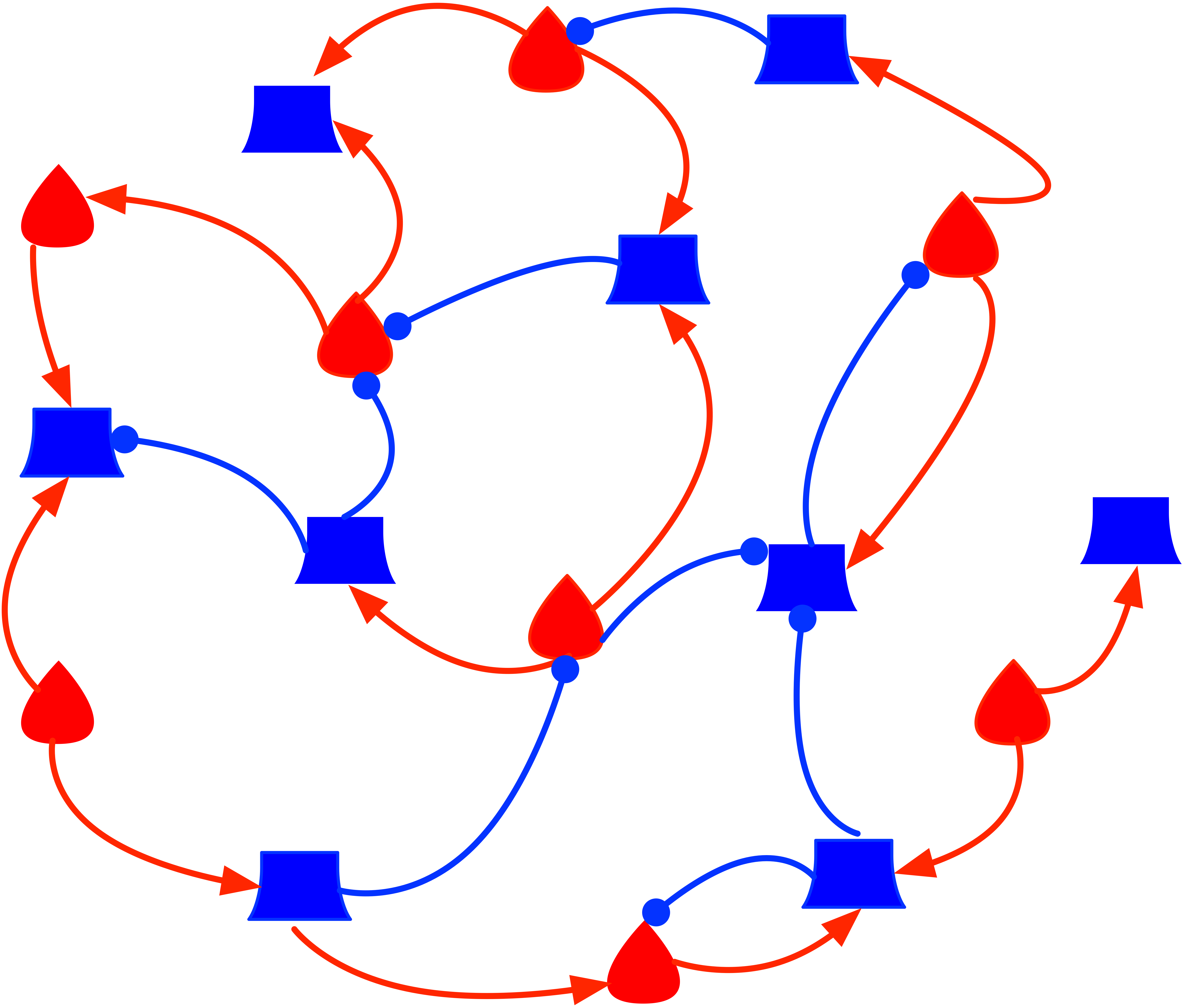}
  \caption{Graph-theoretic model of a brain network. Excitatory
    neurons and connections are shown in red, while inhibitory neurons
    and populations are shown in blue.}\label{fig:single_brain_layer}
\end{figure}

Regardless of the scale of the brain network model, they all have
basic graph-theoretic elements in common. A brain network is modeled
as a collection of nodes, with nodes representing individual neurons,
populations of neurons, or brain regions, depending on the scale of
the model. Each of these nodes has its own defining properties (e.g.,
consisting of excitatory or inhibitory neurons) and they are connected
to form a network structure as shown in
Figure~\ref{fig:single_brain_layer}. Specific model aspects
(functional forms, parametric constraints, etc.) are then defined in
accordance with the information scale of the model. Our next section
describes the particular assumptions made to describe firing rate
models. The reader familiar with these models can safely skip this discussion.

\subsection{Firing Rate Models}\label{sec:firing_rate_models}

\begin{figure}[htb]
\centering
  \includegraphics[scale=0.15]{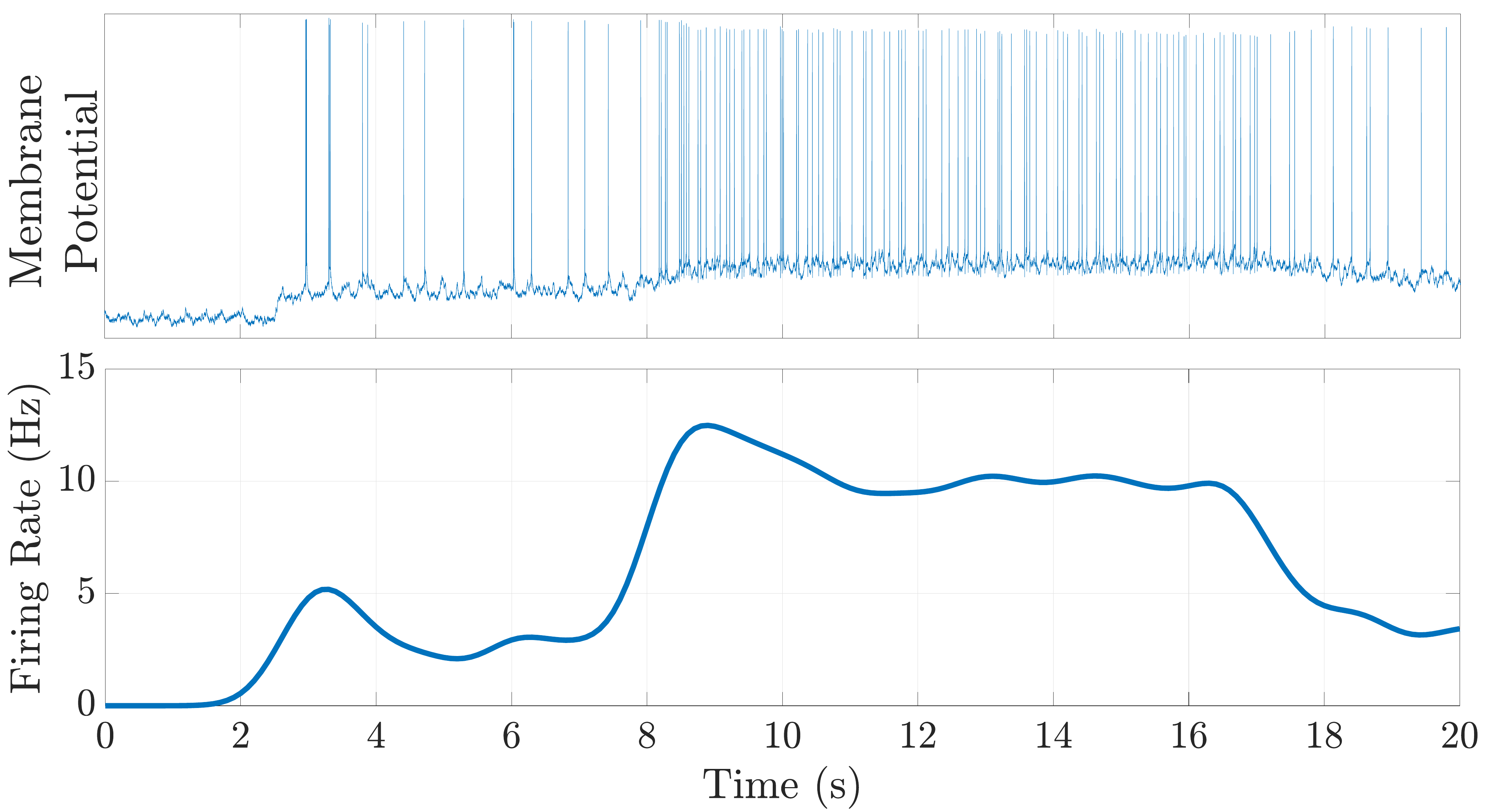}
  \caption{An intracellular recording showing a spike train as is used
    for communication between neurons (top) and the corresponding
    firing rate (bottom), estimated by binning spikes in
    $100^{\mathrm{ms}}$ bins and smoothing using a Gaussian window
    with $500^{\mathrm{ms}}$ standard
    deviation~\cite{EN-JC:21-tacI,DAH-ZB-JC-MAM-KDH-GB:00,DAH-KDH-ZB-JC-AM-HH-AS-GB:09-crcns}.}\label{fig:spike_trains_firing_rates}
\end{figure}

In this section we outline the construction of firing rate models in
the brain, as per~\cite[$\S$ 7]{PD-LFA:01}. At the level of neurons,
brain dynamics consist of a series of spikes, or action
potentials, being transmitted between neurons, see
Figure~\ref{fig:spike_trains_firing_rates}. The spike train is
transmitted from one neuron to another at a \emph{synapse}, and as
such the two neurons are referred to as the \emph{pre-synaptic} and
\emph{post-synaptic} neurons. The sequence of spikes (both input and
output signals) transmitted between neurons is defined by a neural
response function, $\rho(t)$, modeled as an impulse train of the form
$ \rho(t) = \sum_{k} \delta(t-t_k)$, with $\delta$ denoting the Dirac
delta function.

In many areas of the brain, the spike trains defined by the neural
response function appear to be highly random, and observations have
little trial-to-trial reproducibility, which makes accurate spike
train models difficult to construct. Replacing the neural response
function with the average firing rate provides more trial-to-trial
reproducibility (see Figure~\ref{fig:spike_trains_firing_rates}),
along with providing some other benefits. First, spiking models can
only accurately predict spike trains sequences if all inputs into a
neuron are known. Given the complexity of the brain with billions of
neurons, knowing this is highly unlikely.
Second, 
the probability of any two randomly selected neurons being connected
is low. Hence, the construction of a network model that has a high
degree of connectivity while maintaining this property requires using
a large number of nodes. Therefore it is standard practice to instead
model a single node in a network model as the average response of a
population of neurons.
%
%
This allows for a less sparse network model.
When using spike trains, this practice makes it difficult to describe
what the average response of the population is.  The use of
firing rates instead allows us to specify the average response simply
as the average firing rate of the neurons within the population.

We next briefly explain how the firing rate model is constructed.
First, we determine how the total synaptic input of a neuron depends
on the firing rates of its pre-synaptic afferents. Consider a pair of
pre- and post-synaptic neurons, with firing rates given by
$x_{\pre}(t)$ and $x_{\post}(t)$. Then, the firing rate of the
pre-synaptic neuron generates the synaptic input into the
post-synaptic neuron in the form of an electrical current, denoted
$I_{\post}(t)$. Assuming the synapse has fast dynamics, $I_\post(t)$
is approximately proportional to $x_\pre(t)$ with proportionality
constant $w_{\post,\pre}$, where $w_{\post,\pre}$ is known as the
synaptic weight.  The pre-synaptic neuron is excitatory if
$w_{\post,\pre} > 0$ and is inhibitory if $w_{\post,\pre} < 0$. As
such, an excitatory neuron increases the activity of its
out-neighbors while an inhibitory neuron decreases it. We note that
excitation and inhibition is a property of neurons, rather than
synapses, so a neuron either excites or inhibits all of its
out-neighbors, but not a combination (this is known as \emph{Dale's
  law}).  The synaptic current of a neuron that receives multiple
synaptic inputs follows a superposition law, with
\begin{align}\label{eq:I_post_sum}
  I_\post(t) = \sum_j w_{\post,j}x_j(t),
\end{align}
where the sum is taken over the neurons providing inputs.

\begin{figure}
\centering
  \includegraphics[width = 0.4\linewidth]{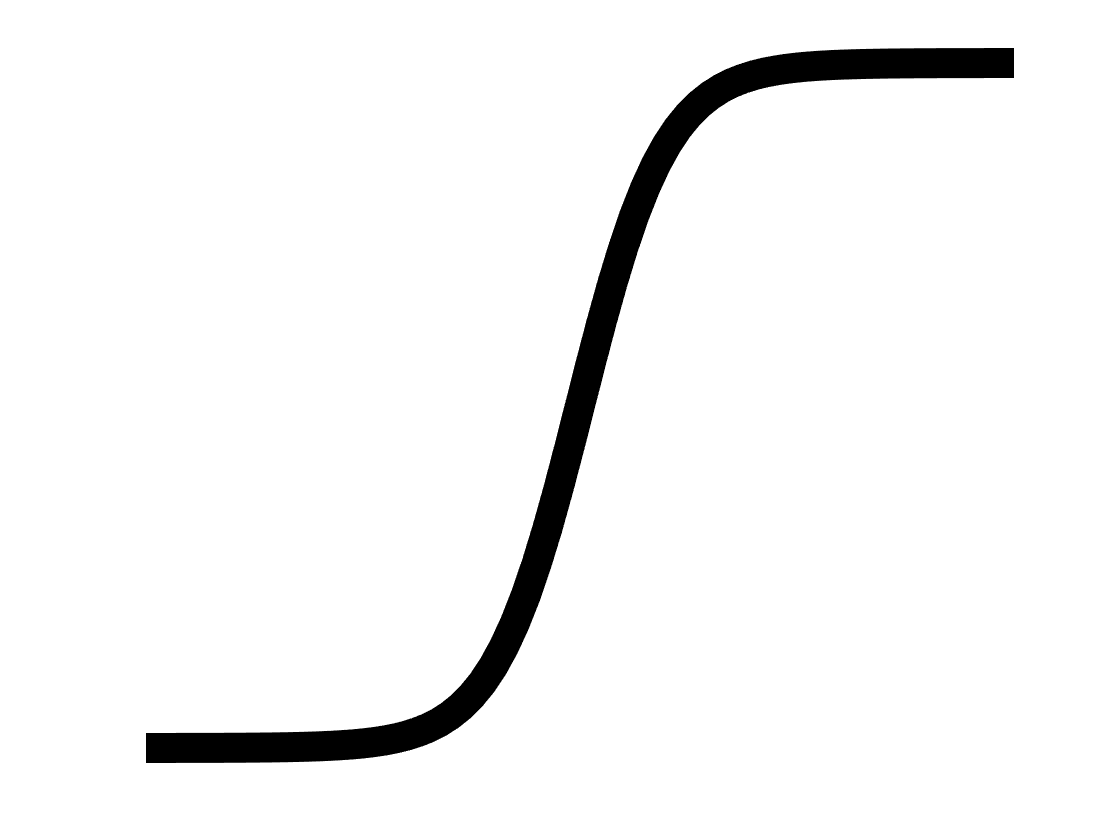}
  \includegraphics[width = 0.4\linewidth]{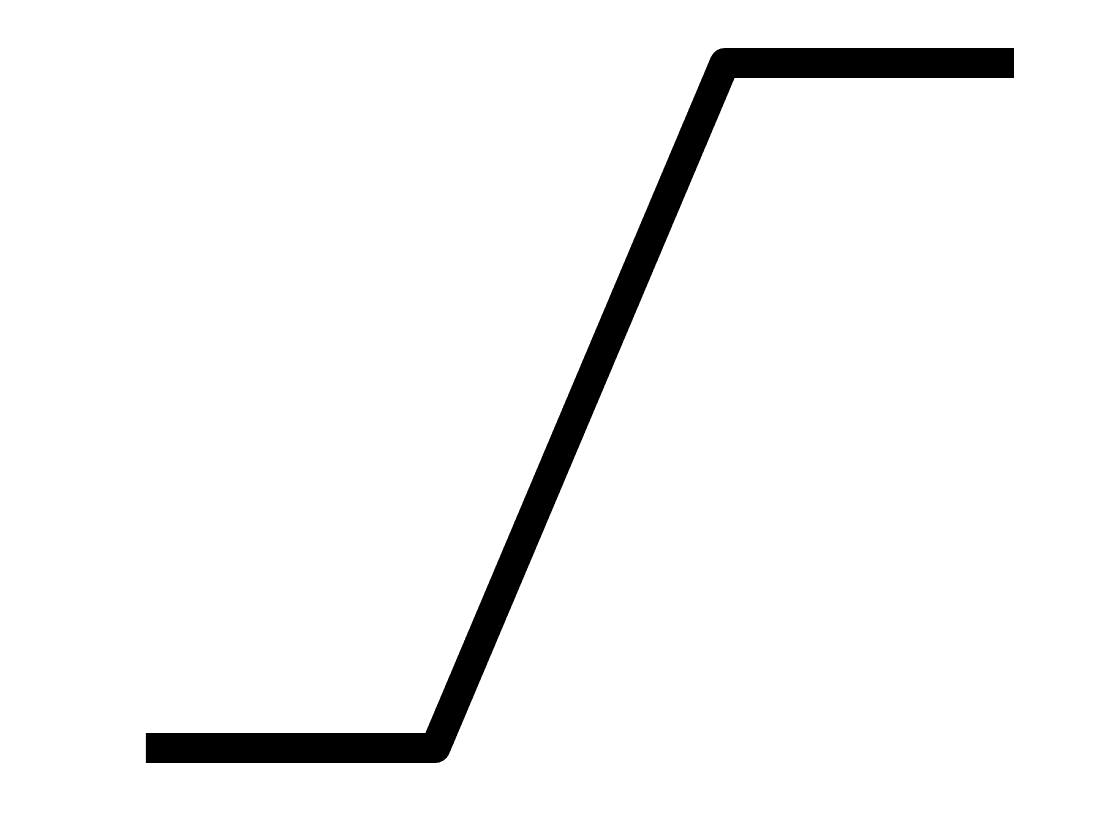}
  \caption{The sigmoidal (left) and linear-threshold (right) activation functions are commonly used for defining firing rate models.}\label{fig:activation_functions}
\end{figure}

Next, we model how the firing rate of the post-synaptic neuron depends
on the synaptic input as $ x_\post(t) = F(I_\post(t))$, where
$F(\cdot)$ is a nonlinear ``activation'' function. While a variety of
functions can be used for $F$, the most commonly used are sigmoidal
and linear-threshold functions, shown in Figure~\ref{fig:activation_functions}.
%
Both functions are bounded below, as required for the interpretation
that a negative firing rate is physiologically impossible. In
addition, they are both bounded above, which is relevant for
stabilizing the network against excessively high firing rates. Here,
we consider a linear-threshold activation function, giving
$x_\post = [I_\post]_0^m$, with $m$ being the upper bound on the
firing rate.
The resulting firing rate dynamics from this choice of activation
function is
\begin{align}\label{eq:firing_rate_model_ind_neurons}
  \tau\dot{x}_\post(t) = -x_\post + [I_\post(t)]_0^m,
\end{align}
where $\tau$ is a timescale constant indicating a ``lag'' between the
change in the synaptic input and the change in the firing rate. For
simplicity, we assume that $I_\post(t)$ is measured in Hz, as such it
has a multiplicative constant converting it from the traditional
current unit of amperes. In addition, this makes the synaptic weights
$w_{\post,j}$ dimensionless constants.

While this derivation of a firing rate model as described above uses
individual neurons, it is common to replace the individual neurons
with populations of neurons with similar activation patterns,
resulting in a firing rate model at the mesoscale. In this case, the
firing rates $x_\pre$ and $x_\post$ represent the average firing rate
of the population of neurons. Finally, to move from the dynamics of a
single pair of neurons (or populations of neurons) to the dynamics of
a brain network, we take the following steps. We consider a network
with $n$ nodes and let $\x \in \R^n$ represent the firing rates of the
nodes. Then, combining the synaptic weights $w_{i,j}$ into a matrix
$\W$ and using~\eqref{eq:I_post_sum}
and~\eqref{eq:firing_rate_model_ind_neurons}, we obtain the network
linear-threshold dynamics
\begin{align}\label{eq:network_lin_threshold_dynamics}
  \tau\dot{\x} = -\x + [\W\x + \d(t)]_\zero^\m.
\end{align}
The term $\d(t)$ is added to the synaptic input to model external
inputs to the network, such as un-modeled background activity, inputs
from other parts of the brain or external sources, or non-zero
thresholds. We note that the vector of firing rate upper bounds $\m$
can be modeled as either finite or infinite. While the finite model is
biologically sensible, if one assumes that the activity of the network
does not approach the typical firing rate threshold, then using an
infinite bound (and therefore removing the upper threshold) can be
convenient for network analysis.

\subsection{Modeling Different Brain Regions}
Since distinct functions involve different areas in the brain, it is
important for models to accommodate different structures to describe
multiple phenomena and behaviors. A majority of the literature on
brain networks studies the cortex and cortical
networks~\citep{GTE-CK-NKL-SP:13,QL-AU-BH:17,MK-RN-FG-MK-YI-TA-MS:21,JC-UH-CJH:15}
due to its role in higher-level processes in the brain, including
memory and attention~\citep{MB-BC-MAP:20}.  However, most cortical
regions have inputs from subcortical areas, including the thalamus,
that play critical roles in the many functions undertaken by the
cortex~\citep{SMS-RWG:06}, see Box~\ref{box:thalamus} for details.

\begin{myblock}{The Thalamus: More than a Relay
    Station}\label{box:thalamus}
  The thalamus is a component of many different brain networks and
  thus plays a role in a large number of functions. Traditionally, the
  thalamus has been considered primarily as being a sensory relay to
  the cortex, playing minimal other functional
  roles~\citep{MW-SDV:19}. Despite being known since the late
  $19^{\text{th}}$ century to have additional functions such as a role
  in memory loss~\citep{HG:96}, up until recently the majority of the
  research into thalamic function has studied its function as a
  sensory relay~\citep{SMS-RWG:06,EA-TO:15,YBS-SK:11}.  This view was
  gradually changed by several works, including the pioneering
  work~\citep{SMS-RWG:96}, establishing the thalamus as a
  heterogeneous structure only a small portion of whose nuclei play
  the role of a sensory relay. Following this work, research has shown
  that thalamus plays a role in learning and memory~\citep{ASM:15},
  attention, impulse control and
  decision-making~\citep{ASM-SMS-MAS-RGM-RPV-YC:14,FA-VF-ARM-EJK-EC-WM:18},
  and feedforward inhibitory control of cortical
  regions~\citep{ASM-SMS-MAS-RGM-RPV-YC:14,JMA-HAS:15,LG-SPJ-DEF-MC-MS:05,MMH-LA:16},
  among others.

  Thalamic nuclei can be broadly divided into two categories depending
  on their role in these applications. Nuclei involved mainly in the
  role of the thalamus as a sensory relay are known as
  \emph{first-order} or \emph{specific} and receive their input from
  other subcortical structures. The thalamic nuclei involved in the
  variety of other brain functions associated with the thalamus are
  referred to as \emph{higher-order} or \emph{non-specific} and
  receive their inputs from varying cortical
  regions~\citep{SMS:12}. The higher-order nuclei are then able to
  directly elicit activity in cortical regions based on the modulation
  of inputs to the
  cortex. 
  These divided roles in thalamic function motivate the construction
  of separate models for explaining their functions in brain networks.

  One mechanism for the interaction of thalamus with cortical regions
  is through feedforward inhibitory
  control~\citep{MMH-LA:16,LG-SPJ-DEF-MC-MS:05,SJC-TJL-BWC:07}. As
  such, when studying thalamocortical networks, models need to take
  into account that many of the net impacts from the thalamus onto
  cortical regions are inhibitory,
  while returning connections are both excitatory and inhibitory. In
  the context of modeling the brain from a control-theoretic
  perspective, the thalamus provides an exciting avenue of study as an
  internal controller, modulating and transferring information between
  cortical regions in parallel to cortico-cortical transmissions.
\end{myblock}


Given the various functions in which distinct brain regions are
involved, homogeneous modeling of all brain regions can be overly
simplistic. In order to account for the differences in regional
properties, we can impose restrictions at various levels, such as
functional forms, hyper-parameters, or parameters, on the sub-models
used for different regions. In this work we assume a homogeneous use
of the firing rate model with a linear-threshold functional form as
derived in Section~\ref{sec:firing_rate_models}, and encode regional
heterogeneity into the structure of the synaptic weight matrices
making up the sub-model corresponding to each region. The following is
a description of the constraints we impose on the models of cortical
and thalamic regions, respectively.

\emph{The cortex} is composed of a mix of excitatory and inhibitory
neuron populations and, while excitatory neurons significantly
outnumber inhibitory neurons, both play important roles in the
transmission and processing of
information~\citep{MK-RN-FG-MK-YI-TA-MS:21}. As
such, 
we allow our firing rate model for cortical regions to be composed of
populations of excitatory and inhibitory neurons with arbitrary
numbers and connectivity patterns. We only restrict synaptic weight
matrices of cortical regions such that outgoing connections from each
population are all either excitatory or inhibitory. This is reflected
in the matrices such that each column has either nonnegative or
nonpositive values~\citep{PD-LFA:01}. Within the cortex different stimuli are processed in different areas and at different rates. Regions closer to the sensory areas process information faster than those further away, creating distinct temporal hierarchies for stimuli such as visual and auditory~\citep{SJK-JD-KJF:08}. Within the model these hierarchies are reflected by different neurons having different timescales, $\tau$, which combined with the network topology dictate the location of a neuron within the hierarchy.
%
%

The \emph{thalamus} connects with cortical regions through a series of
parallel pathways, with most thalamic nuclei projecting to a unique
cortical population~\citep{KG-JMD:20}. However, lateral connections
within the thalamus (including both excitatory and inhibitory
populations, the latter of which lying primarily in the thalamic
reticular nucleus) construct the transthalamic pathways between
cortical regions that can lie in different places within a
hierarchical structure in the
cortex~\citep{SMS-RWG:06,SMS:12}. Experimental observations indicate
that along these pathways one of the mechanisms through which the
thalamus and cortex interact is feedforward inhibition mediated by
local
interneurons~\citep{LG-SPJ-DEF-MC-MS:05,SJC-TJL-BWC:07,KD-JT-ZJH-BL:15}.

In particular, these observations show that the cortex receives
  excitatory thalamic input but is inhibited due to connections
  between the thalamic input and inhibitory interneurons both for
  first-order sensory thalamic
  nuclei~\citep{JTP-CKJ-AA:01,HAS:02,EKK-JV:02} and higher-order
  thalamic nuclei~\citep{KD-JT-ZJH-BL:15}.

Given the complexity of thalamic structure, we make the following
simplifying assumptions towards its computational modeling. First, we model the thalamus as a single region within the model that can project to any cortical population. With this assumption we are including multiple nuclei within a single region and lateral connections between the thalamic nuclei are included in the internal dynamics. Second, we model the projections (outgoing connections) of the thalamus onto the cortical
regions as being strictly inhibitory, mimicking the above-cited experimental observations of feedforward inhibition of the cortex by the thalamus 
while also simplifying the
model. 
We note that the connections back from the cortical regions to the
thalamus are allowed to be both excitatory and inhibitory. Finally,
the internal dynamics of the thalamus are restricted only such that
each column has a nonpositive or nonnegative sign, similarly to the
cortical regions\footnote{We note that while in our models we
  restrict the sign pattern of the synaptic weight matrices for
  biological reasons, the mathematical results provided are valid for
  matrices with arbitrary sign patterns.}.

\subsection{Control of Brain
  Models}\label{sec:control_mechanisms_brain_models}
Brain models in general, and the simplified and tractable form of
firing rate models in particular, are natural pathways to the study of
control mechanisms of and for the brain.  Akin to engineered systems,
the types of control utilized in the brain can be (roughly) separated
into feedback and feedforward. Feedback control operates off of
circuits where the populations providing the control input are
directly stimulated by the populations within the network. In these
circuits, the magnitude of the control input is directly dependent on
the activity level in the network. On the other hand, feedforward
control is not dependent upon current
activity levels within the network and is based upon input received from populations of neurons that
may be further from the network.

\begin{figure}
\centering
  \includegraphics[scale=0.5]{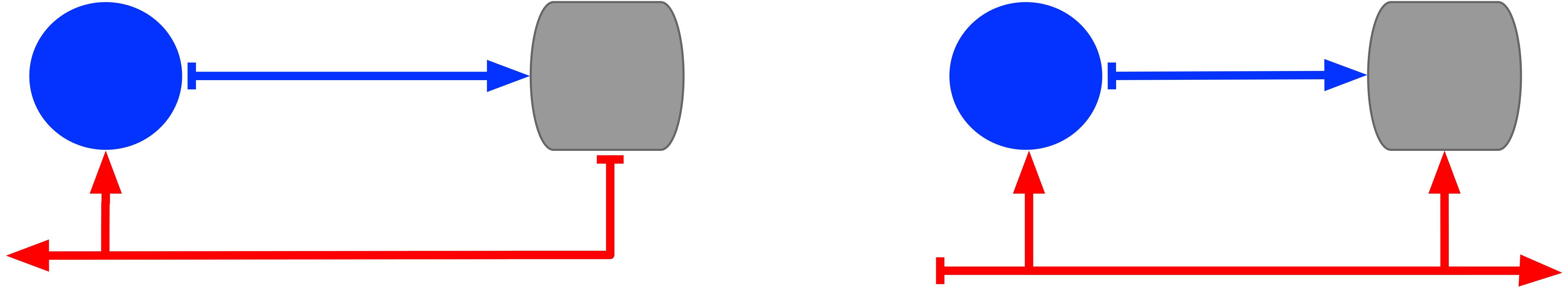}
  \caption{Feedback and feedforward mechanisms of control within brain
    networks.  The left panel shows an inhibitory feedback loop, where
    an excitatory signal (red) from the main neuronal population
    (grey) stimulates an inhibitory interneuron (blue), which in turn
    inhibits the main population. The right panel shows the
    feedforward inhibition mechanism, where an excitatory signal from
    a separate neuronal population stimulates an inhibitory
    interneuron as well as the main population. The interneuron then
    inhibits the main population, typically resulting in its net
    inhibition.}\label{fig:control_mechanisms}
\end{figure}

\emph{Feedback control} is a mechanism based upon the interaction of
two neuronal populations that form a closed loop.  While feedback
exists across the brain, a large component of feedback control occurs
in local feedback loops.
In the feedback circuit, the first population stimulates a second
``control'' population, which in return stimulates the first
population in order to control its dynamics. As the ``control'' neuron
population can be either excitatory or inhibitory, both excitatory or
inhibitory feedback control exists within the brain. However, despite
the existence of more excitatory than inhibitory neurons, the
inhibitory neurons frequently exhibit higher firing rates and are able to
influence the firing rates of other neuronal populations more than
excitatory populations can, cf.~\cite{MK-RN-FG-MK-YI-TA-MS:21}. As
such, inhibitory feedback control is more common than excitatory
feedback~\citep{JSI-MS:11}. Figure~\ref{fig:control_mechanisms}(left)
illustrates a standard inhibitory feedback loop.



\emph{Feedforward control} is more often studied in the context of
(potentially unidirectional) non-local connections between neuronal
populations. In the case of cortical populations, for instance, they
receive afferents from subcortical nuclei (i.e., the thalamus) as well
as cortical populations in distant regions. These long-distance
connections may not form clear feedback loops, but instead provide a
feedforward control input that can modify the dynamics of the
receiving neuronal population. While long-range connections in the
brain are almost universally excitatory, they can indeed induce
feedforward inhibition by exciting inhibitory ``interneurons''
(neurons with only local output connections), which in turn inhibit
their downstream neuronal population. If this two-hop inhibition is
stronger than the direct excitatory afferent received by the
downstream population, as is commonly the case, a net feedforward
inhibition would occur~\citep{JSI-MS:11}.
Figure~\ref{fig:control_mechanisms}(right) illustrates the feedforward
inhibition mechanism.

\section{Analysis of Individual Brain Regions}\label{sec:single-region}
As outlined in the prior section, at any scale the brain is composed
of multiple regions. However, in order to understand the behavior of a
model of the overall network, it is beneficial to start by examining
the function of the model in individual regions. In this section, we
examine the dynamical systems modeling of GDSA, epileptic seizure
activity
%
 %
and declarative memory using linear-threshold dynamics (only) of a
single region. In the next section, we address more complex scenarios
involving networks of networks modeling the interconnection of
multiple brain regions.  The applications considered here relate to
three main properties of the linear-threshold rate dynamics: selective
inhibition and recruitment in GDSA relates to stabilizability,
epileptic seizure activity
 %
%
relates to bifurcations, and our models of declarative memory are
based on multistability of neural dynamics.

We consider a brain region composed of $n$ nodes, with each node
representing a population of either excitatory or inhibitory neurons,
and governed according to the linear-threshold dynamics
in~\eqref{eq:network_lin_threshold_dynamics}.  The network structure,
cf.  Figure~\ref{fig:single_brain_layer}, is encoded by the synaptic
weight matrix~$\W$.  We begin by formalizing the problem of selective
inhibition and recruitment within the framework of linear-threshold
firing rate dynamics.

\subsection{Goal-driven Selective Attention in a Single Brain
  Region}\label{subsec:selective_inhib_single}
\begin{figure}[tbh]
\centering
  \includegraphics[scale=0.5]{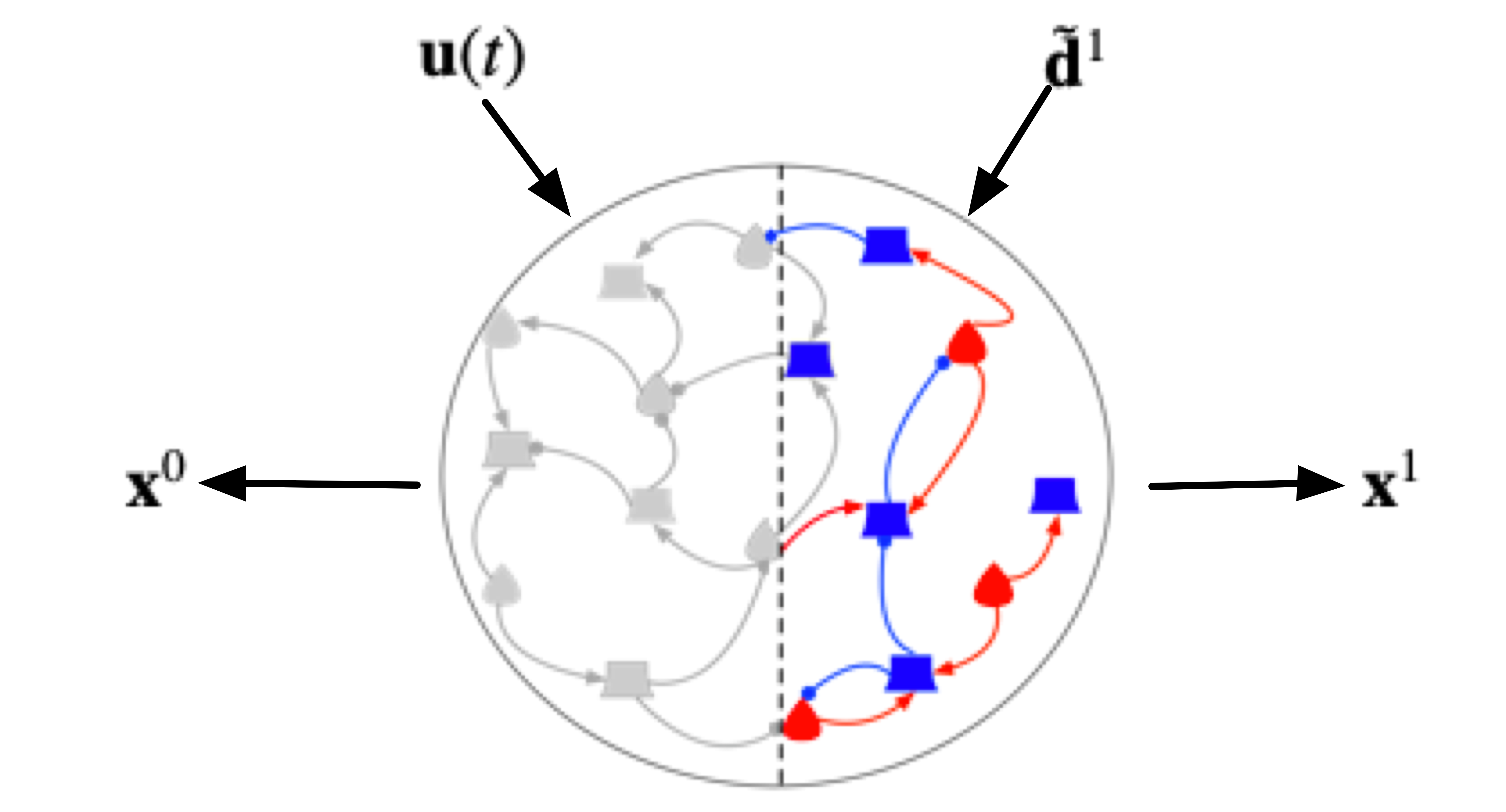}
  \caption{Brain region divided into task-irrelevant (grey) and
    task-relevant (red and blue) neuron populations. The
    task-irrelevant nodes make up $\x^0$ in the partition of the
    state, while the task-relevant nodes form $\x^1$. The control
    input $\u(t)$ is used to selectively inhibit the task-irrelevant
    populations, while the input $\tilde{\d}^1$ recruits the
    task-relevant populations to an
    equilibrium.}\label{fig:inhibition_recruitment_with_inputs}
\end{figure}

Selective inhibition and recruitment is the process of identifying
task-irrelevant and task-relevant stimuli, and suppressing the
task-irrelevant ones while processing the task-relevant ones. Since
different stimuli are, to a first approximation, processed by distinct
populations of neurons~\citep{RD-JD:95,JM-RD:85,BCM:93,NL:05} (a fact
closely related to the sparseness of the neural code,
see~\cite{FP:03,MI-AR-DK-TR-JMR:17}), this can be rephrased as
inhibiting the populations of neurons associated with the
task-irrelevant stimuli and recruiting the populations associated with
task-relevant stimuli. As such, for the purpose of considering GDSA
from a model-based perspective, we partition both the state variables
$\x \in \R^n$ and the synaptic weight matrix $\W$ based upon the
task-irrelevant and task-relevant nodes, as follows
\begin{align}\label{eq:partition_state_variables_and_matrix}
  \x(t) = \begin{bmatrix}
    \x^0(t) \\
    \x^1(t)
  \end{bmatrix} \qquad \W = \begin{bmatrix}
    \W^{00} & \W^{01} \\
    \W^{10} & \W^{11}
  \end{bmatrix}.
\end{align}
Then, inhibiting the task-irrelevant stimuli corresponds to driving
$\x^0$ to $\zero$, whereas processing the task-relevant stimuli
corresponds to driving $\x^1$ to a desired attractor $\x^1_*$. While
the ensuing framework is generalizable to arbitrary attractors, for
simplicity of exposition, in our treatment we consider $\x^1_*$ to be
a desired equilibrium. Accordingly, we also partition the input
$\d(t)$ from~\eqref{eq:network_lin_threshold_dynamics} as
\begin{align*}
	\d(t) = \B\u(t) + \tilde{\d}(t),
\end{align*}
where
\begin{align*}
  \B = \begin{bmatrix}
    \B^0 \\
    \zero
  \end{bmatrix} \qquad \tilde{\d}(t) = \begin{bmatrix}
    \zero \\ \tilde{\d}^1(t)
  \end{bmatrix},
\end{align*}
resulting in the dynamics
\begin{align}\label{eq:full_lin_thresh_dynamics_for_selective_inhibition}
  \tau{\begin{bmatrix}
      \dot\x^0(t) \\
      \dot\x^1(t)
    \end{bmatrix}} = -\begin{bmatrix}
    \x^0(t) \\
    \x^1(t)
  \end{bmatrix} + \left[\begin{bmatrix}
      \W^{00} & \W^{01} \\
      \W^{10} & \W^{11}
    \end{bmatrix}\begin{bmatrix}
      \x^0(t) \\
      \x^1(t)
    \end{bmatrix} + \begin{bmatrix}
      \B^0 \\
      \zero
    \end{bmatrix}\u(t) + \begin{bmatrix}
      \zero \\ \tilde{\d}^1(t)
    \end{bmatrix} \right]_\zero^\m.
\end{align}
This decomposition allows us to separate the inhibition of the
task-irrelevant populations from the recruitment of the task-relevant
populations. The term $\B^0\u(t)$ allows for the inhibition of the
task-irrelevant populations through either an external control source
or connections with another controlling neuron population. Meanwhile,
the term $\tilde{\d}^1(t)$ allows for the recruitment of the
task-relevant populations by representing the information pathways
between different brain regions along with any additional unmodeled
activity impacting the equilibrium to which the task-relevant
components are recruited. Figure~\ref{fig:inhibition_recruitment_with_inputs} illustrates the
discussion above.

With these decompositions in place, the problem of selective
inhibition and recruitment of the network can now be formulated
mathematically as follows: under what conditions on the network
structure can we find control $\u(t)$ such that the network dynamics
are stabilizable to $(\zero,\x_1^*)$?

\subsubsection{Stability of Linear-Threshold Networks as a Function of
  Interconnection Structure}\label{sec:stability-conditions}
The ability to achieve selective inhibition and recruitment depends on
the stabilizability of the network through feedforward and feedback
control mechanisms.  Understanding stabilizability, in turn, requires
understanding the stability properties of the dynamics, and how the
interconnection structure of the network affects it.  Our ensuing
discussion tackles this point.

For simplicity, we assume that the input term is constant, $\d(t) =
\d$, before generalizing the discussion to time-varying $\d(t)$
subsequently. The stability properties of the dynamics are closely
related to its network structure, which in the linear-threshold brain
model is encoded into the synaptic connectivity matrix $\W$. As such,
we introduce the following notions  of matrix classes.

\begin{definition}\label{def:matrix_classes}
  A matrix $\W \in \R^{n \times n}$ is
  \begin{itemize}
  \item \emph{absolutely Schur stable} if $\rho(|\W|) < 1$ where $\rho(\cdot)$ is the spectral radius;
  \item \emph{totally $\mathcal{L}$-stable} ($\W \in \mathcal{L}$) if
    there exists $\bf{P} = \bf{P}^\top > \zero$ such that for all
    $\sigma \in \{0,1\}^n$
    \begin{align*}
      (-\eye + \W^\top \Sigma)\bf{P} + \bf{P}(-\eye + \Sigma \W) < 0,
    \end{align*}
    where $\Sigma = \mathrm{diag}(\sigma)$;
  \item \emph{totally Hurwitz} ($\W \in \mathcal{H}$) if all its
    principal submatrices are Hurwitz;
  \item a \emph{P-matrix} ($\W \in \mathcal{P}$) if all its principal
    minors are positive.
  \end{itemize}
\end{definition}

\begin{figure}[tbh]
  \centering
  \includegraphics[scale=0.4]{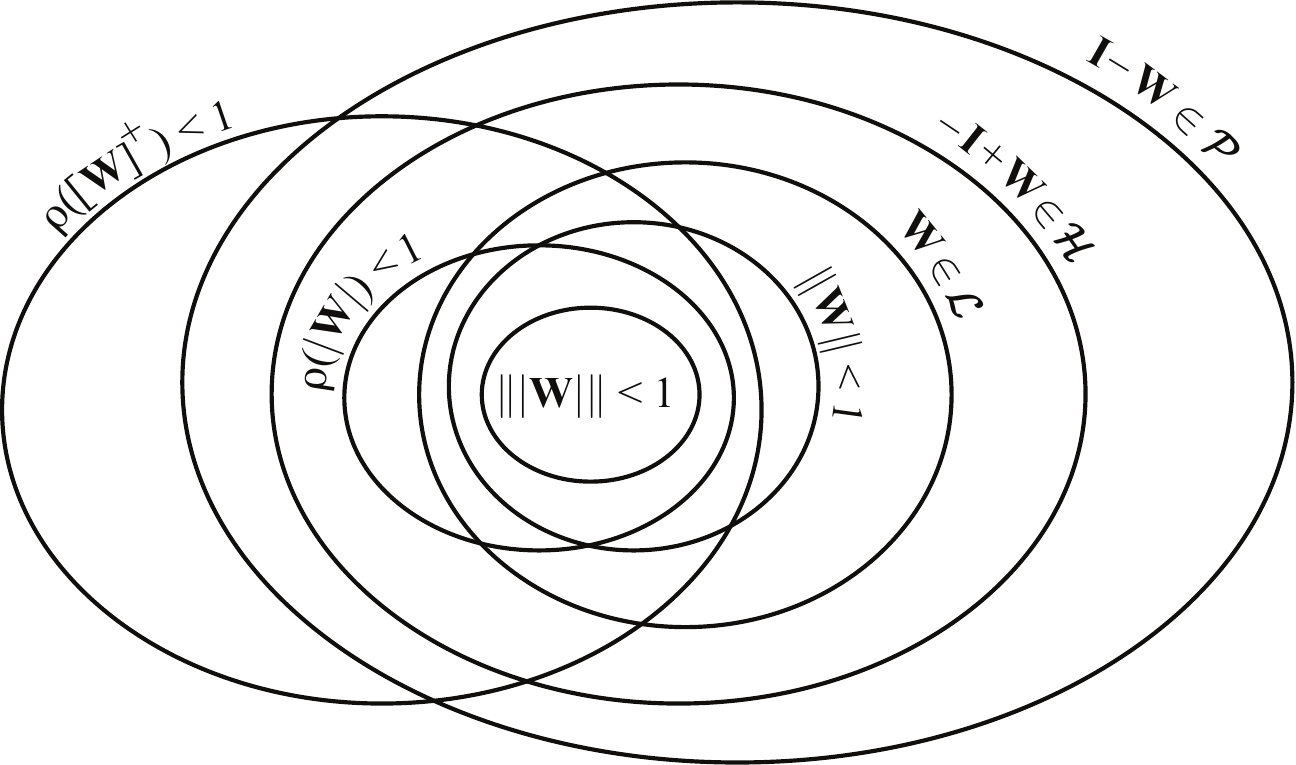}
  \caption{Inclusions satisfied by the different matrix classes in
    Definition~\ref{def:matrix_classes}.}\label{fig:matrix_class_inclusions}
\end{figure}

We note that these matrix classes are related to each other through a
variety of inclusions, as shown in
Figure~\ref{fig:matrix_class_inclusions}.

Using these matrix classes, we can provide conditions for both the
existence and uniqueness of equilibria (EUE) and global exponential
stability (GES) of the dynamics
\begin{align*}
  \tau\dot{\x} = -\x + [\W\x + \d]_\zero^\m.
\end{align*}
In particular, following~\cite{EN-JC:21-tacI}, the dynamics have a
unique equilibrium point if and only if $\eye-\W \in \mathcal{P}$, are
locally asymptotically stable if and only if $-\eye + \W \in
\mathcal{H}$, and are GES if $\W \in
\mathcal{L}$. Further,~\cite{EN-JC:21-tacI} conjectures that $-\eye +
\W \in \mathcal{H}$ is necessary and sufficient for GES. One challenge
with using these results, however, is that determining that matrices
are P-matrices or totally Hurwitz is computationally challenging,
especially as the scale of the network increases. As such, the
inclusions in the matrix classes from
Figure~\ref{fig:matrix_class_inclusions} take on importance as they
can be used to construct more conservative conditions (such as the
conditions based on the spectral radius and norm of $\W$) for EUE and
GES that are computationally tractable.

\subsubsection{Achieving Selective Inhibition and Recruitment through
  Feedforward and Feedback}~\label{subsubsec:selective_inhib_feedback_feedforward}
Given the importance of feedforward and feedback control for the
brain, cf. Section~\ref{sec:control_mechanisms_brain_models}, we
consider achieving selective inhibition and recruitment using both
mechanisms. Here, we characterize first each mechanism separately and
later discuss the advantages of the combinations thereof.

We begin by considering feedforward inhibition. In this case, the brain region under consideration is being inhibited
by a separate (not explicitly modeled here) brain region to an
activity/inactivity pattern of its choice through the control
input~$\u(t)$. Since the linear-threshold activation function is
unaffected by excessive inhibition due to its thresholding at zero,
selective inhibition through feedforward inhibition is possible by
using a sufficiently strong inhibitory input. Utilizing this, and the
conditions for stability of a linear-threshold dynamics described in
Section~\ref{sec:stability-conditions}, we can formalize the following
conditions.

\begin{theorem}\longthmtitle{Selective Inhibition and Recruitment
    through Feedforward Inhibition~\citep[Theorem
    V.2]{EN-JC:21-tacI}}\label{thm:feedforward_selective_inhib}
  Consider a brain region modeled with the linear-threshold dynamics
  in~\eqref{eq:full_lin_thresh_dynamics_for_selective_inhibition}. Suppose
  that $\dim(\u(t)) \geq \dim(\x^0)$ and that
  $ \range([\W^{00} \ \ \W^{01}]) \subseteq \range(\B^0)$.  Then, for
  any constant $\tilde{\d}^1$, there exists a constant feedforward
  control input $\u(t) = \bar{\u}$ that stabilizes the dynamics to the
  unique equilibrium of the form $(\zero,\x^*)$ if and only if
  $\W^{11}$ is such that the internal dynamics
  \begin{align*}
    \tau\dot{\x}^1 = -\x^1 + [\W^{11}\x^1 + \tilde{\d}^1]_\zero^{\m^1}
  \end{align*}
  is GES to a unique equilibrium.
\end{theorem}

We discuss the interpretation of the conditions of this result at the
end of the section, in parallel with the feedback inhibition case.  We
note that in Theorem~\ref{thm:feedforward_selective_inhib} the
potential for selective inhibition and recruitment is predicated on
the stability of the uncontrolled portion of the dynamics. This
illustrates that the properties required for selective inhibition and
selective recruitment can be fully separated. The ability for
selective inhibition is based upon the dynamics for the
task-irrelevant nodes, while selective recruitment is based upon only
the dynamics of the task-relevant nodes, with no intersection. In a
similar fashion, for selective inhibition and recruitment using
feedback inhibition, the conditions are dependent only upon the
structure of the task-relevant component of the synaptic weight
matrix, and are given as follows.

\begin{theorem}\longthmtitle{Selective Inhibition and Recruitment
    through Feedback Inhibition~\citep[Theorem
    V.3]{EN-JC:21-tacI}}\label{thrm:feedback_selective_inhib}
  Consider a brain region modeled with linear-threshold dynamics
  in~\eqref{eq:full_lin_thresh_dynamics_for_selective_inhibition}. Let
  the input $\u(t)$ be given by the linear feedback
  $ \u(t) = \K\x(t)$, where $\K$ is a constant control gain and
  suppose $\dim(\u(t)) \geq \dim(\x^0)$. Further, assume that
  $\range([\W^{00} \ \ \W^{01}]) \subseteq \range(\B^0)$. Then, there
  almost always\footnote{That is, the set of allowable matrix pairs
    $(\W,\B)$ for which no satisfactory matrix $\K$ exists, is of
    measure zero. }  exists a $\K$ such that
  \begin{enumerate}[i)]
  \item $\eye - (\W+\B\K) \in \mathcal{P}$ if and only if $\eye -
    \W^{11} \in \mathcal{P}$;
  \item $-\eye + (\W+\B\K) \in \mathcal{H}$ if and only if
    $-\eye+\W^{11} \in \mathcal{H}$;
  \item $\W+\B\K \in \mathcal{L}$ if and only if $\W^{11} \in \mathcal{L}$;
  \item $\rho(|\W+\B\K|) < 1$ if and only if $\rho(|\W^{11}|) < 1$;
  \item $\norm{\W+\B\K} < 1$ if and only if $\norm{[\W^{10}~\W^{11}]}
    < 1$.
  \end{enumerate}
\end{theorem}

The interpretation of each of the results in
Theorem~\ref{thrm:feedback_selective_inhib} is as follows.  First,
recalling that selective inhibition and recruitment occurs when the
system is stabilizable to a unique equilibrium $(\zero,\x^*)$,
Theorem~\ref{thrm:feedback_selective_inhib}$i)$ guarantees the
existence and uniqueness of the equilibrium $\x^*$ for the
task-relevant components. However, this condition does not guarantee
the equilibrium is stable, so it is not sufficient for selective
inhibition and
recruitment. Theorem~\ref{thrm:feedback_selective_inhib}$ii)$ then
guarantees local asymptotic stability (and potentially GES) of the
equilibrium point $\x^*$. The remaining conditions utilize the matrix
inclusions given in Figure~\ref{fig:matrix_class_inclusions} to
provide sufficient conditions for stabilizability of the task-relevant
conditions to the
equilibrium~$\x^*$.
Theorem~\ref{thm:feedforward_selective_inhib}$iii)$ guarantees GES of
the task-relevant components of the dynamics to~$\x^*$, thus providing
a sufficient condition, when combined with the assumptions of the
result, that selective inhibition and recruitment is achievable
through stabilization to $(\zero,\x^*)$. Condition $iv)$ does not
guarantee GES, as the matrix class $\rho(|\cdot|) < 1$ is not a subset
of the class of $\mathcal{L}$-stable matrices, but it is more
computationally tractable and does guarantee local asymptotic
stability. Finally, Theorem~\ref{thrm:feedback_selective_inhib}$v)$ is
computationally tractable, and is sufficient for guaranteeing GES of
the equilibrium $\x^*$, allowing for stabilizability to
$(\zero,\x^*)$.

To conclude this section, we discuss the assumptions in
Theorems~\ref{thm:feedforward_selective_inhib}
and~\ref{thrm:feedback_selective_inhib}. Both the feedforward and
feedback case require an assumption on the relationship between the
range of the task-irrelevant component of the synaptic weight matrix
and the task-irrelevant component of the inhibitory control
matrix~$\B^0$.  In both results, the purpose of this condition is to
guarantee the existence of a control allowing the result to hold.  In
the feedforward inhibition result, it guarantees the existence of the
constant control $\u$ used to achieve inhibition, while in the
feedback inhibition result, it guarantees the existence of the
matrix~$\K$. The second assumption is that the dimension of the control
is at least as large as the number of task-irrelevant nodes in the
system. Intuitively, this result guarantees that the control can
inhibit every task-irrelevant node in the system (it would not be
possible to achieve this if the dimension was smaller).

\subsection{Epileptic Seizures through Bifurcations in a Single Brain Region}\label{subsec:single-region-epilepsy}

Epilepsy is a neurological disease characterized by chronic unprovoked
seizures. While epilepsy can be caused by many different factors and
there are multiple types of seizures, seizures typically correspond to
a sudden change from healthy to unhealthy activity. As such, when
studying epilepsy from a dynamical systems perspective, the emergence
of seizures can be modeled through bifurcations in the
model~\citep{FLdS-WB-SNK-JP-PS-DNV:03,RAS-RGS-SST:12}. In this
section, we look at phrasing epileptic events using bifurcations in
the model of a single brain region. In particular, using
linear-threshold dynamics~\eqref{eq:network_lin_threshold_dynamics},
consider an excitatory-inhibitory pair (i.e., here $n=2$) as follows,
\begin{align}\label{eq:LTN_epilepsy}
  \tau\dot{\x} = -\x + [\W\x + \u]_\zero^\m,
\end{align}
where the synaptic weight matrix $\W$ and input $\u$ can be written as
\begin{align}\label{eq:EI_pair_matrices}
  \W = \begin{bmatrix}
    a & -b \\
    c & -d
  \end{bmatrix} \qquad \u = \begin{bmatrix}
    u_1 \\ u_2
    \end{bmatrix},
\end{align}
with $a,b,c,d \in \Rplus$ and $u_1,u_2 \in \R$. Since we are studying a single region, we are considering focal seizures, but in Section~\ref{subsec:epilepsy_spreading} we will consider the idea of generalized seizures through the spread of epileptic behavior across networks. 

Electroencephalography (EEG) is one of the most commonly used tools
for measuring and viewing brain activity, particularly for the
diagnosis and study of epilepsy. As such, when taking a dynamical
systems approach to studying epilepsy, it is common to assume that the
activity (output) of the dynamical system represents (abstracted and
simplified) EEG signals~\citep{CJS:05}. During both healthy and
unhealthy activity a variety of behaviors appear in the EEG
measurements. However, a number of types of waveforms have been commonly observed during epileptic seizures~\citep{EN-FHLdS:05}.
%
These waveforms can then be used as a basis to approximate an EEG response. One can then relate each of these waveform types to specific properties of a dynamical system
being used to model EEG signals~\citep{JT-FW-PC-OF:11}. The types of
waveforms and the dynamical system properties we use to model them are
summarized as follows.

\begin{figure}[tbh]
	\centering
  \includegraphics[width = 0.6\linewidth]{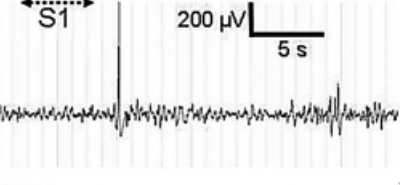}
	\includegraphics[width = 0.6\linewidth]{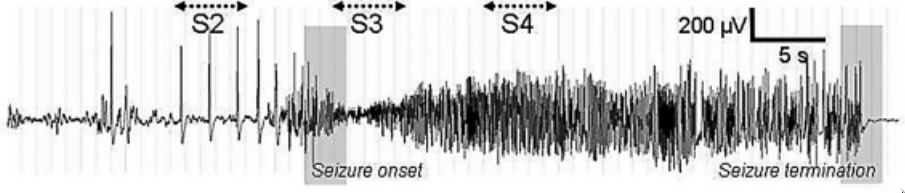}
  \includegraphics[width = 0.6\linewidth]{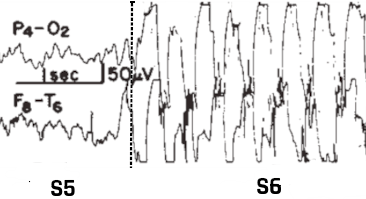}
  \caption{EEG recording of an epileptic
    waveform~\citep{PC:96}. Sections of the waveforms are labeled
    according to the activity types described in
    Remark~\ref{remark:waveform_types}. S1 corresponds to healthy
    background activity, S2 to low frequency spikes, S3 to irregular
    high frequency oscillations and S4 to quasi-sinusoidal low-frequency oscillations. S5 and S6 correspond to high frequency oscillations and slow waves,
    respectively~\citep{EN-FHLdS:05}.}\label{fig:epileptic_waveform_types}
\end{figure}

%
%
\begin{remark}\longthmtitle{Waveforms Observed in EEG
    Recordings}\label{remark:waveform_types}
    The following waveform types are commonly observed in EEG signals, and we discuss them based on the seizure shown in Figure~\ref{fig:epileptic_waveform_types}, with regions divided based on the qualitative changes in behavior.

  %
  %
  \begin{enumerate}
  \item Background activity (S1 in
    Figure~\ref{fig:epileptic_waveform_types}): 
    This is characterized by low-amplitude fluctuations or oscillations around a mean-centered steady state value. This activity generally lies in the theta band ($4-7$ Hz)~\citep{JT-FW-PC-OF:11}. In terms of a dynamical systems representation we model the background activity as the network having a equilibrium. In this case the model behavior will stay in the area of the equilibrium, fluctuating based upon noise in the system~\citep{JT-FW-PC-OF:11}.

  \item Epileptic Spiking (S2): These are isolated non-rhythmic spikes and are commonly observed sporadically prior to seizures during otherwise healthy background activity~\citep{JT-FW-PC-OF:11}. In modeling these from a dynamical systems perspective, they appear when the system is multistable, with one of the equilibria being the origin. The spikes then appear when the system activity moves into the orbit of the non-zero equilibrium~\citep{FC-AA-FP-JC:21-csl}.

  \item Irregular low amplitude oscillations (S3): These significantly higher frequency oscillations than those seen in background activity, with a dominant frequency in the range of $20-40$ Hz~\citep{JT-FW-PC-OF:11}. In the seizure shown in Figure~\ref{fig:epileptic_waveform_types} these appear at the onset of the seizure, with the brain activity suddenly and significantly increasing in frequency from the background activity and spikes seen in S1 and S2. We have two approaches to modeling this with a dynamical system. First, as oscillations they can be modeled as limit cycles in the dynamics~\citep{JT-FW-PC-OF:11}. The other option, due to the irregularity in the oscillations, is to model them as the dynamics being multistable, with the irregular oscillations appearing as the movement of the system between the orbits~\citep{FC-AA-FP-JC:21-csl}.

  \item Quasi-sinusoidal oscillations and rhythmic spiking (S4): 
      These are higher amplitde oscillations that begin at a high frequency but slow down into a rhythmic spiking pattern. The rhythmic spiking typically lies in the alpha band, with a dominant frequency near $10$ Hz~\citep{JT-FW-PC-OF:11}. In Figure~\ref{fig:epileptic_waveform_types} these appear as the seizure develops, with the activity slowing down from the the oscillations seen in S3. These are modeled using stable limit cycles in the dynamics~\citep{JT-FW-PC-OF:11}.

  \item Slow waves (S6): These are very low-frequency ($1-2.5$ Hz~\citep{EN-FHLdS:05}) high amplitude waves with intermittent spikes that typically only appear during sleep, but are also observed in
  epileptic patients both preceeding and following seizures~\citep{YH-ET:15}. We choose to model these using a combination of an unstable equilibrium and a limit cycle, with the spikes appearing as the system moves between these states~\citep{FC-AA-FP-JC:21-csl}.
    %
    %
  \end{enumerate}
  \end{remark}

With this relation between EEG waveforms and dynamical systems
properties we can now frame the problem of discussing seizures using
bifurcations and linear-threshold dynamics. First, what type of
bifurcations can occur in the model~\eqref{eq:LTN_epilepsy} and what
components of a seizure do they correspond with? Second, under what
structural conditions on $\W$ and input conditions on $\u$ can EEG
waveforms be achieved that qualitatively resemble what is observed
during epileptic seizures?

\subsubsection{Bifurcations in LTN Dynamics}
In order to relate bifurcations in the LTN model of a brain network to
the EEG activity preceding, during, and following epileptic seizures,
we first need to understand how and when
bifurcations occur in the dynamics. In this section we will explain
both the types of bifurcations that occur in LTN dynamics and
structural conditions under which they occur. We begin with a closer
look at the structure of the LTN dynamics.

One of the benefits of the linear-threshold network model is that the
nonlinearities in the model are switched-linear.
In particular, the dynamics form a piecewise-affine state-dependent
switched system, which allows for applying linear analysis within each
switching region. Further, this form allows for an elegant
characterization of bifurcations occurring in the network, as the
qualitative changes in behavior occur as network equilibria move
across switching boundaries.

At each point in time, each node in the network can be in one of three
states: inactive ($(\W\x+\u)_i \leq 0$), linearly active
($(\W\x + \u)_i \in (0,\m_i)$), or saturated
($(\W\x + \u)_i \geq \m_i$). Therefore at any point in time the
network state can be associated with a switching index
$\sigma = \sigma(\x) \in \{0,\ell,s\}^n$, where $0,\ell,s$ correspond
to a node being inactive, linearly active and saturated,
respectively. The indices then define the switching regions for the
linear-threshold dynamics as follows
\begin{align*}
  \Omega_\sigma = \left\{ \x \in [\zero,\m] ~\bigg|~
  \begin{cases}
    (\W\x+\u)_i \leq 0 & \text{if~} \sigma_i = 0,
    \\
    0 \leq (\W\x+\u) \leq \m_i & \text{if~} \sigma_i = \ell,
    \\
    (\W\x+\u)_i \geq \m_i & \text{if~} \sigma_i = s
  \end{cases}\right\}.
\end{align*}
%
%

%
%
Within each switching region $\Omega_\sigma$, we have that
$[\W\x + \u]_\zero^\m = \Sigma^\ell(\W\x + \u) + \Sigma^s \m$, in
which $\Sigma^{\ell}$ and $\Sigma^{s}$ are diagonal matrices with
$\Sigma^\ell_{ii} = 1$ if $\sigma_i = \ell$ and $\Sigma^{s}_{ii} = 1$
if $\sigma_i = s$. Applying this to the
dynamics~\eqref{eq:LTN_epilepsy}, we get the
piecewise-affine form take the form
\begin{align}\label{eq:piecewise_affine_LTN_epilepsy}
  \tau\dot{\x} = (-\eye + \Sigma^{\ell}\W)\x + \Sigma^\ell \u + \Sigma^s \m.
\end{align}

With the piecewise-affine dynamics, each input $\d$ and switching
region $\sigma$ corresponds with a unique equilibrium
candidate\footnote{In order to work with the piecewise-affine form of
  the dynamics we require the assumption that $\det(\W) \neq 0$ and
  $\det(-\eye + \Sigma^{\ell}\W) \neq 0$ for all
  $\sigma \in \{0,\ell,s\}^2$. This assumption is required for having
  well-defined equilibrium candidates, but is not a restriction as the
  set of matrices not satisfying the assumption has measure
  zero.~\citep{EN-JC:21-tacI}} $\x^*_\sigma(\d)$ given by
\begin{align*}
  \x_\sigma^* = (\eye - \Sigma^\ell \W)^{-1}(\Sigma^\ell\u + \Sigma^s\m).
\end{align*}
Each equilibrium candidate is then an equilibrium of the system if
$\x^*_\sigma$ actually belongs to~$\Omega_\sigma$. It is of interest to note that the switching regions $\Omega_\sigma$ are dependent on both the system structure and input. As such, as the input to the system varies the switching regions themselves are changing, which differs from standard state-dependent switched systems. These changing regions result in the dynamics exhibiting richer behavior but correspondingly complicates the analysis. With this form of
the dynamics in hand we are ready to discuss bifurcations in the
linear-threshold dynamics for the purpose of modeling seizures.

In the excitatory-inhibitory linear-threshold
network~\eqref{eq:LTN_epilepsy}, bifurcations arise due to the
piecewise-affine form of the network and are characterized as a
function of the network input, $\u$~\citep{FC-AA-FP-JC:21-csl}, which
is referred to as the bifurcation parameter. Bifurcations then have
the opportunity to occur when equilibrium candidates from multiple
switching regions coincide for a given value of the bifurcation
parameter. Such bifurcations are formalized as follows.

\begin{definition}\longthmtitle{Bifurcations in Linear-Threshold
    Networks~\citep{FC-AA-FP-JC:21-csl}}
  A bifurcation parameter $\u$ is called a \emph{bifurcation
    candidate} for the linear-threshold
  dynamics~\eqref{eq:LTN_epilepsy} if
  $\x_{\sigma_1}^*(\u) = \x_{\sigma_2}^*(\u)$ for
  $\sigma_1 \neq \sigma_2$. A \emph{boundary equilibrium bifurcation
    (BEB)} occurs at a bifurcation candidate $\u$ if
  $\x_{\sigma_1}^*(\u) \in \Omega_{\sigma_1}$ and
  $\x_{\sigma_2}^*(\u) \in \Omega_{\sigma_2}$.
\end{definition}

Now, while we have determined when a bifurcation exists and occurs
within the linear-threshold network, there are multiple types of
bifurcations. These correspond with different changes in dynamical
behavior of the network, and as such can be related to transitions
between different waveform types in the EEG. The types of BEBs are
defined as follows.

\begin{definition}\longthmtitle{Types of Boundary Equilibrium
    Bifurcations~\citep{FC-AA-FP-JC:21-csl}}\label{def:types_BEBs}
  If a boundary equilibrium bifurcation occurs at input $\u$ it is
  called a
  \begin{enumerate}
  \item a persistent BEB (P-BEB) if the number of equilibria is
    constant in a neighborhood of $\u$;
  \item a non-smooth fold BEB (NSF-BEB) if the number of equilibria is
    not constant in a neighborhood of $\u$;
  \item a Hopf bifurcation if it is locally a NSF-BEB such that a
    limit cycle emerges globally.
  \end{enumerate}
\end{definition}

With these types of bifurcations in hand, we are now able to classify
qualitative changes in the behavior of the linear-threshold
dynamics. However, to understand the dynamics, and to be able to use
it to model seizure behavior, it is desirable to have
conditions such that each type of bifurcations occur.

We will operate with only a single bifurcation parameter, rather than
the two inputs that are part of the model. That is, we will vary the
parameter $u_1$ to achieve bifurcations while maintaining $u_2$ to be
a constant\footnote{Note that all of the following results could be
  determined using $u_2$ as the bifurcation parameter while keeping
  $u_1$ constant.}. This is called a codimension $1$ bifurcation, and
occur more frequently in biological systems than higher-order
bifurcations~\citep{EMI:00}. The following set of inequalities provide
conditions for the types of bifurcations in
Definition~\ref{def:types_BEBs} to occur.

\begin{theorem}\longthmtitle{Conditions for Boundary Equilibrium
    Bifurcations~\citep{FC-AA-FP-JC:21-csl}}\label{thrm:conditions_BEBs}
  Consider an excitatory-inhibitory pair governed by the
  linear-threshold dynamics~\eqref{eq:LTN_epilepsy} with synaptic
  weight matrix~\eqref{eq:EI_pair_matrices}. Let $\u$ be the input and
  assume that $u_2$ is constant, while $u_1$ is the bifurcation
  parameter. Suppose that
  \begin{align*}
    -\m_1 c < u_2 < (1 + d)\m_2.
  \end{align*}
  The following inequalities result in four different bifurcation
  behaviors in the system:
  \begin{subequations}
    \begin{align}
      a &< 1, \label{eq:bifurcation_1}
      \\
      (a-1)(d+1) &< bc, \label{eq:bifurcation_2}
      \\
      a &< d+2. \label{eq:bifurcation_3}
    \end{align}
  \end{subequations}
  The bifurcation possibilities are as follows:
  \begin{enumerate}
  \item[(A)] If~\eqref{eq:bifurcation_1} is satisfied, there exists a
    unique equilibrium for every input $\v$ and all bifurcations are
    P-BEB.
  \item[(B)] If~\eqref{eq:bifurcation_1} and~\eqref{eq:bifurcation_2}
    are not satisfied, then the system has either one or three
    equilibria. Any bifurcations involving the $\Omega_{\ell\ell}$
    region and one other region are P-BEB. Any other bifurcations are
    NSF-BEB.
  \item[(C)] If\eqref{eq:bifurcation_1} is not satisfied
    while~\eqref{eq:bifurcation_2} and~\eqref{eq:bifurcation_3} are,
    then bifurcation candidates involving $\Omega_{00}$ or
    $\Omega_{ss}$ and one other region are P-BEB. Any other
    bifurcations are NSF-BEB.
  \item[(D)] If~\eqref{eq:bifurcation_2} is satisfied
    while~\eqref{eq:bifurcation_1} and~\eqref{eq:bifurcation_3} are
    not, then the bifurcations are the same as in case $(C)$ except a
    Hopf bifurcation occurs at $\u_{00}^{\ell0}$ and at
    $\u_{ss}^{\ell s}$.
  \end{enumerate}
\end{theorem}

From this result we see that bifurcation behavior in linear-threshold
networks is highly dependent on the structure of the network in
addition to the actual bifurcation parameter. As such, when
considering how to represent seizure behavior using the
linear-threshold model it is important to ensure that the specific
network considered has a structure that permits the wide variety of
behavior observed in the EEG of epileptic patients.

\subsubsection{Constructing Seizure Behavior with Linear-Threshold
  Bifurcations}
In order to use bifurcations in the network to represent seizure
behavior, we need to relate the types of bifurcations given in
Definition~\ref{def:types_BEBs} with the EEG waveforms from
Remark~\ref{remark:waveform_types}. This is done by matching the
dynamical systems properties of the EEG waveform types
from~\citep{JT-FW-PC-OF:11} with the properties of each bifurcation
types. Table~\ref{table:bifurcations_and_waveforms}  proposes a
relationship between the types of bifurcations in the network and the
types of waveforms that make up both healthy and epileptic EEG
behavior.


\begin{table}[tbh]
  \centering
  \begin{tabular}{| c| c| c| c|}
    \hline
    Network Structure & Initial Behavior & Bifurcation & Resulting Behavior \\
    \hline
    (A) & Healthy background & P-BEB & Healthy background \\
    \hline
    (B) & Healthy background & P-BEB & Healthy background \\
    \hline
    (C) & Healthy background	& NSF & Spikes \\
                      & Spikes & NSF & Healthy background \\
    \hline
                      & Healthy background & NSF & Spikes \\
    
    (D) & Spikes & Hopf & Oscillations \\
    
                      & Oscillations & Hopf & Slow waves \\
    
                      &Slow waves & NSF & Healthy background\\
    \hline
  \end{tabular}
  \caption{Relation between the type of bifurcation and the change in
    EEG waveform behavior for networks satisfying the structural
    conditions in Theorem~\ref{thrm:conditions_BEBs} as defined by
    Remark~\ref{remark:waveform_types}~\citep{FC-AA-FP-JC:21-csl}.}\label{table:bifurcations_and_waveforms}
\end{table}

From Table~\ref{table:bifurcations_and_waveforms}, we can see that
bifurcations in the excitatory-inhibitory network governed by
linear-threshold dynamics can result in all of the waveform types that
appear in the EEG from Remark~\ref{remark:waveform_types}.  Note,
however, that due to the structural constraints on bifurcation types
provided in Theorem~\ref{thrm:conditions_BEBs}, not all
excitatory-inhibitory pairs exhibit the range of EEG behaviors discussed in Remark~\ref{remark:waveform_types}, and as such are are not useful for exhibiting epileptic behavior.
%
%
This allows, in
at least two ways, a realistic representation of modeling epileptic
dynamics using LTN models. First, it does not make sense for any
arbitrarily constructed network to be a reasonable model for the
brain. As such the existence of structural requirements for a network
to be able to exhibit specific behavior is expected. Second, seizures
generally begin in specific areas of the brain and then either remain
there (focal seizures) or spread throughout the brain (generalized
seizures)~\citep{WTB:10}. This, together with the fact that not all
people exhibit epileptic symptoms further explains the existence of
specific structural conditions for the emergence of epileptic dynamics
in the model. 

\begin{figure}[tbh]
\centering
  \includegraphics[width = 0.8\linewidth]{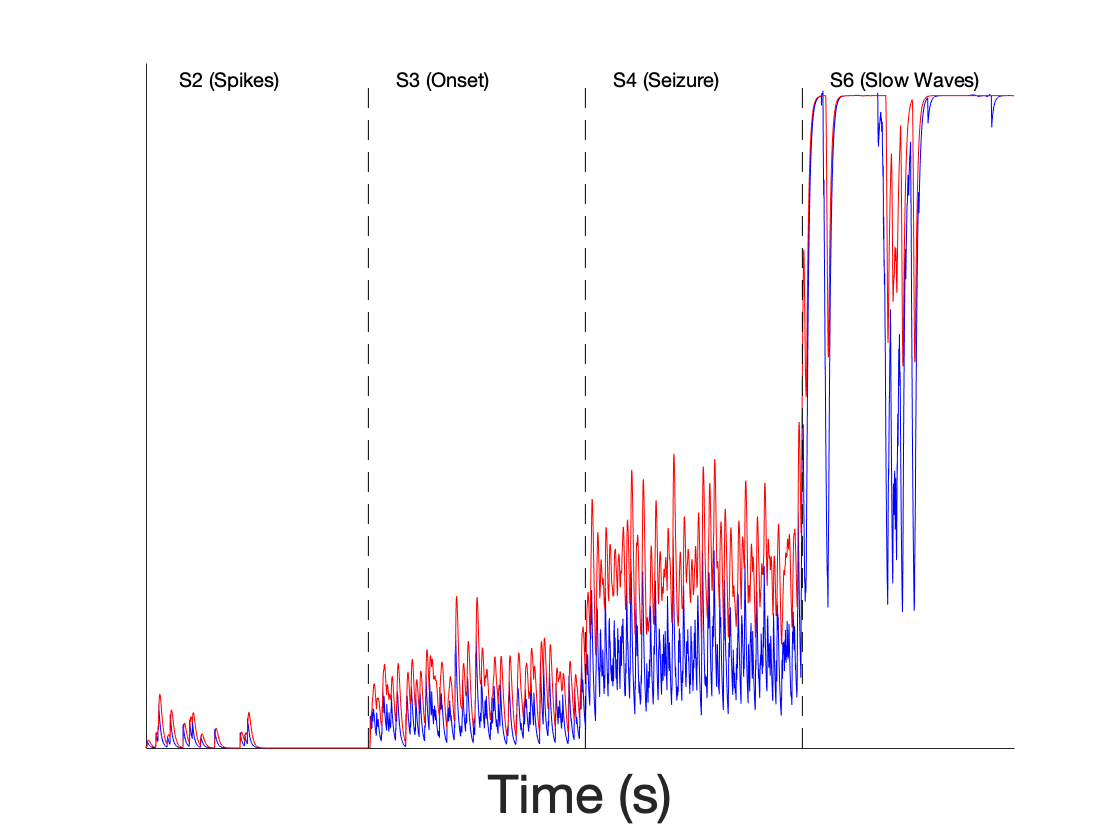}
  \caption{Basic epileptic waveforms
    (cf. Figure~\ref{fig:epileptic_waveform_types}) replicated from an
    excitatory-inhibitory pair with LTN dynamics, as shown
    in~\eqref{eq:noisyLTN}.}\label{fig:bifurcation_epilepsy_example}
\end{figure}

We conclude this section with an example of using an
excitatory-inhibitory pair governed by the linear-threshold dynamics
to replicate the seizure behavior seen in
Figure~\ref{fig:epileptic_waveform_types} through bifurcations. We
will use the relations between bifurcation and waveform types from
Table~\ref{table:bifurcations_and_waveforms} to construct the desired
waveforms. We further extend the deterministic linear-threshold
dynamics in~\eqref{eq:LTN_epilepsy} as follows
\begin{align}\label{eq:noisyLTN}
  \tau\dot{\x} = -\x + [\W\x + \v + \w]_\zero^\m.
\end{align}
%
%
Here the input $\u$ is divided into components $\v$ and $\w$, where $\w$ is
a white noise input used to simulate the noise content of EEG
signals~\citep{JWB-JC-KRP:13}. We construct the synaptic weight matrix
$\W$ such that the structural conditions of part (D) in
Theorem~\ref{thrm:conditions_BEBs} are satisfied. In
Figure~\ref{fig:bifurcation_epilepsy_example} we show how the network
exhibits each of the waveform types by moving through bifurcations by
changing the bifurcation parameter $v_1+w_1$ while the input $v_2+w_2$
is fixed at $0$. Four main behaviors appear in the network,
corresponding with the regions seen in
Figure~\ref{fig:epileptic_waveform_types}.

If the input to the network satisfies $v_1 + w_1 < 0$, the network has
a stable equilibrium point at zero which is corresponding to healthy
background activity and is not shown in the plot.
%
%
Approaching $v_1 + w_1 > 0$ is the first bifurcation as when
$v_1 + w_1 > 0$ the system has a unique limit cycle.
%
%
This is a NSF-BEB bifurcation and
results in the spikes shown in section S2 of the graph. Further
increasing the bifurcation parameter reaches a Hopf bifurcation and
gives the small oscillations in S3, and further increasing of the
input results in the higher magnitude oscillations of S4 and S5. At
the boundary of S5 and S6 we reach a second Hopf bifurcation to move
from high amplitude oscillations to the slow waves seen in S6. Further
increasing of the bifurcation parameter would result in a NSF-BEB
bifurcation that would result in a stable equilibrium point at the
upper threshold of the network and correspond to healthy activity.

%
%

\subsection{Declarative Memory in Single Region LTR Models}
Declarative memory is the process of encoding and storing data in the brain, while allowing for the conscious retrieval of that data~\citep{AB:20}. In early dynamical system representations of memory, such as~\cite{JJH:82,GEH-TJS:86}, memories are encoded and saved as attractive equilibrium points, with retrieval of the memory being represented by the system state reaching a particular equilibrium. As such, the ability of a model to encode and retrieve multiple memories is dependent on the ability of the model to admit multiple stable equilibria.

In this section we will model memory with a threshold-linear dynamical
system, that is, the linear-threshold model without an upper
saturation bound.
%
%
While the saturating linear-threshold model is a more accurate
representation as derived in Section~\ref{sec:firing_rate_models}, the
upcoming results are not readily generalizable to the saturating
model. For modeling memory, we utilize the idea of the support for
defining memories. For a state variable $\x \in \R^n$, let the support
of the vector be the subset of nodes with non-zero activity, that is
\begin{align*}
  \supp(\x) = \{i \in \{1,\dots,n\} ~|~ \x_i \neq 0 \}.
\end{align*}

We now provide a definition of a memory in this model.

\begin{definition}\longthmtitle{Memory}\label{def:memory}
  Consider a brain network composed of $n$ nodes with symmetric
  synaptic weight matrix $\W$ governed by the threshold-linear
  dynamics as follows
  \begin{align}\label{eq:TLN_dynamics}
    \tau\dot{\x} = -\x + [\W\x + \d]_\zero^\infty.
  \end{align}
  A set $\sigma \subseteq \{1,\dots,n\}$ is a memory in the network if
  there exists an input $\d$ such that there is a stable equilibrium
  point $\x^*$ with $\supp(\x^*) = \sigma$.
\end{definition}

The reason for using a memory as a set rather than as a single
equilibrium point as in~\cite{JJH:82} is input independence. Unlike
the earlier models where equilibrium points remained the same for
small changes in the input, the equilibrium points of the
threshold-linear model often change with any change in the input. This
is undesirable, as inputs are supposed to provide cues for what memory
should be invoked, rather than changing the memory itself. Using the
support of the vector instead of the equilibrium point provide the
desired invariance under similar inputs, as there is always a small
enough change in the input such that the support remains the same for
the resulting equilibrium point(s).

In what follows, we address two problems related to memory in this
model. First, under what conditions on the network structure $\W$ does
the network admit the ability to encode multiple memories?  Second,
for a given set $\sigma \subseteq \{1,\dots,n\}$ and network structure
$\W$, does there exist an input $\d$ such that $\sigma$ represents a
memory that can be encoded and retrieved by the network?

\subsubsection{Multiple Memories in Symmetric Threshold-Linear
  Networks}
The ability of a network to encode multiple memories is directly
related to the ability of the network to admit multiple stable
equilibria. In particular, a network must admit stable equilibrium
points with different supports in order to encode multiple
memories. As such, we introduce the following notion of permitted and
forbidden sets to distinguish whether or not a memory can be encoded
on a given set.

\begin{definition}\longthmtitle{Permitted and Forbidden
    Sets~\citep{RHRH-HSS-JJS:03}}
  Consider a brain network composed of $n$ nodes with symmetric
  synaptic weight matrix $\W$ governed by the threshold-linear
  dynamics~\eqref{eq:TLN_dynamics}. A set of neurons
  $\sigma \subseteq \{1,\dots,n\}$ is called \emph{permitted} if for
  some input $\d$ there exists a stable equilibrium point $\x^*$ with
  $\supp(\x^*) = \sigma$. A set of neurons is called \emph{forbidden}
  if it is not permitted.
\end{definition}

From this, we have that a memory can only be encoded on a permitted
set, and with memories defined by Definition~\ref{def:memory}, it
requires multiple distinct permitted sets to encode multiple
memories. Due to this requirement, the existence of multiple stable equilibria is not sufficient for the encoding of multiple memories. This is since the equilibria could form a continuum lying only on a line or surface, which have the same support, and as such are part of the same permitted set.
In this case, we can have a network with only one permitted set, and as
such only one possible memory. In order to have multiple permitted
sets, and as such, memories, we introduce the property of conditional
multiattractiveness.

\begin{definition}\longthmtitle{Conditionally
    Multiattractive~\citep{RHRH-HSS-JJS:03}}
  A network governed by the threshold-linear
  dynamics~\eqref{eq:TLN_dynamics} is \emph{conditionally
    multiattractive} if there exists an input $\d$ such that the set
  of stable equilibrium points is disconnected.
\end{definition}

If a network is conditionally multiattractive, then we can guarantee
the ability to encode multiple memories, as the disconnected stable
equilibrium points will lie in distinct permitted sets. As noted
above, a network being multiattractive differs slightly from being
multistable. In particular, all multiattractive networks are
multistable, but not all multistable networks are multiattractive, by
virtue of excluding line and surface attractors.

Before addressing conditions for multiattractiveness in
threshold-linear dynamics we return to addressing convergence to an
equilibrium through network interconnection structure. In
Section~\ref{subsec:selective_inhib_single} we noted that there exists
a unique equilibrium point for each input if and only if $\eye - \W \in
\mathcal{P}$. However, due to the desire for multiple in this section
we desire a slightly more general condition. For this we introduce the
notion of copositivity.

\begin{definition}\longthmtitle{Copositive~\citep{RHRH-HSS-JJS:03}}
  A symmetric matrix $\W \in \R^{n \times n}$ is called copositive if
  $\x^\top\W \x > 0$ for all $\x \in
  \Rpluseq^n\backslash\{\zero\}$. An equivalent condition is that all
  positive eigenvectors of all submatrices of $\W$ have positive
  eigenvalues.
\end{definition}

With the definition of copositivity in hand, we are able to provide a
condition guaranteeing convergence to an equilibrium for
threshold-linear networks with symmetric synaptic weight matrices
$\W$. In particular, the symmetric threshold-linear network is
guaranteed to converge to an equilibrium (which is dependent on the inputs and initial
conditions) if and only if the matrix $\eye - \W$ is
copositive~\citep{RHRH-HSS-JJS:03}.
%
%
This condition is more general than the conditions provided for the
matrix classes in Figure~\ref{fig:matrix_class_inclusions}, as the set
of P-matrices is a subset of the copositive matrices.
%
While the copositivity condition provides a condition for convergence
to an equilibrium, it does not guarantee the existence of
multiattractiveness. The following result provides additional
structural conditions on the synaptic weight matrix to guarantee
multiattractiveness.

\begin{theorem}\longthmtitle{Conditional Multiattractiveness in a Brain Network~\citep{RHRH-HSS-JJS:03}}\label{thrm:multiattractiveness}
  Consider a threshold-linear brain network defined by symmetric
  synaptic weight matrix $\W$. If $\eye - \W$ is copositive the
  following statements are equivalent:
  \begin{enumerate}
  \item The matrix $\eye - \W$ is not positive semidefinite.
  \item There exists a forbidden set.
  \item The network is conditionally multiattractive.
  \end{enumerate}
\end{theorem}

This result can be interpreted as follows. The goal of the result is
to provide conditions such that the network is conditionally
multiattractive, and as such can encode multiple memories. By the
first assumption in the result, that the matrix $\eye - \W$ is
copositive, we are guaranteed that for all inputs and initial
conditions the dynamics converges to an equilibrium.
From Theorem~\ref{thrm:multiattractiveness}i) we get that a negative
eigenvalue, which prevents the equilibrium point corresponding to all
neurons being active from being stable, guaranteeing the forbidden
set. This negative eigenvalue then allows for guaranteeing the
conditional multiattractiveness of the network, as it allows for the
construction of a separating hyperplane between stable equilibrium
points guaranteed by the copositivity of $\eye - \W$, giving
disconnected equilibrium points, and the ability to encode multiple
memories in the network.

We finish this section with a discussion of some of the additional
assumptions we have added in this section. First, we have used the
non-saturating threshold-linear model rather than the linear-threshold
model which we derived for brain modeling in
Section~\ref{sec:brain_modeling}. While being slightly less accurate,
if it is assumed that the brain activity being considered is not
approaching the saturation threshold, then the two models (bounded and
unbounded) become equivalent. The use of this model is due to the
requirement of input independence with memories defined in the manner
of stable equilibrium points. If instead we used the saturating model,
then for any desired support we could construct a stable equilibrium
by choosing an input that is sufficiently large in the active neurons
and sufficiently negative in the inactive neurons to give an
equilibrium point at the saturation threshold and zero. Since this is
unrealistic, it motivates the use of threshold-linear models. 
%
%
  Second, we have made the assumption that the synaptic
weight matrix is symmetric, which then does not generally satisfy
Dale's Law (except in cases where all neuron populations are
excitatory or inhibitory). This assumption simplifies the analysis for
multiple equilibrium points, but can prevent the existence of behavior
such as oscillations~\citep{PD-LFA:01}. However, under assumptions of
a recurrently connected network and having fast-acting inhibitory
neurons, the symmetric model can typically be a valid
approximation~\citep{RHRH-HSS-JJS:03}.

\subsubsection{Permitted Memories in Threshold-Linear Networks}
In the prior section we provided conditions such that a network has
the ability to encode and retrieve multiple memories. However, it is
of interest to be able to determine whether the network structure
permits the encoding of a memory in a given set of nodes. As such we
are interested in determining if a given set of nodes is permitted or
forbidden. The piecewise-affine nature of the threshold-linear
dynamics and the relation between permitted sets and stable
equilibrium points motivates the following result for determining if a
set is permitted or forbidden.

\begin{theorem}\longthmtitle{Conditions for Permitted and Forbidden
    Sets~\citep{CC-AD-VI:12}}\label{thrm:conditions_permitted}
  Consider a threshold-linear network~\eqref{eq:TLN_dynamics} with
  synaptic weight matrix $\W$. A set $\sigma \subseteq \{1,\dots,n\}$
  is permitted if and only if the matrix $(-\eye + \W)_\sigma$, the
  principal submatrix corresponding with the indices in $\sigma$, is
  stable. Further, the set $\sigma$ is forbidden if and only if
  $(-\eye + \W)_\sigma$ is unstable.
\end{theorem}

This result gives a simple test for whether or not a set of neurons
allows for the encoding of a memory. It follows from the fact that
each possible memory (and hence support) corresponds with an
individual switching region in the threshold-linear dynamics. The
existence of a stable equilibrium point with that support can then be
determined using the same conditions as for a standard linear system,
that is the stability of the synaptic weight matrix.

It is important to note that unlike the results in the prior section
on multiattractiveness, this result does not require the synaptic
weight matrix to be symmetric nor $\eye - \W$ to be
copositive. However, due to these omissions, it does not directly
apply with Theorem~\ref{thrm:multiattractiveness}, in that the
existence of a forbidden set does not imply conditional
multiattractiveness. If we do enforce those conditions again we obtain
the following result.

\begin{corollary}\longthmtitle{Set Operations on Permitted and
    Forbidden
    Sets~\citep{RHRH-HSS-JJS:03}}\label{cor:subsets_permitted}
  Consider a threshold-linear~\eqref{eq:TLN_dynamics} brain network
  with symmetric synaptic weight matrix $\W$ such that $\eye - \W$ is
  copositive. Then all subsets of permitted sets are permitted and all
  supersets of forbidden sets are forbidden.
\end{corollary}

This result gives a way to determine a variety of permitted and
forbidden sets based off the knowledge of just one permitted (or
forbidden) set. In concert with
Theorem~\ref{thrm:conditions_permitted}, it makes it possible to
determine many permitted or forbidden sets with minimal computations
on the stability of matrices. In particular, by determining small
forbidden sets, which requires minimal computational power, node sets
can be ruled out immediately when looking for larger permitted sets,
which requires more computational power due to considering larger
matrices. The opposite use of this result is that if a single
permitted set is found, then a large number can be constructed using
subsets of the original set.

\begin{figure*}[tbh]
  \centering
  \includegraphics[width=0.3\linewidth]{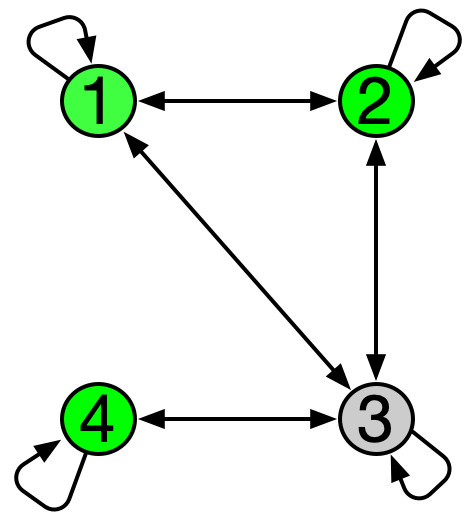}
  \includegraphics[width =
  0.3\linewidth]{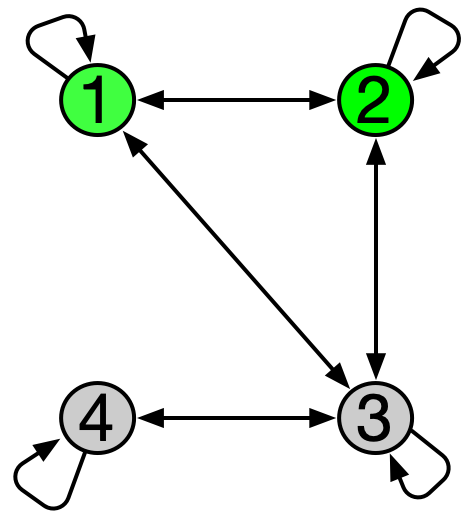}
  \includegraphics[width =
  0.3\linewidth]{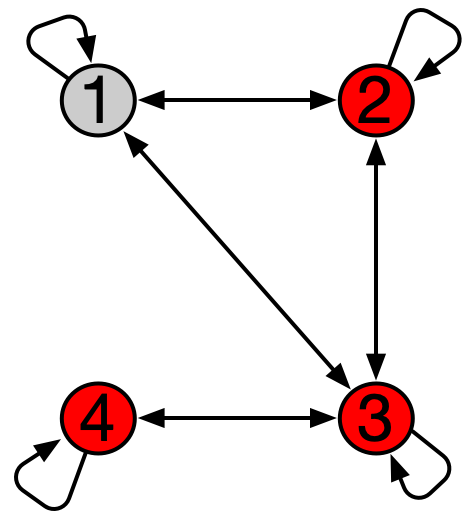}
  \caption{The network~\eqref{eq:memory_example_network} with two
    permitted sets and one forbidden set highlighted. We see that a
    memory can be encoded on the set $\{1,2,4\}$ and no memories can
    be encoded on the set $\{2,3,4\}$ from the first and third images,
    respectively. The middle image illustrates that a memory can be
    encoded on the node set $\{1,2\}$, which can be determined both by
    Theorem~\ref{thrm:conditions_permitted} and by
    Corollary~\ref{cor:subsets_permitted}.}\label{fig:permitted_forbidden_sets}
\end{figure*}

We finish this section with an example of a network able to encode
multiple memories. Consider a threshold-linear brain network with four
nodes and symmetric weight matrix $\W$ given by
\begin{align}\label{eq:memory_example_network}
  \W = \begin{bmatrix}
         0.8 & 0.2 & -0.5 & 0 \\
         0.2 & 0.3 & -0.2 & 0 \\
         -0.5 & -0.2 & 0.4 & -0.4 \\
         0 & 0 & -0.4 & 0.9
       \end{bmatrix}.
\end{align}
This matrix satisfies $\eye - \W$ copositive, while also satisfying
that it is not positive semidefinite. As such, by
Theorem~\ref{thrm:multiattractiveness} there exists a forbidden set
and the network can encode multiple memories. In
Figure~\ref{fig:permitted_forbidden_sets} we show the network with two
permitted and one forbidden set highlighted, as determined by the
conditions in Theorem~\ref{thrm:conditions_permitted}. Using
Corollary~\ref{cor:subsets_permitted} we can determine that all
subsets of $\{1,2,4\}$ are able to encode memories, while all supersets
of $\{2,3\}$ are unable to encode memories.

\section{Analysis of Interconnected Brain Regions}\label{sec:multi-region}

While studying the brain using models of individual brain regions can
be useful, one of the defining characteristics of the brain is its
interconnected structure between regions that can have vastly
different properties. Almost all brain functions are based on the
interaction and transmission of information between different areas,
with other actions performed through the transmission of information
into other parts of the nervous system. As such it is of interest to
expand our study to models composed of multiple brain regions. We
maintain the model of linear-threshold firing rate dynamics for each
region and connect them to create larger networks of networks, as
described in Box~\ref{box:construction_multi-region_networks}. These
networks can be formed with a variety of different topologies which
appear based on network location and application.

\begin{myblock}{Construction of Multi-Region Brain
    Networks}\label{box:construction_multi-region_networks}
  Multi-region brain networks are constructed through the
  interconnection of the dynamics of individual brain regions. This
  occurs through the definition of sensory information processing
  pathways along connections within the structure of the
  network. Consider a single brain region, denoted by $\mathcal{N}_1$,
  and defined by the dynamics
  \begin{align*}
    \tau_1 \dot{\x}_1 &= -\x_1 + [\W_{1,1}\x_1 +
                        \d_1(t)]_\zero^{\m_1}.
  \end{align*}
  With the internal dynamics of the brain region given by the synaptic
  weight matrix $\W_{1,1}$, we define the processing pathways between
  regions in the input term $\d_1(t)$. Let $\mathcal{N}_2$ denote a
  second brain region, which we assume is connected to
  $\mathcal{N}_1$, forming a small network motif. Then, synaptic
  weight matrices $\W_{1,2}$ and $\W_{2,1}$ represent the information
  processing pathways between the two regions, defining the structure
  of the overall network. The input term $\d_1(t)$ for $\mathcal{N}_1$
  is then given by
  \begin{align*}
    \d_1(t) = \W_{1,2}\x_2 + \B_1\u_1(t) + \c_1,
  \end{align*}
  where $\u_1(t)$ is the control term for the region, and $\c_1$
  corresponds to nonzero bias terms and any unmodeled background
  activity. Then, defining the dynamics of $\mathcal{N}_2$ with the
  same linear-threshold activation function, and defining $\d_2(t)$
  analogously, the dynamics for the overall network composed of
  $\mathcal{N}_1$ and $\mathcal{N}_2$ is
  \begin{align*}
    \tau_1 \dot{\x}_1 &= -\x_1 + [\W_{1,1}\x_1 + \W_{1,2}\x_2+
                        \B_1\u_1(t) + \c_1]_\zero^{\m_1}
     \\
    \tau_2 \dot{\x}_2 &= -\x_2 + [\W_{2,2}\x_2 + \W_{2,1}\x_1+
                        \B_2\u_2(t) + \c_2]_\zero^{\m_2}.
  \end{align*}
  This process can then be used iteratively to construct brain
  networks of arbitrary size and connectivity structure, which we will
  subsequently use to study GDSA and epileptic seizures over
  interconnected brain networks.
\end{myblock}

In this section, we will expand our treatment of both GDSA and
epileptic seizures to interconnected brain networks. For GDSA we will
consider two network topologies, hierarchical thalamocortical networks
and star-connected thalamocortical networks, and illustrate that
selective inhibition and recruitment can be achieved in the
interconnected networks based on properties of the linear-threshold
dynamics. For modeling epileptic seizures, we will consider networks
composed of interconnected excitatory-inhibitory pairs, and examine
how oscillations occur and spread throughout the network governed by
linear-threshold dynamics.

\subsection{GDSA in Interconnected Brain Networks}

Goal-driven selective attention occurs based on interactions and the
transmission of information between areas in the brain. Here, we
expand our treatment of selective inhibition and recruitment to
networks composed of multiple brain regions. A key part of
interconnected networks is that brain regions are not necessarily
homogeneous and, as such, it is important to consider types with
different characteristics. In particular, we consider networks
composed of both cortical and thalamic regions and, using the
construction of interconnected linear-threshold brain networks from
Box~\ref{box:construction_multi-region_networks}, will examine the
following topologies:
\begin{itemize}
\item A \emph{hierarchical thalamocortical topology}, as shown in
  Figure~\ref{fig:network_topologies}(a). In this topology, we
  consider a series of cortical regions forming a hierarchy, where
  each region is connected only to regions directly above and
  below. The thalamus is represented by a single region, composed of multiple higher-order thalamic nuclei, 
%
%
  which accepts inputs from cortical regions, before
  modulating them and transmitting the information back to other
  cortical regions. This models transthalamic pathways that function
  parallel to the direct cortico-cortical connections~\citep{SMS:12}.
  By connecting to each
  cortical region in the hierarchy, this topology allows for the
  indirect connection of any two cortical regions through the
  thalamus. Such a topology has been implicated in higher-level brain
  functions, such as learning and decision-making, where the thalamus
  supports the transfer of information across different areas of the
  prefrontal cortex~\citep{ASM:15}.

\item A \emph{star-connected thalamocortical topology}, as shown in
  Figure~\ref{fig:network_topologies}(b). In this topology, the
  thalamus node is a first-order thalamic nuclei, which transmits
  sensory or other information from subcortical structures to cortical
  regions. Here the thalamus is operating mostly as a relay. Such
  topologies arise in all sensory systems, with the exception of the
  olfactory system. Examples of first-order thalamic nuclei that
  result in this topology are the ventral posterior nucleus in the
  somatosensory system and the lateral geniculate nucleus in the
  visual system~\citep{EJR-JWG-SMS:05,MCS-SWM-JT-RCS-MW-AJP-FQY-DAL:10}. 
\end{itemize}

\tikzset{
mynode/.style={
text width=0.06\textwidth
}}
 \begin{figure*}[htb!]
   \centering
   \qquad
   \subfigure[]{
     \begin{tikzpicture}
       \node[mynode, circle, draw, line width=0.6pt, inner sep=0pt] (1)
       {\includegraphics[width=28pt]{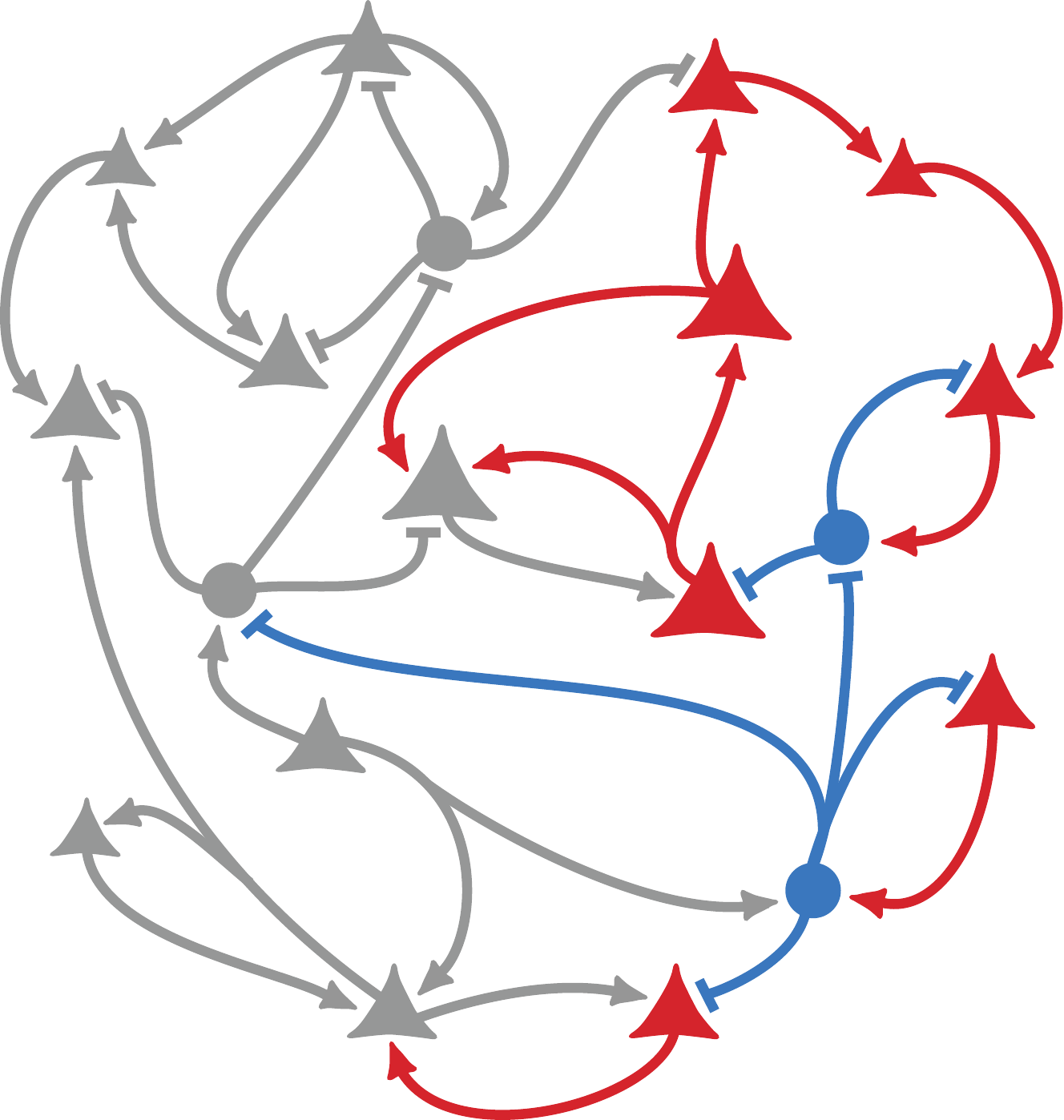}};
       \node[mynode, above of=1, yshift=30pt, circle, draw, line
       width=0.6pt, inner sep=-0.4pt] (2)
       {\includegraphics[width=28pt]{EI_network_structure_half_inhibited}};
       \node[mynode, below of=1, yshift=-30pt, circle, draw, line
       width=0.6pt, inner sep=-0.4pt] (3)
       {\includegraphics[width=28pt]{EI_network_structure_half_inhibited}};
       \draw[latex-latex, line width=0.3pt, shorten <=1pt, shorten
       >=1pt] (1.90) to (2.270);
       \draw[latex-latex, line width=0.3pt, shorten <=1pt, shorten
       >=1pt] (1.270) to (3.90);
       \node[below of=3, yshift=5pt] {$\vdots$};
       \node[above of=2, yshift=0pt] {$\vdots$};
       \node[left of=2, xshift=-20pt, yshift=10pt, scale=0.8] (i-1)
       {Subnetwork $i - 1$};
       \node[left of=1, xshift=-20pt, yshift=10pt, scale=0.8] (i) {Subnetwork $i$};
       \node[left of=3, xshift=-20pt, yshift=10pt, scale=0.8] (i+1) {Subnetwork $i +
         1$};
       \node[mynode, right of=1, xshift=30pt, circle, draw, line
       width=0.6pt, inner sep=-0.4pt] (i-big)
       {\includegraphics[width=28pt]{EI_network_structure_half_inhibited}};
       \draw[latex-latex,shorten <= 1pt,shorten >= 1pt] (1) to (i-big);
       \draw[latex-latex,shorten <= 1pt,shorten >= 1pt] (2) to (i-big);
       \draw[latex-latex,shorten <= 1pt,shorten >= 1pt] (3) to (i-big);
       \node[above of=i-big, xshift=0pt, yshift=-1pt, scale=0.8] (i1)
       {\parbox{40pt}{\centering  Thalamus}};
     \end{tikzpicture}
   }
   \qquad
   \subfigure[]{\begin{tikzpicture} \node[mynode, circle, draw, line
       width=0.6pt, inner sep=-0.3pt] (1)
       {\includegraphics[width=28pt]{EI_network_structure_half_inhibited}};
       \node[above of=1, xshift = 25pt, yshift = -5pt, scale=0.8]{Thalamus};
       \node[mynode, above of=1, yshift=30pt, circle, draw, line
       width=0.6pt, inner sep=-0.4pt] (2)
       {\includegraphics[width=28pt]{EI_network_structure_half_inhibited}};
       \node[above of=2, xshift = 0pt, yshift = -5pt, scale=0.8]{Subnetwork $1$};
       \draw[latex-latex, line width=0.3pt, shorten <=1pt, shorten
       >=1pt] (2.270) to (1.90);
       \node[mynode, left of=1, xshift=-30pt, circle, draw, line
       width=0.6pt, inner sep=-0.4pt] (3)
       {\includegraphics[width=28pt]{EI_network_structure_half_inhibited}};
       \node[below of=3, xshift = 0pt, yshift = 5pt, scale=0.8]{Subnetwork $4$};
       \draw[latex-latex, line width=0.3pt, shorten <=1pt, shorten
       >=1pt] (3.0) to (1.180);

       \node[mynode, right of=1, xshift=30pt, circle, draw, line
       width=0.6pt, inner sep=-0.4pt] (4)
       {\includegraphics[width=28pt]{EI_network_structure_half_inhibited}};
       \draw[latex-latex, line width=0.3pt, shorten <=1pt, shorten
       >=1pt] (1.0) to (4.180);
       \node[below of=4, xshift = 0pt, yshift = 5pt, scale=0.8]{Subnetwork $2$};

       \node[mynode, below of=1, yshift=-30pt, circle, draw, line
       width=0.6pt, inner sep=-0.4pt] (5)
       {\includegraphics[width=28pt]{EI_network_structure_half_inhibited}};
       \node[below of=5, xshift = 0pt, yshift = 0pt, scale=0.8]{Subnetwork $3$};
       \draw[latex-latex, line width=0.3pt, shorten <=1pt, shorten
       >=1pt] (5.90) to (1.270);
     \end{tikzpicture}
 }
 \caption{Topologies for thalamocortical multi-region brain networks
   considered in the paper: (a) multilayer hierarchical
   thalamocortical network, where each layer is connected directly to
   the thalamus, as well as the layers directly above and below
   it; (b) star-connected thalamocortical network, where each cortical
   layer is connected to the thalamus layer only.  In both plots,
   task-relevant excitatory and inhibitory nodes are depicted in red
   and blue, respectively, and (transiently silenced) task-irrelevant
   nodes are depicted in grey.}\label{fig:network_topologies}
\end{figure*}

As discussed in Section~\ref{subsec:selective_inhib_single}, achieving
selective inhibition and recruitment is dependent on the
stabilizability of the network.  When considering an interconnected
network, the stabilizability of the overall network is dependent on
the network timescales and interconnections, in addition to the
internal dynamics of each region. As such, the results using direct
feedback and feedforward inhibition in order to achieve selective
recruitment and inhibition,
cf. Theorems~\ref{thm:feedforward_selective_inhib}
and~\ref{thrm:feedback_selective_inhib}, are not directly
applicable. However, properties of the linear-threshold dynamics can
be used to show that selective inhibition and recruitment is possible
in the interconnected networks of both topologies, as discussed next.

\subsubsection{Hierarchical Networks}\label{sec:hierarchical-networks}
The brain has been known to have a hierarchical organization for
decades, both in terms of structure and
function~\citep{NT:50,ARL:70}. One of these hierarchies is based upon
function, in which primary sensory and motor areas are placed at the
bottom of the hierarchy, while high-level processing areas such as the
prefrontal cortex lie at the top~\citep{DM-RL-AF-KE-ETB:09}.
   Sensory information is processed as it moves up the
hierarchy, while decisions are made at the top and information is then
transmitted back down the hierarchy to perform desired actions or
shape sensory perception. It is in the top-down direction of this
hierarchy that GDSA occurs, where the higher-level layers instruct the
lower layers as to which information is relevant, thus requiring
further processing for performing the desired actions, and which
information should be suppressed and prevented from further
processing.

In addition to being sorted based upon the direction of information
flow, a hierarchy of timescales also exists between brain regions,
which closely aligns with the former. In particular, as one moves up
the hierarchy, the regional dynamics become
slower~\citep{BG-EE-GH-AG-AK:12,UH-JC-CJH:15,SJK-JD-KJF:08}.  This
separation of timescales is important for the ability of the network
to perform selective inhibition and recruitment. To illustrate this,
recall that the goal of GDSA is to have the activation level of the
task-irrelevant components of the network converge to zero, while the
activation level of the task-relevant components converge to a desired
(non-zero) steady-state pattern, e.g., an equilibrium~$\x^*$.
%
%
In the case of a network of networks, the determination of such
equilibria at each layer is not trivial, as the layer interconnection
makes it dependent upon the inputs of layers higher in the hierarchy.
The timescale separation plays a key role in reasoning about such
equilibria. This is because, for a given layer, the state of the ones
above it can be considered as constant (since they evolve on a slower
time scale), whereas the state of the ones below it can be described
as a static nonlinear function dependent upon the state of the current
layer (since they evolve on a faster time scale).

Formally, one defines an equilibrium map, which we denote
$h_i^1(\cdot)$ (where the superindex $1$ indicates that this is for the task-relevant nodes), for each layer~$i$.
%
%
This map, given the inputs from the higher layers in the network,
returns the set of equilibria for network layer~$i$.
 One can set the state of the task-irrelevant nodes to zero
(as this will be taken care of by selective inhibition) and shrink the
state of each layer to strictly include the task-relevant
components. The equilibrium maps require a constant input representing
the state of the layers higher in the network, whose existence can be
justified through the timescale separation between layers in the
hierarchical network and using singular perturbation theory. Under the
assumption that the timescale separation is infinitely large, and
assuming global asymptotic stability of each layer, the state of a
given layer appears as a constant at the timescale of the lower layers
in the hierarchy. Conversely, at the timescale of layers above, the
state of a lower layer becomes a static nonlinear function (i.e., the
equilibrium map) of the states of the higher layers of the
hierarchy. Utilizing this, the equilibrium map takes its inputs to be
the state (modified through interconnection matrices) of any layers
higher in the network that are directly connected. In the hierarchical thalamocortical topology this is either one or two inputs. The first input is the cortical layer directly above the considered layer. Then, if the current layer is below the thalamus in the hierarchy, the state of the thalamus will also be an input to the equilibrium map.
We are then able to recursively
define equilibrium maps from the bottom layer up to the top of the
hierarchy.

Utilizing these equilibrium maps, and the goal equilibrium they define
for selective recruitment, we next discuss conditions for achieving
selective inhibition and recruitment in the cortical and
thalamocortical hierarchical networks.  The hierarchical
thalamocortical network, shown in
Figure~\ref{fig:network_topologies}(a), is composed of $N$ cortical
regions connected to the regions directly above and below and one
thalamic region connected to all of the cortical regions. In order to represent the thalamus as a single region we assume that the relevant thalamic neurons have comparable timescales.
%
  %
The regions form a temporal (and functional) hierarchy, encoded by
timescales satisfying $\tau_1 \gg \tau_2 \gg \dots \gg \tau_N$. The
thalamus region, forming parallel transthalamic pathways lies in the
same hierarchy. The exact position of the thalamus is dependent on the
cortical regions in the hierarchy, but it can lie at any
point~\citep{JAH-SM-KEH-JDW-HC-AB-PB-SC-LC-AC-AF-DF-NG-CRG-NG-PAG-AMH-AH-RH-JEK-LK-XK-JL-PL-YL-JL-SM-MTM-MN-LN-SWO-BO-ES-SAS-WW-QW-YW-AW-JWP-ARJ-CK-HZ:19}. Without loss of
generality we assume the thalamus lies at a point in the middle of the
hierarchy, with $\tau_a \gg \tau_T \gg \tau_{a+1}$ for some
$a \in \{1,\dots,N\}$. The results that follow remain valid if the thalamus is at an endpoint (top or bottom) of the hierarchy, with slight changes in the resulting expressions.
%
%
Using the linear-threshold firing rate
model~\eqref{eq:network_lin_threshold_dynamics}, we obtain the
following dynamics for the hierarchical network.
\begin{align}
  \tau_i\dot{\x}_i
  &= -\x_i + [\W_{i,i}\x_i(t) + \W_{i,i-1}\x_{i-1}(t)
    + \W_{i,i+1}\x_{i+1}(t) + \W_{i,T}\x_T +
    \B_i\u_i(t) + \c_i)]_\zero^{\m_i}, \qquad i \in
    \{1,\dots,N\} \cr 
    \tau_T\dot{\x}_T
  &= -\x_T + [\W_T\x_T +
    \sum_{i=1}^N \W_{T,i}\x_i +
    \B_T\u_T(t) +
    \c_T]_\zero^{\m_T}. 
\end{align}
From this dynamics we can define equilibrium maps for each layer in
the cortical hierarchy, each giving the set of possible equilibrium
points as a function of the constant $\c$. Note that these equilibrium
maps are for the task-relevant portion of the dynamics only and, as
such, are not impacted by the control input which acts only on the
task-irrelevant components. Due to the interconnected form of the
dynamics, the maps are defined recursively, with expressions dependent
on the location of the region in the hierarchy. For the bottom layer
of the network, the equilibrium map is defined as
\begin{align}
  h_{N}^1(\c) = \{\x_N^1 ~|~ \x_N^1 = [\W_{N,N}^{11}\x_N^1 +
  \c]_\zero^{\m_N^1} \}. 
\end{align}
The equilibrium maps further up the hierarchy become notationally more
complicated due to their recursive nature, but all maintain the form
\begin{align}\label{eq:equilibrium_maps}
  h_{i}^1(\c) = \{\x_i^1 ~|~ \x_i^1 = [\W_{i,i}^{11}\x_i^1 + \sum_{j = i+1,T}
  \W_{i,j}^{11}h_j^1(\mathbf{y}) + \c]_\zero^{\m_i^1}\}, 
\end{align}
%
%
where, for an arbitrary layer~$i$, the sum is over the layers below
the current layer that are directly connected, potentially including the
thalamus. Here $\mathbf{y}$ is an input into the
%
%
equilibrium maps below layer $i$ dependent on the state of the higher layers in
the network. In these maps, the impact from the layers higher in the
hierarchy on the current layer's equilibrium appear in the input term
$\c$, while the impact from the lower networks comes from the
appearance of the lower-level equilibrium maps.

The equilibrium maps can be written in a switched-affine form, denoted
by
\begin{align*}
  h_i^1(\c) = \F_{\lambda}\c + \bf{f}_{\lambda}
\end{align*}
for switching regions $\lambda$, based on the piecewise-affine form of
the linear-threshold dynamics (details on this form can be found
in~\citep{EN-JC:21-tacII}). This representation of the equilibrium map
gives rise to a gain matrix $\bar{\F}_i$, defined as the entry-wise
maximum of the matrix $|\F_\lambda|$ over all switching regions. These
gain matrices are relevant to the results providing conditions for
selective inhibition and recruitment to be achieved in the
interconnected network, as given next.

\begin{theorem}\longthmtitle{Selective Inhibition and Recruitment in
    Hierarchical Thalamocortical
    Networks}\label{thrm:thalamocortical_hierarchy_conditions}
  Consider a thalamocortical hierarchical network governed by a
  linear-threshold dynamics and suppose subnetwork $\mathcal{N}_1$ has
  bounded trajectories. If
  \begin{align*}
    \rho(|\W_{i,i}^{11}| +
    |\W_{i,i+1}^{11}|\bar{\F}_{i+1}|\W_{i+1,i}^{11}|) < 1,
  \end{align*}
  for layers below the thalamus,
    \begin{align*}
      \rho(|\W_{i,i}^{11}| +
      |\W_{i,i+1}^{11}|\bar{\F}_{i+1}|\W_{i+1,i}^{11}| +
      |\W_{i,T}^{11}|\bar{\F}_T|\W_T^{11}|) < 1,
  \end{align*}
  for layers above the thalamus, and
  \begin{align*}
    \rho(|\W_{T}^{11} + \sum_{i=a+1}^N |\W_{T,i}^{11}|\bar{\F}_i|\W_{i,i}^{11}|) < 1,
  \end{align*}
  then there exists $\K_i$ and $\bar{\u}_i(t)$ such that using the
  feedback-feedforward control $\u_i(t) = \K_i\x_i(t) + \bar{\u}_i(t)$,
  %
  %
  for all $i$, gives
  \begin{align*}
      \begin{cases}
        \x_i^0(t) \to \zero & \forall i \in \{2,\dots,N,T\}
        \\
        \x_i^1(t) \to h_i^1(\W_{i,i-1}^{11}\x_{i-1}^1(t) + \c_i^1) &
        \mathrm{for~layers~above~the~thalamus}
        \\
        \x_i^1(t) \to h_i^1(\W_{i,i-1}^{11}\x_{i-1}^1(t) +
        \W_{i,T}^{11}\x_T^1(t) + \c_i^1) & \mathrm{
          for~layers~below~the~thalamus}
        \\
        \x_T^1(t) \to h_T(\sum_{i=1}^a \W_{T,i}^{11}\x_i^1(t) + \c_T).
        & \mathrm{ for~the~thalamus}
  \end{cases}
  \end{align*}
  as the timescales satisfy $\frac{\tau_i}{\tau_{i-1}} \to 0$ and
  $\frac{\tau_T}{\tau_{a+1}} \to 0$ for all $i \in \{2,\dots,N\}$.
\end{theorem}

This result can be interpreted as follows. The conditions involving
the spectral radius at each layer involve two components: the internal
dynamics of layer $\mathcal{N}_i$, and the impact of its
interconnections with the next layer in the hierarchy
($\mathcal{N}_{i+1}$) and the thalamus. Due to the timescale
separation between layers in the hierarchy, the connections from the
next layer in the hierarchy and the thalamus act as static
nonlinearities due to their continual ``instantaneous" convergence to
their respective (and moving) equilibrium points. Therefore, the sums
combine the pathways that $\mathcal{N}_i$ has to impact its state, and
the spectral radius condition gives an upper bound on the combined
effects of these pathways. If the spectral radius conditions hold,
reduced-order dynamics for the task-relevant components in the network
constructed by replacing connections from lower layers in the
hierarchy with their equilibrium values are GES~\citep[Lemma
IV.2]{MM-JC:24-tcns}, which allows for the convergence to the desired
equilibrium for GDSA. The result $\x_i^0(t) \to \zero$ shows the
task-irrelevant components of the dynamics converging to zero, with
the remaining terms give convergence to a steady state, as expected. The control $\u_i(t)$ used to achieve this convergence is a feedback-feedforward control where each component plays a different role. The feedback term $\K_i\x_i$ is used to manage the impact of the layers of the network below layer $i$, while the feedforward term $\bar{\u}_i(t)$ controls the inputs from higher in the network and the term $\c_i^0$. Meanwhile, the value of the steady state the task-relevant components converge to is dependent on the terms $\c_1^1,\dots,\c_N^1,\c_T^1$ and the interconnection between the network layers. In general it is difficult to specify values such that the network converges to a specific steady state, but this problem has been approached with a reservoir computing approach~\citep{MM-JC:24-ojcsys}.

We now give some intuition on the remaining components of
Theorem~\ref{thrm:thalamocortical_hierarchy_conditions}. First, the
gain matrices $\bar{\F}_i$ provide a worst-case bound on how much the
equilibrium point of a layer $\mathcal{N}_i$ can change based on one
unit of change in the state of layer $\mathcal{N}_{i-1}$. Due to this,
its use in the spectral radius conditions covering pathways for
modifying system states makes the bound conservative. Second, the
condition $\tau_i/\tau_{i-1} \to \zero$ corresponds to the timescale
separation between the layers becoming infinitely large. This
separation, creating completely distinct timescales between each
layer, provides the base for being able to apply a generalized
Tikhonov-style singular perturbation argument~\citep{VV:97} and
establish the stabilizability of the hierarchical interconnected
linear-threshold network. It is important to note that, while the
result calls for an infinite timescale separation, in brain pathways
the ratios between successive layers can be on the order of
$1/1.5 \sim 1/2.5$, which we have empirically found to typically be
sufficient for selectively inhibiting and recruiting the system with a
small degree of tracking error~\citep{EN-JC:21-tacII}.


Finally, note that the conditions of
Theorem~\ref{thrm:thalamocortical_hierarchy_conditions} for the
stabilizability of interconnected linear-threshold networks,
illustrated through the ability to achieve selective inhibition and
recruitment, depend only on matrices of size $\R^{n_i \times n_i}$,
where $n_i$ is the size of layer $\mathcal{N}_i$. This corresponds with the analysis of the result occurring at the level of individual regions with the infinite timescale separation. The other method for analysis of the network is considering it as one large network with finitely different timescales. Taking this approach then leads to determining if selective inhibition and recruitment is possible through a calculation related to the full matrix of the network. In this case, if selective inhibition is not possible, we are not able to pinpoint where the problem in the network might be. As such it is preferable to take the approach of Theorem~\ref{thrm:thalamocortical_hierarchy_conditions} as it provides greater interpretability by being able to determine smaller regions of the network that result in not being able to achieve selective inhibition and recruitment. In addition, in Theorem~\ref{thrm:thalamocortical_hierarchy_conditions} we provide conditions based on the spectral radius of relevant matrices, which is efficient to calculate but conservative. In~\citep{MM-JC:24-tcns} more general conditions are provided which involve determining whether matrices are $\mathcal{L}$-matrices or totally Hurwitz. These conditions are much more difficult to confirm for larger matrices, providing further incentive for the hierarchical approach used here.

\subsubsection{Star-Connected Networks}\label{sec:star-networks}
A star-connected network, as shown in
Figure~\ref{fig:network_topologies}(b), arises when we model the
thalamus as a relay center, as is common in the studies of sensory
processing~\citep{SMS:12,AAM-RDM-BR-DHH-HHH:17,RWG-SMS:02}.
In this context, the thalamus often receives an input
signal from a subcortical area and sends outputs to one or more
cortical regions. In this simplified model, the cortical regions are
not directly connected and, hence, their timescale differences do not
play an important role. 

We model the (subcortical) region that provides the input to the
thalamus with a linear-threshold firing rate model in the same manner
as the other regions in the network, except for an explicit control
term modeling the sensory input to the region and represented by
a time-varying input signal $\c_1(t)$.  Accordingly, we model the
dynamics for a star-connected thalamocortical network with $N-1$
cortical regions as
\begin{align}\label{eq:star_connected_dynamics}
  \tau_1\dot{\x}_1 &= -\x_1 + [\W_{1,1}\x_1 + \W_{1,T}\x_T +
                     \c_1(t)]_\zero^{\m_1} \notag
  \\
  \tau_i \dot{\x}_i &= -\x_i + [\W_{i,i}\x_i + \W_{i,T}\x_T + \B_i\u_i
                      + \c_i]_\zero^{\m_i} \notag
  \\
  \tau_T\dot{\x}_T &= -\x_T + [\W_{T}\x_T + \sum_{i=1}^N \W_{T,i}\x_i
                     + \B_T\u_T + \c_T]_\zero^{\m_T},
\end{align}
%
%
where $i \in \{2,\dots,N\}$, $\x_1$ denotes the state vector of the
subcortical region providing sensory input to the thalamus and the terms $\c_2,\dots,\c_N,\c_T$ are unmodeled background activity that also shape the desired steady-state convergence point.

Due to the lack of temporal hierarchy in this network topology, the
singular perturbation method used earlier for the hierarchical
architectures no longer applies. First, with the lack of timescale
separation, the equilibrium of the network is no longer determined
through recursive equilibrium maps and, instead, must be computed
concurrently for all subnetworks. In particular, the network
equilibrium at time $t$ is given by the solution to the following set of nonlinear
equations
\begin{align*}
  \x_1^1 &= [\W_{1,1}^{1,\all}\x_1 + \W_{1,T}^{1,\all}\x_T +
  \c_1^1(t)]_\zero^{\m_i^1},\\ 
  \x_i^1 &= [\W_{i,i}^{1,\all}\x_i + \W_{i,T}^{1,\all}\x_T +
  \c_i^1]_\zero^{\m_i^1}, \qquad i = 2,\dots,N, \notag
  \\
  \x_T^1 &= [\W_T\x_T + \sum_{i=1}^N \W_{T,i}\x_i +
  \c_T^1]_0^{\m_T^1},
\end{align*}
where $\x_i^0 = \zero$ for all $i \in \{1,\dots,N,T\}$.
%
%

For considering selective inhibition and recruitment in the
star-connected topology, we instead use results on the stability of
slowly varying nonlinear systems. These rely on assuming that the input
signal to the subcortical region, $\c_1(t)$, has a bounded-rate
derivative (that is, there exists finite $\alpha$ such that $\norm{\dot{\c}_1(t)} < \alpha$). In addition, we allow for a slightly weaker notion of
selective inhibition and recruitment in which the convergence to zero
and the equilibrium, respectively, is within a constant
$\epsilon$. The following result provides conditions such that the
thalamocortical network with a star-connected topology can achieve
this notion of selective inhibition and recruitment.

\begin{theorem}\label{thrm:star_connected}
  Consider an $N$-layer star-connected thalamocortical network with
  $N-1$ cortical regions $\mathcal{N}_2,\dots,\mathcal{N}_N$, thalamic
  layer $\mathcal{N}_T$ and subcortical input layer
  $\mathcal{N}_1$. Suppose the following hold for all values of
  $\c_i \in \R^{n_i}$, $i \in \{2,\dots,N\}$ and $\c_T \in \R^{n_T}$:
  \begin{enumerate}
  \item The input layer $\mathcal{N}_1$ has no nodes to be inhibited,
    $\rho(\W_{1,1}|) < 1$, and the input $\c_1(t)$ lies in a compact
    set and has a bounded rate derivative; 
  \item For each $i \in \{2,\dots,N\}\cup\{T\}$, the matrix
    $\W_{i,i}^{11}$ satisfies $\rho(|\W_{i,i}^{11}) < \alpha_i$, with
    $\alpha_i < 1$;
  \item The matrix of task-relevant interconnections
    \begin{align*}
      \bar{\W}^{11}
      &= \begin{bmatrix}
           \W_{1,1} & \zero & \dots & \zero & \W_{1,T}^{11}
           \\
           \zero & \W_{2,2}^{11} & \dots & \zero & \W_{2,T}^{11}
           \\
           \vdots & \dots & \ddots & \vdots & \vdots
           \\
           \zero & \zero & \dots & \W_{N,N}^{11} & \W_{N,T}^{11}
           \\
           \W_{T,1}^{11} & \W_{T,2}^{11} & \dots & \W_{T,N}^{11} &
                                                                   \W_{T}^{11}
         \end{bmatrix},
    \end{align*}
    with Schur decomposition\footnote{For a matrix $\W$ we consider
      the Schur decomposition
      $\W = \mathbf{Q}^\top(\mathbf{D}+\mathbf{N})\mathbf{Q}$ where
      $\mathbf{Q}$ is unitary, $\mathbf{D}$ is diagonal, and
      $\mathbf{N}$ is upper triangular with a zero
      diagonal~\citep{KEC:86}.}
    $\bar{\W}^{11} = \mathbf{Q}^\top(\mathbf{D}_{\bar{\W}^{11}} +
    \mathbf{N}_{\bar{\W}^{11}})\mathbf{Q}$, is such
    $\alpha + \max(\delta,\delta^{1/p}) < 1$, where $p$ is the
    dimension of $N_{\bar{\W}^{11}}$ and
    \begin{align*}
      \alpha &= \max_{i \in \{2,\dots,N\}\cup\{T\}}\{\alpha_i\},
               \qquad \delta = \gamma
               \sum_{j=1}^{p-1}\norm{N_{\bar{\W}^{11}}}^j
      \\
      \gamma &= \max\{\sum_{i=1}^{N-1}\W_{i,t}^{11}\W_{i,T}^{11^\top},
               \sum_{i=1}^{N-1}\W_{T,i}^{11}\W_{T,i}^{11^\top} \} 
    \end{align*}
  \end{enumerate}
  Then there exist equilibrium maps $\x_i^*(t)$ and $\epsilon > 0$ such
  that
  \begin{align*}
    \begin{cases}
      \lim_{t \to \infty}\norm{\x_1(t) - \x_1^*(\c_1(t))} < \epsilon &
                                                                       \mathrm{(Selective~Recruitment~of~Driving~Layer)} 
      \\
      \intertext{and for all layers $\{\mathcal{N}_i\}_{i=2}^{N}$ and
      $\mathcal{N}_T$,}   \quad \lim_{t \to
      \infty} \norm{\x_i^0(t)} < \epsilon; & \mathrm{(Selective~Inhibition)}
      \\
      \quad \lim_{t \to \infty} \norm{\x_i^1(t)
      - \x_i^{1^*}(\c_1(t),\dots,\c_N,\c_T)} < \epsilon & \mathrm{(Selective~Recruitment)}
    \end{cases}
  \end{align*}
  Further, if $\norm{\dot{\c}_1(t)} \to 0$ as $t \to \infty$ then $\epsilon = 0$.
\end{theorem}

The result in Theorem~\ref{thrm:star_connected} can be interpreted as
follows: the star-connected network can achieve selective inhibition
and recruitment if each layer of the network dynamics is independently
stable and the magnitude of the thalamocortical and corticothalamic
connections does not exceed a determined stability margin. This bound
on the magnitude of the corticothalamic feedback in the network aligns
with neuroscientific observations: it has been seen that enhanced
corticothalamic feedback can result in pathological
behavior~\citep{KG-JMD:20}. In particular, strong corticothalamic
feedback has been found to coincide with epileptic loss of
consciousness in absence seizures due to over-inhibition of the
cortical regions~\citep{GKK:01}.

\begin{myblock}{Remote Synchronization in Star-Connected
    Thalamocortical Networks}
  A phenomenon of vast classical and recent interest in neuroscience
  is that of synchronization between brain
  regions~\citep{SRC-BV:17,GB-AD:04}. In particular, \emph{remote
    synchronization} refers to the condition where regions in a
  network synchronize despite a lack of direct links between
  them~\citep{LVG-AC-AF-LF-JG-MF:13}. Remote synchronization is
  frequently studied in networks of oscillators, such as the Kuramoto
  model~\citep{HS-YK:86} of neural population phase oscillators. In
  particular, \cite{YQ:19} studies remote synchronization in
  star-connected networks where synchronization can only happen
  remotely between peripheral nodes due to the lack of direct
  connections between them.


  Remote synchronization is also commonly observed in the brain
  between distant cortical regions, often in the context of functional
  connectivity analysis~\citep{VV-PH:14}. Given the thalamus' hub-like
  connectivity to cortical regions (similar to the above star
  topology), the thalamus has been suggested to play a key role in remote
  synchronization within the brain~\citep{LG-CM-AV:10}. Interestingly,
  it is shown in~\cite{VN-MV-MC-ADG-VL:13} that remote synchronization
  is dependent on symmetry between the outer regions in a
  star-connected topology, a phenomenon that is also shown to hold for
  remote synchronization of both linear-threshold and Kuramoto
  oscillator star-connected thalamocortical networks
  in~\cite{MM-JC:24-tcns} and~\cite{YQ:19}, respectively.
\end{myblock}

\subsubsection{Comparison of Selective Inhibition and Recruitment
  Across Interconnected Topologies}

From the discussion in Sections~\ref{sec:hierarchical-networks}
and~\ref{sec:star-networks}, we can see that the linear-threshold rate
dynamics allow for the network to achieve selective inhibition and
recruitment in multiple interconnected network topologies. In the two
topologies considered, 
the thalamus serves significantly different roles: those of first order
vs. higher order nuclei~\citep{SMS:12}. This, in turn, results in
different network properties across the two topologies, including
different conditions for achieving selective inhibition. In
particular, the hierarchical network is strongly dependent on the
interconnection properties between the cortical regions, whereas the
star-connected network is more reliant on the internal dynamics of
each layer.

However,
a star-connected topology can also arise directly from the
hierarchical thalamocortical network in the case of damage to the
network. If the hierarchical thalamocortical network incurs injury
such that one or more of the information pathways between cortical
regions are damaged, the network topology becomes locally akin to the
star topology of Figure~\ref{fig:network_topologies}(b). This
possibility underscores the importance of the transthalamic pathways
between cortical regions and the ability to achieve selective
inhibition and recruitment in a star-connected network. Indeed, if we
consider a strictly cortical hierarchy as in~\citep{EN-JC:21-tacII},
the removal of any information pathway would disconnect the underlying
chain-like
topology. 
In the following example, we illustrate achieving selective inhibition
and recruitment in a hierarchical thalamocortical network along with
the star-connected network formed by damage to the corticocortical
connections.

\begin{figure}[tbh]
  \centering \includegraphics[width =
  0.7\linewidth]{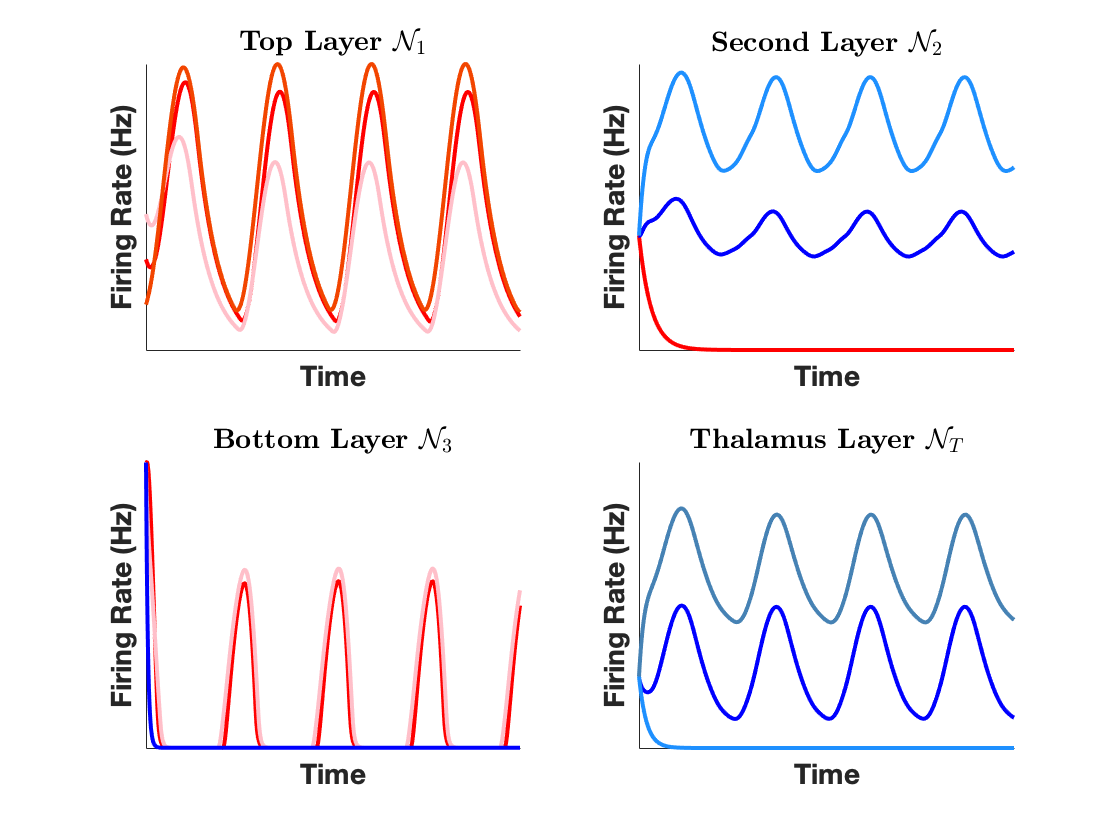}
  \caption{Selective inhibition in a three-layer hierarchical
    thalamocortical network. In the second, third, and thalamic
    layers, one node is selectively inhibited while the other two are
    recruited. }\label{fig:hierarchical_example}
\end{figure}

We consider a hierarchical network composed of three cortical layers
and one thalamic layer. Each layer is composed of three neurons, with
the top layer being purely excitatory and the thalamus being purely
inhibitory. The second layer is composed of two inhibitory nodes and
one excitatory, while the third contains two excitatory and one
inhibitory nodes. We drive the top layer with an oscillatory input and
aim to selectively inhibit and recruit nodes in the remaining three
layers. The parameters for the network are as follows.
\begin{align*}
  \W_1 &= \begin{bmatrix}
    0.4461  &  0.1125  &  0.4637 \\
    0.1213 &   0.1750  &  0.0257 \\
    0.0648  &  0.1435  &  0.2963
  \end{bmatrix} \qquad \W_2 = \begin{bmatrix}
    0.4005 &  -0.3816 &  -0.3963 \\
    0.1165 &  -0.4132 &  -0.1645 \\
    0.4662 &  -0.2867 &  -0.1117
  \end{bmatrix}  \\
  \W_{21} &= \begin{bmatrix}
    0.1668  &  0.3416 &  0.4702 \\
    0.1148  &  0.4811 &  0.0029 \\
    0.4681  &  0.2190 &  0.3052 
  \end{bmatrix} \qquad \W_{12} = \begin{bmatrix}
    0.0814 &  -0.2511 &  -0.0235 \\
    0.4192 &  -0.4997 &  -0.1068 \\
    0.0838 &  -0.1777 &  -0.1989
  \end{bmatrix} \\
  \W_{T,1} &= \begin{bmatrix}
    0.4782  &  0.1382  &  0.4817 \\
    0.2865  &  0.3112  &  0.0430 \\
    0.4249  &  0.2942  &  0.2502
  \end{bmatrix} \qquad \W_{32} = \begin{bmatrix}
    0.4136 & -0.2379 &  -0.4003 \\
    0.3379 & -0.1995 &  -0.0525 \\
    0.1245 & -0.2997 &  -0.4107
  \end{bmatrix} \\
   \W_T &= \begin{bmatrix}
   -0.3482 &  -0.4450 &  -0.0570 \\
   -0.2599 &  -0.1651 &  -0.1555 \\
   -0.0295 &  -0.1149 &  -0.1142 
  \end{bmatrix}  \qquad  \W_3 = \begin{bmatrix}
    0.4205 &  0.2861 &  -0.3789 \\
    0.1773 &  0.3504 &  -0.1946 \\
    0.2150 &  0.3712 &  -0.2147
  \end{bmatrix}   \\
   \W_{1,T} &= \begin{bmatrix}
   -0.1184 &  -0.4869 &  -0.4300 \\
   -0.3511 &  -0.4862 &  -0.2009 \\
   -0.1877 &  -0.3218 &  -0.3160
  \end{bmatrix} \qquad \W_{23} = \begin{bmatrix}
    0.1562 &  0.1452 &  -0.3074 \\
    0.2923 &  0.2013 &  -0.4956 \\
    0.4150 &  0.4310 &  -0.1018
  \end{bmatrix} \\
    \W_{2,T} &= \begin{bmatrix}
   -0.4961 &  -0.4507 &  -0.0542 \\
   -0.2012 &  -0.4977 &  -0.0181 \\
   -0.3294 &  -0.3266 &  -0.3090
  \end{bmatrix} \qquad  \W_{T,3} = \begin{bmatrix}
    0.2228  &  0.1519   -0.3992 \\
    0.4220  &  0.2416   -0.4937 \\
    0.0981  &  0.1689   -0.0795
  \end{bmatrix} \\
  \W_{T,2} &= \begin{bmatrix}
    0.2608 &  -0.4422 &  -0.0742 \\
    0.0451 &  -0.2195 &  -0.3099 \\
    0.4523 &  -0.3909 &  -0.1303
  \end{bmatrix} \qquad   \W_{3,T} = \begin{bmatrix}
   -0.2836 &  -0.3313 &  -0.4810 \\
   -0.4810 &  -0.2617 &  -0.2701 \\
   -0.3731 &  -0.1299 &  -0.0151
  \end{bmatrix} \\
  \c_1 &= \c_2 = \c_T = \begin{bmatrix}
    0.5 \\ 1.25 \\ 2
  \end{bmatrix} \qquad \c_3 = \begin{bmatrix}
    2 \\ 1.25 \\ 2 
  \end{bmatrix}
\end{align*}
The threshold is set to be $\m = 10$ and timescales are $\tau_1 = 3$, $\tau_2 = 1.625$, $\tau_3 = 0.25$, and $\tau_T = 1$. These parameters satisfy the conditions of Theorem~\ref{thrm:thalamocortical_hierarchy_conditions} and in Figure~\ref{fig:hierarchical_example} we show that we can achieve selective
inhibition and recruitment for the hierarchical network with the aim
of inhibiting one node in each of the layers other than the top
layer. The remaining nodes are recruited to oscillatory equilibrium
trajectories based on the equilibrium
maps~\eqref{eq:equilibrium_maps}. Note that they are non-constant
trajectories since we drive the top layer in an oscillatory manner.

\begin{figure}[tbh]
  \centering \includegraphics[width =
  0.7\linewidth]{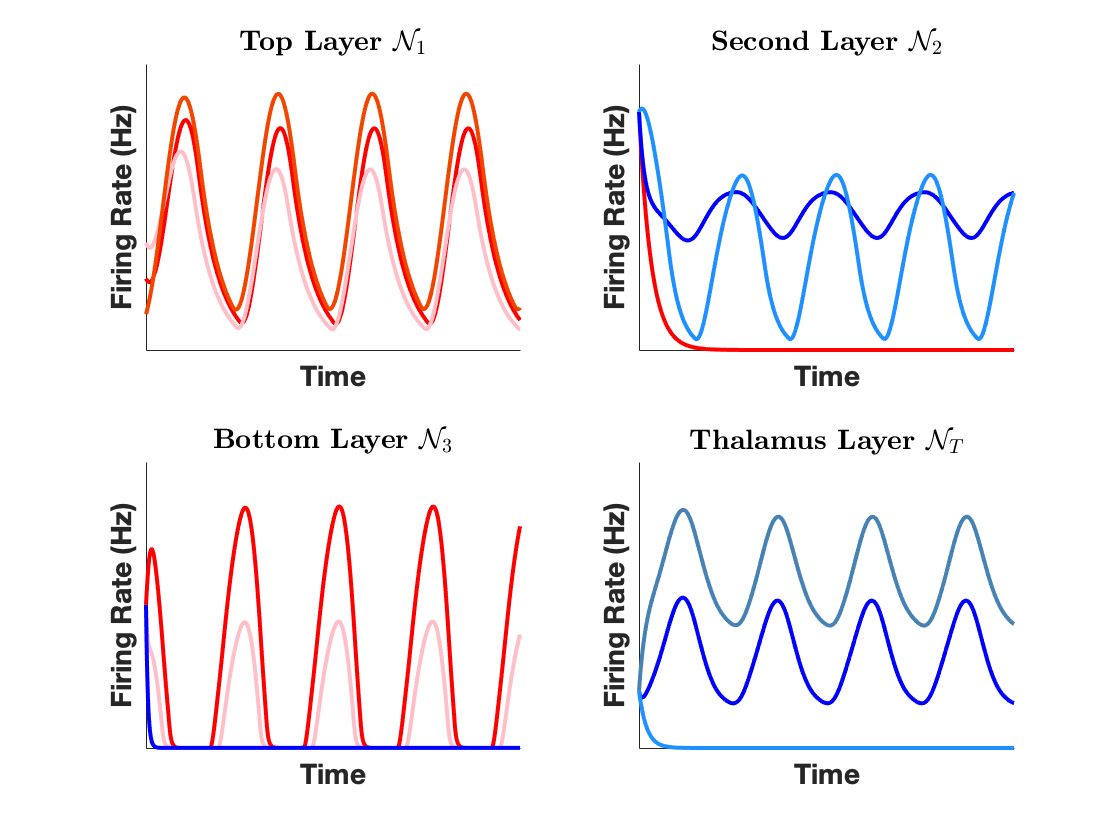}
  \caption{Selective inhibition in a three-layer star-connected
    thalamocortical network formed by removing the corticocortical
    connections from a hierarchical network. In the second, third, and
    thalamic layers, one node is selectively inhibited while the other
    two are recruited. As in the original network, selective
    inhibition and recruitment is achieved, but the recruitment is to a
    different set of equilibria trajectories, as
    expected.}\label{fig:star_connected_example}
\end{figure}

We next consider the scenario where the corticocortical connections in
the network are severed, resulting in a star-connected network. Note
that a hierarchy of timescales still remains between the layers. In
Figure~\ref{fig:star_connected_example}, we see that we can still
achieve selective inhibition and recruitment, but the equilibrium
trajectories to which the task-relevant nodes are recruited are
different than the original network. This is expected as the
equilibrium maps are dependent on the interconnections between the
cortical regions, which are now all zero.

%
%

This example illustrates that the linear-threshold dynamics are
stabilizable in interconnected networks with conditions dependent on
the topologies of the network. This matches the fact that different
regions of the brain have different topologies but are still able to
exhibit similar phenomena.

\subsection{Epilepsy in Interconnected Linear-Threshold
  Networks}\label{subsec:epilepsy_spreading}

We have already analyzed the emergence of epileptic seizures through
bifurcations, representing a sudden change from healthy to unhealthy
dynamics, in the linear-threshold model for a single brain region.
However, more complex patterns emerge during an epileptic seizure when
looking at networks of multiple brain regions. One such pattern is the
excessive synchrony of activity between regions. Further, recalling
from Section~\ref{subsec:single-region-epilepsy} that specific
oscillatory waveforms in the EEG appear during
seizures~\citep{RDT:03,PC:96}, epileptic behavior can appear when
there are extensive synchronous oscillations throughout the
brain~\citep{OD-AV-TJO-NJ-IES-MdC-PP:18}. It has been suggested that
the \textit{broad spread} of synchronized pathological oscillations is
a critical factor in generalized seizures~\citep{SSS:02}, whereas
hyper-synchronized oscillations confined to a local brain area do not
necessarily give rise to a
seizure~\citep{MS-MB-BHB-KL-WRM-FBM-BL-JvG-GAW:10}.
In the following, we discuss epileptic behavior through the modeling
of oscillation spreading throughout an interconnected network of
excitatory-inhibitory pairs in a synchronous manner, following the
exposition
in~\citep{AA-FC-FP-JC:22-ojcsys}. We study excitatory-inhibitory pairs as they aresimple but non-trivial networks that exhibit rich dynamical behavior, and allow for the construction of large high-dimensional systems through the interconnection of many pairs.
%
%
Box~\ref{box:construction_EI_pair_networks}
provides a construction of the networks considered in this section.

\begin{myblock}{Networks of Coupled Excitatory-Inhibitory (E-I)
    Pairs}\label{box:construction_EI_pair_networks}
  Coupled networks of E-I pairs are defined by the interconnection of
  the dynamics of individual E-I pairs, modeling information
  processing pathways between regions. We construct a network of $N$
  coupled E-I pairs as follows. Each E-I pair has dynamics defined by
  synaptic weight matrix $\W_i$, input $\u_i$ and threshold $\m_i$:
  \begin{align*}
    \W_{i} = \begin{bmatrix}
               a_i & -b_i \\
               c_i & -d_i
    \end{bmatrix} \qquad \u_i = \begin{bmatrix}
      u_{i}^{E} \\ u_{i}^{I}
    \end{bmatrix} \qquad \m_i = \begin{bmatrix}
      m_{i}^{E} \\
      m_{i}^{I}
    \end{bmatrix}.
  \end{align*}
  The coupling between E-I pairs $i$ and $j$ is then defined by weight
  matrix $\W_{ij}$ of the form
  \begin{align*}
    \W_{ij} =
    \begin{bmatrix}
      w_{ij}^{EE} & -w_{ij}^{EI} \\
      w_{ij}^{IE} & -w_{ij}^{II}
    \end{bmatrix}.
  \end{align*}
  The weights $w_{ij}^{EE},w_{ij}^{EI},w_{ij}^{IE}$ and $w_{ij}^{II}$
  are nonnegative values that define the excitatory-excitatory,
  excitatory-inhibitory, inhibitory-excitatory and
  inhibitory-inhibitory connections, respectively. Then, the coupled
  E-I pair network has dynamics for the excitatory and inhibitory
  components of node $i$ given by
  \begin{align}\label{eq:interconnected_EI_pair_equations}
    \dot{x}_i^{E}
    &= -x_{i}^{E} + [a_i x_i^{E} - b_i x_{i}^{I} +
      \sum_{j \neq i} (w_{ij}^{EE} x_j^{E} +
      w_{ij}^{EI} x_j^{I}) + u_i^{E}]_0^{m_i^{E}}
      \cr 
      \dot{x}_i^{I} &= -x_{i}^{I} + [c_i x_i^{E} - d_i x_{i}^{I} +
                       \sum_{j \neq i} (w_{ij}^{IE} x_j^{E} +
                       w_{ij}^{II} x_j^{I}) + u_i^{I}]_0^{m_i^{I}}. 
  \end{align}
     This notation for networks of coupled
  excitatory-inhibitory pairs will be used in studying oscillation
  spreading through these networks in the context of epilepsy.
\end{myblock}

As oscillations are a key part of the modeling of epileptic seizures
in this approach, we first discuss conditions such that oscillations
appear in individual linear-threshold networks, before considering
their synchronization and spreading between interconnected regions.

\subsubsection{Oscillations in Linear-Threshold Networks}
Oscillations can arise in linear-threshold brain networks under a
variety of conditions.
Instead of using perfectly periodic trajectories (as in a limit cycle)
as the defining property of an oscillatory system, we use lack of
stable equilibria. This choice is motivated by the hypotheses in the
Poincar\'e-Bendixson theorem~\citep{EAC-NL:55}
%
%
for establishing the existence of limit cycles in planar systems, but
(1) relaxes the (unrealistic) need for exact periodicity and allows
for chaotic oscillations that better match biological neural
oscillations, and (2) opens the door to theoretical analyses in
higher-dimensional systems.  As such, we put forth the following
definition of oscillatory behavior.

\begin{definition}\longthmtitle{Oscillatory and Inactive
    Nodes~\citep{AA-FC-FP-JC:22-ojcsys}}
  Consider a linear-threshold
  network~\eqref{eq:network_lin_threshold_dynamics} composed of $n$
  nodes. We say that the $i$th node of the system is
  \emph{oscillatory} if for all solutions of network dynamics,
  $\x_i(t)$ does not converge to a constant value as $t \to
  \infty$. Furthermore, a non-oscillatory node is said to be
  \emph{inactive} if for all network solutions, $\x_i(t) \to 0$ as
  $t \to \infty$.
\end{definition}

As noted earlier, we are interested in the spreading of oscillations
in networks of interconnected excitatory-inhibitory pairs. This begins
by investigating the appearance of oscillations in an individual
excitatory-inhibitory pair. For that, consider the linear-threshold
dynamics~\eqref{eq:network_lin_threshold_dynamics} with constant input
$\u(t) = \u$ for all $t \in \R$ and synaptic weight matrix
\begin{align*}
  \W = \begin{bmatrix}
         a & -b \\
         c & -d
  \end{bmatrix},
\end{align*}
where $a,b,c,d \in \Rpluseq$.  The existence of oscillations (which
coincide with limit cycles in this 2D system) can be analytically
characterized as follows.

\begin{theorem}\longthmtitle{Oscillations in Excitatory-Inhibitory
    Pairs~\citep{EN-RP-JC:22-auto}}\label{thrm:oscillations-EI-pairs} 
  Consider an excitatory-inhibitory pair governed by the
  linear-threshold dynamics~\eqref{eq:network_lin_threshold_dynamics}. All
  network trajectories (except those with an initial condition at an
  unstable equilibrium) converge to a limit cycle if and only if
  \begin{subequations}\label{eq:independent_oscillations_EI}
    \begin{align}
      d+2 &< a, \\
      (a-1)(d+1) &< bc, \\
      (a-1)m_1 &< bm_2, \\
      0 < u_1 &< bm_2 - (a-1)m_1, \\
      0 < (d+1)u_1 - bu_2 &< [bc-(a-1)(d+1)]m_1.
    \end{align}
  \end{subequations}
\end{theorem}

These necessary and sufficient conditions for limit cycles in E-I
pairs are the basis for the analysis of the spreading of synchronous
oscillations throughout a network of interconnected pairs in upcoming
sections. We note that conditions for oscillations have been
characterized for linear-threshold networks with a variety of other
structures, a discussion of which can be found in
Section~\ref{sec:discussion-conclusions}. In the following we discuss the spreading of
oscillations in relation to epileptic seizures utilizing E-I pairs and Theorem~\ref{thrm:oscillations-EI-pairs}. The exact characterization of oscillations in E-I pairs allows for a more complete analysis of the interconnected networks than if we instead used more general topologies.


%
%

\subsubsection{Spreading Oscillations and Large-Scale Synchrony}

\begin{figure}[tbh]
  \centering
  \includegraphics[width=0.8\linewidth]{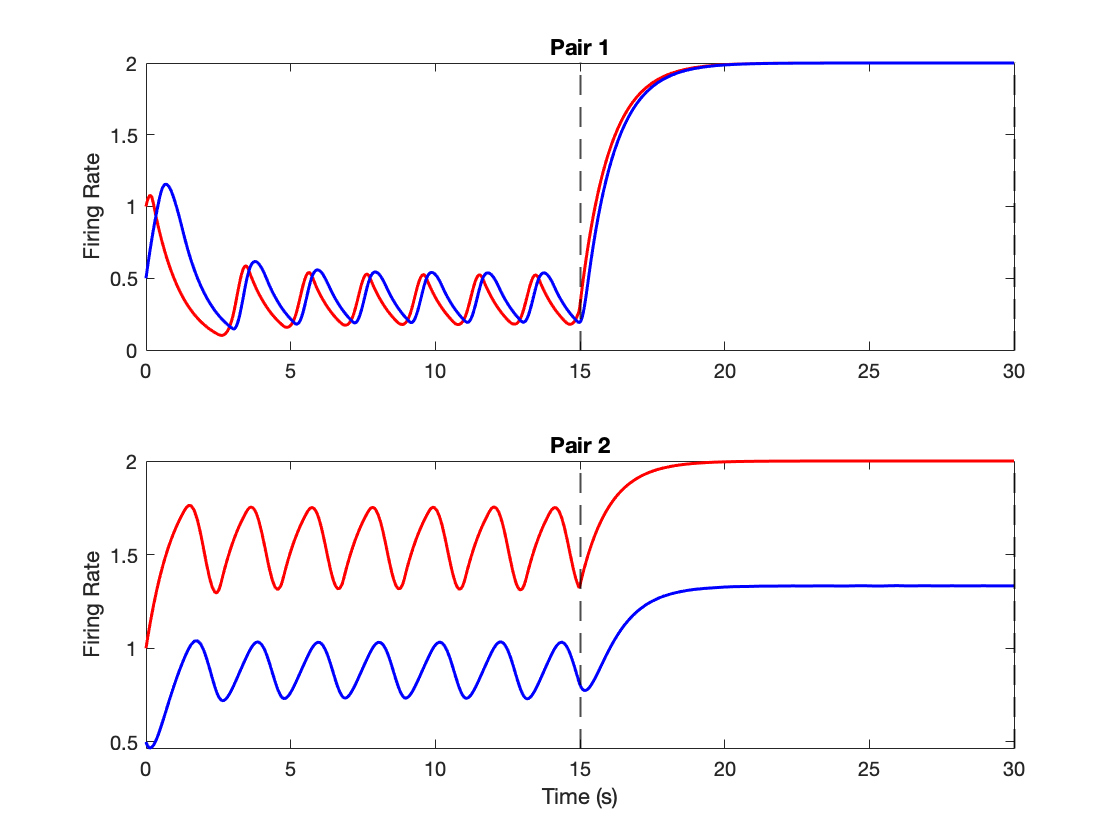}
  \caption{Illustration of oscillations stopping in a network
    following the interconnection of E-I pairs. Here the excitatory
    nodes are denoted by red lines and the inhibitory nodes by blue
    lines. Both pairs satisfy the conditions of
    Theorem~\ref{thrm:oscillations-EI-pairs} and independently
    oscillate, but at $t=15$ when the pairs are connected, the
    oscillatory behavior ends in the entire
    network.}\label{fig:losing_oscillations}
\end{figure}

As generalized seizures occur with synchronous oscillations across
multiple brain regions~\citep{LGD-RAW-WG-DC-OCS-JLPV:05}, we wish to understand
parameters such that oscillations occur in interconnected networks as
defined in Box~\ref{box:construction_EI_pair_networks}. While
Theorem~\ref{thrm:oscillations-EI-pairs} provides conditions such that
we have oscillations in a single region, in an interconnected network,
even if these conditions are satisfied for each pair, the nodes in
that region may no longer oscillate. Conversely, it could also be
the case that that a node that in an individual network would not be
oscillating may oscillate in the interconnected network. For instance,
Figure~\ref{fig:losing_oscillations} shows a network of two E-I pairs
governed by the linear-threshold dynamics with synaptic weight
matrices and controls
\begin{align*}
  \W_1 = \begin{bmatrix}
           4 & -6 \\
           5 & -1
  \end{bmatrix} \qquad \W_2 = \begin{bmatrix}
    5 & -8 \\
    3 & -2
  \end{bmatrix} \qquad \u_1 = \begin{bmatrix}
    1 \\ -1
  \end{bmatrix} \qquad \u_2 = \begin{bmatrix}
    1 \\ 0
  \end{bmatrix},
\end{align*}
with a threshold value of $\m = 2$.
Both of these networks individually satisfy the conditions of
Theorem~\ref{thrm:oscillations-EI-pairs} and oscillate per se, see the
first 15s in the plots of
Figure~\ref{fig:losing_oscillations}. However, after interconnecting
them via their excitatory nodes with interconnection matrices
\begin{align*}
  \W_{12} = \W_{21} = \begin{bmatrix}
                        3 & 0 \\ 0 & 0
  \end{bmatrix},
\end{align*}
the oscillations in the network stop, as displayed in the last 15s in
the plots of Figure~\ref{fig:losing_oscillations}.

\begin{figure}[tbh]
\centering
  \includegraphics[width = 0.8\linewidth]{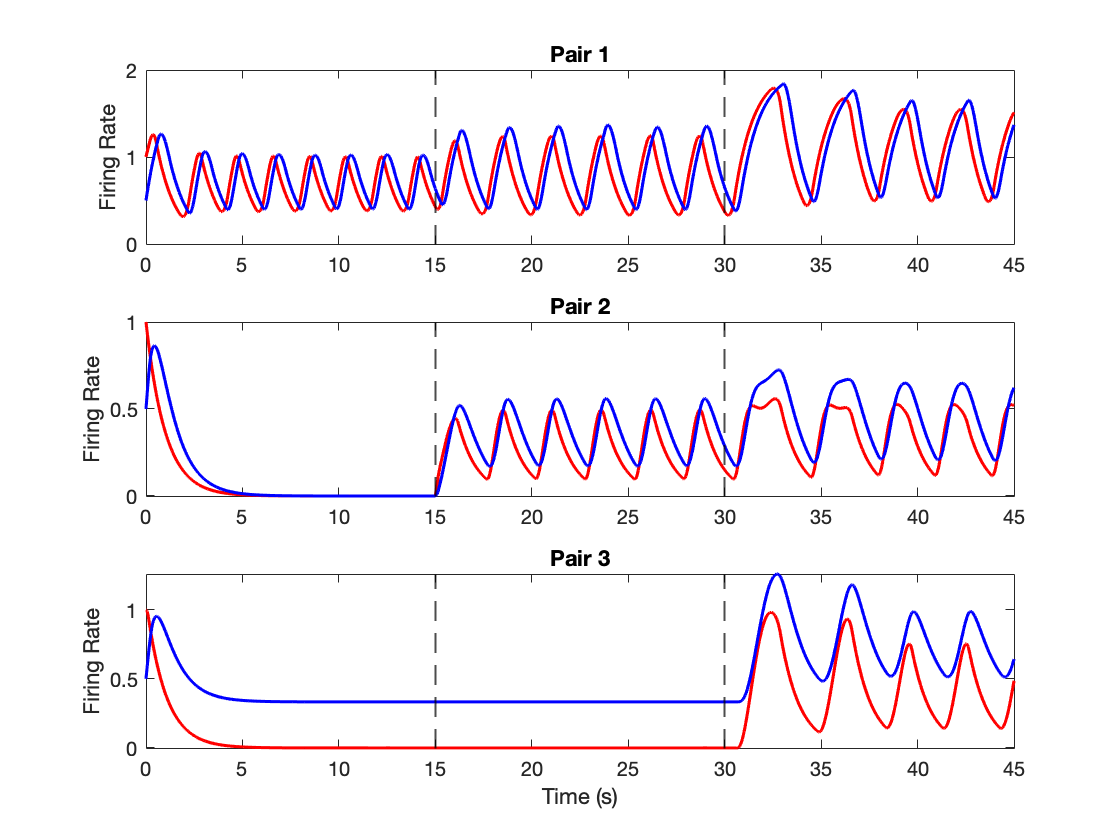}
  \caption{Illustration of oscillations spreading through a network of
    excitatory-inhibitory pairs upon interconnections. Here the
    excitatory nodes are denoted by red lines and the inhibitory nodes
    by blue lines. Initially the three pairs are distinct and only one
    oscillates, but as they are interconnected at $t=15$ and $t = 30$,
    oscillations spread from the first pair to the
    others.}\label{fig:spreading_oscillations}
\end{figure}

Figure~\ref{fig:spreading_oscillations} illustrates the opposite
phenomena, whereby upon the interconnection of excitatory-inhibitory
pairs, nodes that were not oscillating can begin to oscillate. Here we
consider a network composed of three pairs, defined by the following
synaptic weight matrices and controls
\begin{align*}
  \W_1 &= \begin{bmatrix}
            4 & -6 \\
            5 & -1
  \end{bmatrix} \qquad \W_2 = \begin{bmatrix}
    1 & -5 \\
    4 & -2
  \end{bmatrix}
  \qquad \W_3 = \begin{bmatrix}
    3 & -4 \\
    3 & -2
  \end{bmatrix} \\
  \u_1 &= \begin{bmatrix}
    2 \\ -2
  \end{bmatrix} \qquad \u_2 = \begin{bmatrix}
    0 \\ 0
  \end{bmatrix} \qquad \u_3 = \begin{bmatrix}
    0 \\ -1
  \end{bmatrix},
\end{align*}
and a threshold value of $2$. The first pair, governed by $\W_1$
oscillates on its own as per the conditions of
Theorem~\ref{thrm:oscillations-EI-pairs}, while the other two do
not. In Figure~\ref{fig:spreading_oscillations} we show first the
unconnected networks, before connecting pairs $1$ and $2$ and finish
by connecting pair $3$ to the network using the following
interconnection matrices, where $\W_{ij}$ defines the connections from
pair $j$ to pair $i$.
\begin{align*}
  \W_{12} &= \begin{bmatrix}
    2 & 0 \\ 0 & 0
  \end{bmatrix}\qquad \W_{21} = \begin{bmatrix}
                                  2 & 0 \\ 0 & 0
  \end{bmatrix}\qquad \W_{13} = \begin{bmatrix}
    2 & 0 \\ 0 & -1
  \end{bmatrix} \\
  \W_{31} &= \begin{bmatrix}
    2 & 0 \\ 0 & 0
  \end{bmatrix}\qquad \W_{23} = \begin{bmatrix}
    0 & 0 \\ 0 & 0
  \end{bmatrix}\qquad \W_{32} = \begin{bmatrix}
    2 & 0 \\ 0 & 0
  \end{bmatrix}
\end{align*}

We are able to see that pair $1$ maintains its oscillations throughout
the interconnections, while both of the other pairs move from
non-oscillating to oscillating after connecting, despite not
satisfying the conditions of
Theorem~\ref{thrm:oscillations-EI-pairs}. This example illustrates the
ability of the network to spread (putatively pathological)
oscillations. 
We are thus interested in determining properties of the individual
regions that allow for maintaining either the oscillatory or inactive
behavior upon interconnection of regions. In order to do so, we
condense the interconnection terms in the dynamics for the coupled E-I
pairs in Box~\ref{box:construction_EI_pair_networks} to the following,
\begin{align}\label{eq:combined_inputs_coupled_EI}
  \tilde{u}_i^{E}(\x)
  &= u_i^{E} + \sum_{j \neq i} (w_{ij}^{EE}x_{j}^{E} - w_{ij}^{EI}x_{j}^{I}) \cr
    \tilde{u}_i^{I}(\x) &= u_i^{I} + \sum_{j \neq i} (w_{ij}^{IE}x_j^{E} - w_{ij}^{II}x_j^{I}).
\end{align}
 This compiles the interconnections between the E-I pairs into one input term in the
network, providing a similar form to that of an individual uncoupled
E-I pair. Using this change of variables and combining it with
Theorem~\ref{thrm:oscillations-EI-pairs}, we can provide sufficient
conditions for inactivity and oscillations to be maintained upon
network interconnections.

\begin{theorem}\longthmtitle{Robust Behaviors in Coupled
    Networks~\citep{AA-FC-FP-JC:22-ojcsys}}\label{thrm:robust_inactivity}
  Consider a network of $N$ coupled E-I pairs as in
  Box~\ref{box:construction_EI_pair_networks} and define the following
  set
  \begin{align*}
    \mathcal{U}_i = \{(u_i^{E},u_i^{I}) ~|~ \eqref{eq:independent_oscillations_EI}~\mathrm{ are~satisfied} \}.
  \end{align*}.
  If for all $\x$
  \begin{align*}
    \begin{bmatrix}
      \tilde{u}_i^{E}(\x) \\
      \tilde{u}_i^{I}(\x)
    \end{bmatrix} \in \mathcal{U}_i,
  \end{align*}
  then the $i$th pair in the network will oscillate after coupling. If, instead
  \begin{align*}
    u_{i}^E + \sum_{j \neq i} w_{ij}^{EE} m_j^{E} &\leq 0, \\
    u_i^{I} + \sum_{j \neq i} w_{ij}^{IE} m_{j}^{E} &\leq 0,
  \end{align*}
  the $i$th pair in the network does not oscillate after the network
  is connected.
\end{theorem}

%

These results provide an analytical basis for determining when E-I
pairs will either be inactive or oscillate following their
interconnection. Some remarks are then in order. First, we note that
both results are sufficient results and, as such, both the loss and
spreading of oscillations can occur without them being
satisfied. Second, while these results provide conditions such that
oscillations or inactivity can be maintained in the event of
interconnections, they do not directly provide conditions for the
spreading or loss of oscillations following interconnections. In fact,
they only determine if there will be no change after interconnection,
rather than providing conditions on the networks such that
oscillations will spread or stop. Determining direct conditions on the
network parameters so that oscillations spread or disappear in a
network (as shown in the prior examples) is an open problem. The work
in~\citep{AA-FC-FP-JC:22-ojcsys} examines how the spread of
oscillations throughout the network can be controlled through the
modification of network weights, instead of restricting the properties
of the original network itself.

These results and examples have illustrated that a network of couple
E-I pairs with the linear-threshold dynamics can exhibit a variety of
oscillatory behavior. In particular, individual networks can exhibit
limit cycles and, through interconnections, these oscillations can
both spread throughout the network or stop altogether. As generalized
seizures are related to the spread of oscillations between brain
regions~\citep{MS-MB-BHB-KL-WRM-FBM-BL-JvG-GAW:10}, these properties
make the linear-threshold dynamics a good model for the estimation and
control of this dynamical behavior.

 \section{Discussion and Open Avenues}\label{sec:discussion-conclusions}


In the prior sections, we have discussed a variety of properties of
the linear-threshold network dynamics and illustrated their relevance
in modeling brain behavior. This has provided an extensive look at
their use in particular applications. We believe there are further
dynamical properties and control mechanisms that should be studied,
both in their own right, without relation to a specific brain
function, as well as for exploring other avenues for using
linear-threshold networks in modeling the brain.  Here we provide a
discussion on oscillatory properties of the dynamics with different
topologies, paying special attention to the characterization of
interconnections that explain observed behavior in neuronal
populations, and on the potential for systems and control to inform
targeted interventions that leverage the anatomical wiring structure
and explain the mechanisms behind dynamic dimensionality control and
spatial computing in the brain.


\subsection{Oscillatory Behavior in the Brain}
When observing brain networks, oscillations are omnipresent and have
significant ranges in both magnitude and frequency~\citep{GB:06}. In
Section~\ref{sec:multi-region} we discussed oscillations in the
context of epileptic seizures and saw the appearance of oscillations
in EEG measurements in
Figure~\ref{fig:epileptic_waveform_types}. However, oscillations have
been linked with a large number of different cognitive tasks, such as
information processing~\citep{CB-AM-CF:23} and spatial
cognition~\citep{JBC-JRM-SR-MJK:01}.  In our treatment of oscillations
in Section~\ref{sec:multi-region} we limited the network topologies to
individual and coupled E-I pairs. However, for the purposes of
properly understanding the oscillatory properties of the
linear-threshold dynamics and applying the model to other appearances
of oscillations in the brain, it is important to consider other
structures. We do this next.

\subsubsection{Oscillations in Linear-Threshold Dynamics}
In considering oscillations in linear-threshold dynamics, we recall
that due to the difficulties of their exact characterization in
networks of dimension greater than two, we use the proxy of lack of
stable equilibria for oscillations. Indeed, using this
proxy, one can show for a variety of network architectures that the
linear-threshold dynamics admits oscillatory behavior of different
forms, including limit cycles, quasi-periodicity, and chaos. In what
follows we give a summary of oscillatory behavior in three different
network architectures. We note that these network architectures do not
necessarily satisfy Dale's Law, but under various assumptions can
still be used as models for a brain network:
\begin{itemize}
\item \emph{Competitive Networks} are characterized by having their
  weight matrix be a Z-matrix, i.e., one in which all off-diagonal
  elements are non-positive. Such networks have been studied both for
  the linear-threshold dynamics~\citep{MM-TM-JC:23-csl} and the
  unbounded threshold-linear dynamics~\citep{KM-AD-VI-CC:24}. The
  existence of oscillations for such networks have been classified
  using multiple tools: for linear-threshold dynamics, the appearance
  of oscillations with specific support (i.e., set of oscillating
  nodes) has been shown through properties of Z-matrices and subsets
  of the network~\citep{MM-TM-JC:23-csl}; for threshold-linear
  dynamics, the lack of existence of stable equilibria is established
  through graph-theoretic methods~\citep{KM-AD-VI-CC:24}.

\item \emph{Combinatorial Networks} are a special type of competitive
  network in which the weight matrix
  $\W = \{w_{ij}\}_{i,j \in \{1,\dots,n\}}$ is defined by the
  connections in the network and two parameters $\epsilon, \delta$ as follows:
  \begin{align*}
    w_{ij} =
    \begin{cases}
      0 & i = j,
      \\
      -1 + \epsilon & \text{if there is a connection from node $j$ to
                      node $i$},
      \\
      -1 + \delta & \text{if there is no connection from node $j$ to node $i$},
    \end{cases}
  \end{align*}
  where $0 < \epsilon < 1$ and $\delta > 0$. Oscillations in such
  networks have been studied extensively for the threshold-linear
  dynamics~\citep{CP-SM-KM-CC:22}. In a similar fashion to the
  standard competitive networks, in~\citep{KM-AD-VI-CC:24} it is shown
  that oscillations can occur in the combinatorial threshold-linear
  network (CTLN), and can be constructed such that a variety of
  behaviors occur.  Figure~\ref{fig:CTLN_behavior} shows that with
  varying network structures, limit cycles, chaos, and quasi-periodic
  orbits can occur. Beyond having the existence of such behavior,
  in~\citep{CP-SM-KM-CC:22} conditions are given that predict the
  number and type of both dynamic and static attractors based on analysis of sources and sinks in the graph corresponding with the network structure.

  \begin{figure}[htb]
    \centering
    \includegraphics[width=0.8\linewidth]{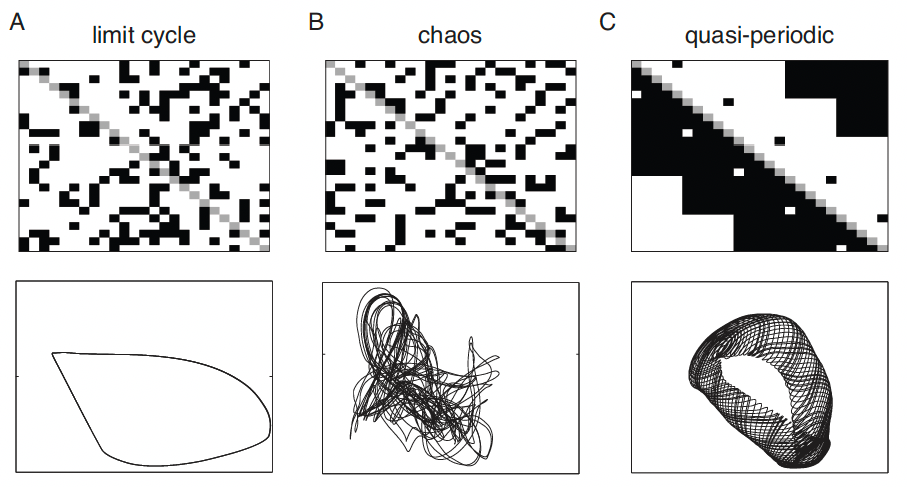}
    \caption{\citep{KM-AD-VI-CC:24} Samples of the possible
      oscillatory behaviors with the CTLN dynamics. The top panels
      illustrate the pattern of the weight matrix, with black squares
      indicating non-zero connections, white squares indicating no
      connections, and diagonal self-connections in grey. The bottom
      panels show two-dimensional projections of the dynamics. From
      this we can see that under appropriate network structure and
      parameter choices, the CTLN dynamics exhibit a wide range of
      oscillatory behavior.}\label{fig:CTLN_behavior}
\end{figure}

\item \emph{Coupled Networks} are those composed of a set of networks
  of a similar form coupled together with additional connections, such
  as the coupled E-I pairs from Section~\ref{sec:multi-region}. The
  oscillatory behavior of such networks can have a variety of
  properties. In particular, for the linear-threshold dynamics on a
  coupled network, in~\citep{EN-JC:19-acc} three different properties
  of the oscillations are analyzed: regularity, synchronization, and
  phase-amplitude coupling.  Using frequency domain techniques, a
  regularity index is defined with a value of $1$ corresponding with
  no oscillations and $\infty$ equal to perfectly regular
  oscillations.  As the interconnection strength increases, the
  regularity of the oscillations reduces and the dynamics exhibit more
  chaotic behavior before reaching a point which guarantees a stable
  equilibrium. Additionally, as the number of coupled oscillators
  increases, the behavior of the oscillations becomes increasingly
  chaotic. Figure~\ref{fig:MLE_chaos} shows that, when using the
  Maximal Lyapunov Exponent (MLE) as a proxy for chaos, as both
  interconnection strength and the size of the network increases. the
  MLE increases until reaching the point at which stable equilibria
  appear and oscillations disappear. It is also shown
  in~\citep{EN-JC:19-acc} that synchronization and phase-amplitude
  coupling in coupled networks increases with the interconnection
  strength between network components.

  \begin{figure}[tbh]
    \centering
    \includegraphics[width = 0.8\linewidth]{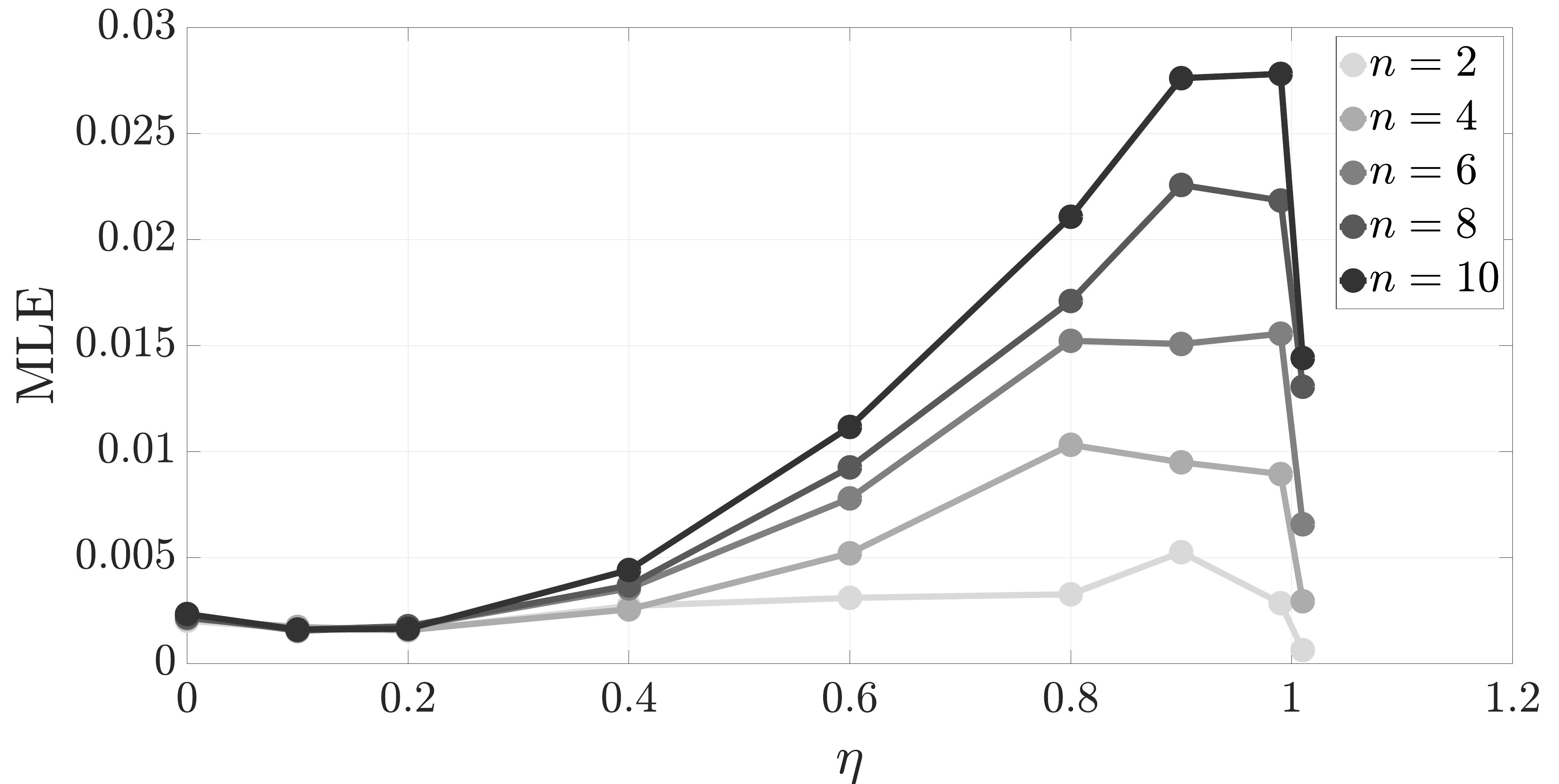}
    \caption{MLE as a function of network size ($n$) and
      interconnection strength between E-I pairs ($\eta$). The MLE
      increases as a function of both parameters until $\eta$ reaches
      a critical value at which point the MLE drops immediately due to
      the lack of oscillations in the network caused by the existence
      of a stable
      equilibrium~\citep{EN-JC:19-acc}.}\label{fig:MLE_chaos}
  \end{figure}

\end{itemize}

For further information we direct the interested reader to the above
references along with~\citep{EN-RP-JC:22-auto} and the references
therein.

\subsubsection{Communication through Coherence (CTC)}

We start our discussion of future avenues for the application of the
linear-threshold dynamics to studying brain functions by examining
communication through coherence.  Communication in the brain
fundamentally relies on the transmission of the response of one neuron
to the inputs arriving from another through the
network~\citep{PF:15}. When observing two brain regions, each
exhibiting oscillatory behavior, that are communicating, it is common
that the regions exhibit a level of
synchronization. \emph{Communication through coherence} is the idea
that there exists an optimal level of synchrony, seen through the
phase difference or a coherence metric, that maximizes the
communication between the regions.

Due to the extensive connectivity in the brain, each region receives
inputs from multiple other regions, and the CTC hypothesis provides a
method to selectively process a single input at a
time~\citep{DM-RV:16}. Within the variety of oscillatory inputs
arriving, the communication from one input will exhibit a level of
synchrony closer to the optimal phase difference for CTC. This input
will then communicate with the brain region to a higher degree than
the competing inputs and will create a response related to that
stimulus. In this way, one input will be managed at a time, dictated by
which input has the optimal phase difference with the activity of the
region under consideration.

It has been hypothesized that the optimal phase difference for CTC is
the same as the stable phase difference, that is, the phase shift that
two oscillators naturally converge to in the steady
state~\citep{EN-JC:19-sfn}. In order to numerically study CTC, it is
then required to be able to compute the stable phase difference, which
requires accurate knowledge of the underlying system, particularly in
the case of the network generating non-sinusoidal waveforms. As
accurate knowledge of the brain network is generally unavailable,
which is exacerbated by the appearance of non-sinusoidal waveforms, it
is desirable to be able to provide a characterization of the stable
phase shift in a model-free fashion.

As a complementary avenue to existing spiking models that have been
used to study CTC~\citep{DM-RV:16,DR-GH:22}, the ability of the
linear-threshold network to generate oscillations, both in individual
and coupled networks~\citep{EN-JC:19-acc,EN-RP-JC:22-auto}, has made
it an attractive option for numerically studying CTC through the
determination of the stable phase shift. \citep{EN-JC:19-sfn} provides
a preliminary analysis of CTC using the linear-threshold dynamics
through an analysis of the stable phase shift for coupled E-I
oscillators. An extended study of CTC using the linear-threshold
dynamics over varying network topologies and properties, however,
remains an avenue for future work.


\subsubsection{Working Memory and Spatial Computing}

Working memory (WM) is a model for the short-term storage and control,
in a top-down fashion, of a small number of
items~\citep{AHVL-JH-SL-TN-BM-AV-FB:17}. Representations of items can
selectively be encoded into, maintained in, and deleted from WM and
then manipulated for the purpose of tasks such as reasoning and
decision making.

WM has been associated with oscillations in the brain and the relation
between types of oscillations in different regions. In particular, WM
has been suggested to rely on coupling between beta oscillations
(those with a frequency between $12$ and $30$ Hz) and gamma
oscillations (those with frequency between $30$ and $100$
Hz)~\citep{EKM-ML-AMB:18}. In this model, gamma activity
corresponds with spiking that encodes and maintains information in
working memory, while top-down information is encoded in beta activity
which inhibits gamma oscillations and thus controls access to the
represented information in WM~\citep{ML-PH-MRW-SLB-EKM:18}.
Through this, when information is being encoded into WM, the beta
activity decreases and high gamma activity is observed.  While
information is deleted, the opposite relation appears. Note also that this interplay between beta and gamma oscillations is interestingly similar to that during selective attention, where the attentional state follows a (theta-frequency) rhythmic switch between a `sampling' state and a `shifting' state~\citep{ICF-SK:19}. The sampling state involves increased beta oscillations in the frontal eye field (FEF, a motor processing area) and gamma activity in the lateral intraparietal cortex (LIP, a sensory processing area) linked to attention suppression and sensory enhancement, respectively. During the shifting state, in contrast, both oscillations are suppressed, leading to reduced visual processing before an attentional shift. Various other similarities and shared mechanisms have also been suggested between working memory (as internal attention) and (external) selective attention, see, e.g., \citep{EA-JJ:01,AG-ACN:12,AK-TE:13,NEM-MGS-ACN:17}.

However, as with any model, this relationship between gamma and beta
waves does not fully explain the neural mechanisms underlying WM. In
particular, the anti-correlated increases and decreases in gamma and
beta behavior provides an `on-off' concept for
WM~\citep{ML-PH-MRW-SLB-EKM:18}. However, with the oscillations
representing large populations of neurons, this model is unable to
explain how individual item representations are manipulated.  One such
hypothesis for how WM encodes such information is called `spatial
computing'. In this setup, the gamma-beta interaction is used to
encode representations, while the spatial movement of this activity
across a cortical network is used for the purpose of manipulating the
information to the relevant
task~\citep{ML-SLB-JR-MWW-TJB-EKM-PH:23}.

The variety of oscillatory behavior illustrated by the
linear-threshold dynamics provides an avenue to model the gamma-beta
interactions in~WM.  In addition, the fact that oscillations can
spread through linear-threshold networks and the availability of
analytical tools to characterize when and how, cf.
Section~\ref{subsec:epilepsy_spreading}, provides an opportunity to
quantitatively model spatial computing.

\subsection{Linear-Threshold Dynamics with Network Plasticity}

Throughout this work we have discussed using the linear-threshold
dynamics as a model for the brain and studied properties of the
dynamics for a given synaptic weight matrix. However, the brain is
plastic~\citep{BK-RG:11,PM-AR:19,JL-NS:05} and the connections between
neurons can both increase and decrease in magnitude, resulting in a
changing synaptic weight matrix. A related, but distinct process that
also occurs in the brain is neurogenesis, namely, the generation of
new neurons in the brain, resulting in networks with additional (at
the macroscale) or modified (at the mesoscale)
nodes~\citep{AK-VP-MAF-SKG-CK:19}.
%
%

Both of these processes and their associated brain functions can be
studied through the lens of linear-threshold dynamics with a synaptic
weight matrix that is itself dynamic. One particular function related
to both neuroplasticity and neurogenesis is
memory~\citep{DNA-JMW:15,WD-JBA-FHG:10,CC-AD-VI:12,RCS-ST-MTdM:16,MF-CH-MMM-SV:09}. In
Section~\ref{sec:single-region} we modeled declarative memory using
the threshold-linear dynamics, with memories being represented by a
set of nodes that permit a stable equilibrium point. With the basis
for memory representation using the threshold-linear dynamics in
place, this opens itself for extension to covering the concept of
neuroplasticity and neurogenesis on memory. The
work~\citep{CC-AD-VI:12} encodes memories in `flexible' networks,
which are those where the synaptic weight matrix can be perturbed by a
small amount. This correlates with the concept of neuroplasticity and
investigating neurogenesis in memory models through the
threshold-linear
dynamics, 
which forms an interesting avenue for future work.



\subsection{Controller Synthesis in Linear-Threshold Networks}
In both the single and multi-region networks considered in this work,
we consider the convergence of the dynamics to either constant
equilibria or trajectories with varying forms. The convergence of the
network to specific trajectories is important for a variety of
problems, including goal-driven selective attention, as discussed in
Section~\ref{sec:multi-region}, or memory recall. However, determining
conditions and control inputs such that the network converges to a
specific non-zero equilibrium or trajectory is difficult. When using
the hierarchical approach of Section~\ref{sec:multi-region}, the
values of the trajectory are given by recursively defined signals, and
choosing parameters to set these maps to specific values is
unrealistic, especially for large systems. On the other hand, if
considering the network as a whole, it can be difficult to fully
understand stability and stabilization issues, and the choice of
parameters might not be particularly robust.

Further, an issue that arises in both approaches is that frequently
the synaptic weight matrix is not known, or only is known as a
(potentially poor) estimate. Machine learning methods can be of use in
approaching the synthesis of controllers that make the network
converge to explicit trajectories. These methods can be used both to
provide a model of how the brain functions internally (such as how
specific signals are achieved in selective attention or memory
recall), or to determine a method for external modulation of the brain
to achieve desired activity (such as deep brain stimulation). One such
approach that has been used in this context is reservoir
computing~\citep{JZK-DSB:23,FD-CCH-AG:22} due to its origins in
computational neuroscience, with specific applications to
linear-threshold networks considered in~\citep{MM-JC:24-ojcsys}. There
are many interesting avenues for future research in the area of
determining explicit signals with machine learning, such as comparing
the use of different learning models and methods.

\section{Conclusions}

The development of computational models with sufficient biological
realism \textit{and} mathematical tractability is one of the key goals
in computational neuroscience.
In this work, we have investigated one such model at the mesoscale,
the linear-threshold rate dynamics. Due to the wide array of functions
the brain performs, any model of the brain that aims to cover multiple
applications must exhibit a rich suite of behaviors. Through
discussions of a variety of behaviors exhibited by the brain, we have
shown the linear-threshold dynamics to be a rich and versatile model.

Through modeling of GDSA in both single and interconnected networks of
different topologies, we showed the linear-threshold dynamics are
stabilizable through simultaneous selective inhibition of
task-irrelevant nodes and selective recruitment of task-relevant
ones. Using a model of declarative memory as sets of neurons that can
be the support sets of stable equilibria, we illustrated that the
unbounded version of the linear-threshold dynamics can permit multiple
stable equilibrium points and are conditionally multiattractive. In
addition, we illustrated that the dynamics admit both bifurcations and
oscillatory behavior in E-I pairs, two dynamical properties that occur
during epileptic seizures.

We finished by extending the discussion of oscillations beyond the E-I
pair network structure and discussing additional dynamical brain
processes such as communication through coherence and spatial
computing. While this is neither a comprehensive list of all
properties of the linear-threshold dynamics nor a complete collection
of brain functions it could be used on, this work highlights the rich
behavior of this family of mesoscopic models, illustrates its
significant utility in modeling brain activity, and showcases the
benefits of adopting a perspective from systems and controls to
analyze how neuronal networks optimize their computational
capabilities.  We hope our exposition will spur further investigations
at the intersection of neuroscience and control to realize the
potential of network and control-theoretic tools in deepening our
understanding of the interplay between structure and dynamics in the
brain, its role in shaping dynamical behavior, performance, and
robustness, and the mechanisms that govern information transfer.



\clearpage
 \bibliographystyle{ieeetr}

\begin{thebibliography}{100}

\bibitem{SJS:12}
S.~J. Schiff, {\em Neural Control Engineering: the Emerging Intersection
  between Control Theory and Neuroscience}.
\newblock Cambridge, MA: MIT Press, 2012.

\bibitem{SC-MYL-JJC-MBW-JDK-KS-PLP-ENB:13}
S.~Ching, M.~Y. Liberman, J.~J. Chemali, M.~B. Westover, J.~D. Kenny, K.~Solt,
  P.~L. Purdon, and E.~N. Brown, ``Real-time closed-loop control in a rodent
  model of medically induced coma using burst suppression,'' {\em
  Anesthesiology: The Journal of the American Society of Anesthesiologists},
  vol.~119, no.~4, pp.~848--860, 2013.

\bibitem{SG-FP-MC-QKT-ABY-AEK-JDM-JMV-MBM-STG-DSB:15}
S.~Gu, F.~Pasqualetti, M.~Cieslak, Q.~K. Telesford, A.~B. Yu, A.~E. Kahn, J.~D.
  Medaglia, J.~M. Vettel, M.~B. Miller, S.~T. Grafton, and D.~S. Bassett,
  ``Controllability of structural brain networks,'' {\em Nature
  Communications}, vol.~6, p.~8414, October 2015.

\bibitem{SVS-PS:18}
S.~V. Sarma and P.~Sacr{\'e}, ``Characterizing complex human behaviors and
  neural responses using dynamic models,'' in {\em Dynamic Neuroscience:
  Statistics, Modeling, and Control} (Z.~Chen and S.~V. Sarma, eds.),
  pp.~177--195, Springer, 2018.

\bibitem{MSM-NJC:20}
M.~S. Madhav and N.~J. Cowan, ``The synergy between neuroscience and control
  theory: the nervous system as inspiration for hard control challenges,'' {\em
  Annual Review of Control, Robotics, and Autonomous Systems}, vol.~3,
  pp.~243--267, 2020.

\bibitem{EN-JC:21-tacI}
E.~Nozari and J.~Cort\'es, ``Hierarchical selective recruitment in
  linear-threshold brain networks. {P}art {I}: Intra-layer dynamics and
  selective inhibition,'' {\em IEEE Transactions on Automatic Control},
  vol.~66, no.~3, pp.~949--964, 2021.

\bibitem{EN-JC:21-tacII}
E.~Nozari and J.~Cort\'es, ``Hierarchical selective recruitment in
  linear-threshold brain networks. {P}art {II}: Inter-layer dynamics and
  top-down recruitment,'' {\em IEEE Transactions on Automatic Control},
  vol.~66, no.~3, pp.~965--980, 2021.

\bibitem{PS-EN-JZK-HJ-DZ-CB-FP-GJP-DSB:20}
P.~Srivastava, E.~Nozari, J.~Z. Kim, H.~Ju, D.~Zhou, C.~Becker, F.~Pasqualetti,
  G.~J. Pappas, and D.~S. Bassett, ``Models of communication and control for
  brain networks: distinctions, convergence, and future outlook,'' {\em Network
  Neuroscience}, vol.~4, no.~4, pp.~1122--1159, 2020.

\bibitem{GD-TOL-JD-AF-RF:15}
G.~Drion, T.~O'Leary, J.~Dethier, A.~Franci, and R.~Sepulchre, ``Neuronal
  behaviors: a control perspective,'' in {\em {IEEE} Conf.\ on Decision and
  Control}, (Osaka, Japan), Dec. 2015.
\newblock Tutorial.

\bibitem{MEB:21}
M.~E. Broucke, ``Adaptive internal model theory of the oculomotor system and
  the cerebellum,'' {\em IEEE Transactions on Automatic Control}, vol.~66,
  no.~11, pp.~5444--5450, 2021.

\bibitem{MEB:22}
M.~E. Broucke, ``Adaptive internal models in neuroscience,'' {\em Foundations
  and Trends in Systems and Control}, vol.~9, no.~4, pp.~365--550, 2022.

\bibitem{CAC:15}
C.~A. Cot{\'e}, ``A dynamic systems theory model of visual perception
  development,'' {\em Journal of Occupational Therapy, Schools, \& Early
  Intervention}, vol.~8, no.~2, pp.~157--169, 2015.

\bibitem{AK-WT:11}
A.~Kukona and W.~Tabor, ``Impulse processing: A dynamical systems model of
  incremental eye movements in the visual world paradigm,'' {\em Cognitive
  Science}, vol.~35, no.~6, pp.~1009--1051, 2011.

\bibitem{MK-KF-RS:87}
M.~Kawato, K.~Furukawa, and R.~Suzuki, ``A hierarchical neural-network model
  for control and learning of voluntary movement,'' {\em Biological
  Cybernetics}, vol.~57, pp.~169--185, 1987.

\bibitem{KVS-MTK-MS-MMC:11}
K.~V. Shenoy, M.~T. Kaufman, M.~Sahani, and M.~M. Churchland, ``A dynamical
  systems view of motor preparation: implications for neural prosthetic system
  design,'' {\em Progress in Brain Research}, vol.~192, pp.~33--58, 2011.

\bibitem{CIC-JBB-MSJ:00}
C.~I. Connolly, J.~B. Burns, and M.~S. Jog, ``A dynamical-systems model for
  {Parkinson's} disease,'' {\em Biological Cybernetics}, vol.~83, pp.~47--59,
  2000.

\bibitem{EJM-SJvA-JWK-PAR:17}
E.~J. M{\"u}ller, S.~J. van Albada, J.~W. Kim, and P.~A. Robinson, ``Unified
  neural field theory of brain dynamics underlying oscillations in
  {Parkinson’s} disease and generalized epilepsies,'' {\em Journal of
  Theoretical Biology}, vol.~428, pp.~132--146, 2017.

\bibitem{AG-ACN:12}
A.~Gazzaley and A.~C. Nobre, ``Top-down modulation: bridging selective
  attention and working memory,'' {\em Trends in Cognitive Sciences}, vol.~16,
  no.~2, pp.~129--135, 2012.

\bibitem{RQ-JK-ES-NF-RT-ERB-LGC:19}
R.~Quentin, J.~King, E.~Sallard, N.~Fishman, R.~Thompson, E.~R. Buch, and L.~G.
  Cohen, ``Differential brain mechanisms of selection and maintenance of
  information during working memory,'' {\em Journal of Neuroscience}, vol.~39,
  no.~19, pp.~3728--3740, 2019.

\bibitem{GLH:15}
G.~L. Holmes, ``Cognitive impairment in epilepsy: the role of network
  abnormalities,'' {\em Epileptic Disorders}, vol.~17, no.~2, pp.~101--116,
  2015.

\bibitem{YH-ET:15}
Y.~H{\"o}ller and E.~Trinka, ``Is there a relation between {EEG}-slow waves and
  memory dysfunction in epilepsy? a critical appraisal,'' {\em Frontiers in
  Human Neuroscience}, vol.~9, p.~341, 2015.

\bibitem{CRB-AZZ:09}
C.~R. Butler and A.~Z. Zeman, ``Recent insights into the impairment of memory
  in epilepsy: transient epileptic amnesia, accelerated long-term forgetting
  and remote memory impairment,'' {\em Brain}, vol.~131, no.~9, pp.~2243--2263,
  2008.

\bibitem{JS:90}
J.~Sully, ``The psycho-physical process in attention,'' {\em Brain}, vol.~13,
  no.~2, pp.~145--164, 1890.

\bibitem{AMT:69}
A.~M. Treisman, ``Strategies and models of selective attention.,'' {\em
  Psychological Review}, vol.~76, no.~3, p.~282, 1969.

\bibitem{ECC:53}
E.~C. Cherry, ``Some experiments on the recognition of speech, with one and
  with two ears,'' {\em The Journal of the Acoustical Society of America},
  vol.~25, no.~5, pp.~975--979, 1953.

\bibitem{RD-JD:95}
R.~Desimone and J.~Duncan, ``Neural mechanisms of selective visual attention,''
  {\em Annual Review of Neuroscience}, vol.~18, no.~1, pp.~193--222, 1995.

\bibitem{LI-CK:01}
L.~Itti and C.~Koch, ``Computational modelling of visual attention,'' {\em
  Nature Reviews Neuroscience}, vol.~2, no.~3, p.~194, 2001.

\bibitem{BCM:93}
B.~C. Motter, ``Focal attention produces spatially selective processing in
  visual cortical areas {V}1, {V}2, and {V}4 in the presence of competing
  stimuli,'' {\em Journal of Neurophysiology}, vol.~70, no.~3, pp.~909--919,
  1993.

\bibitem{MAP-GMD-SK:04}
M.~A. Pinsk, G.~M. Doniger, and S.~Kastner, ``Push-pull mechanism of selective
  attention in human extrastriate cortex,'' {\em Journal of Neurophysiology},
  vol.~92, no.~1, pp.~622--629, 2004.

\bibitem{MM-JC:24-tcns}
M.~McCreesh and J.~Cort\'es, ``Selective inhibition and recruitment in
  linear-threshold thalamocortical networks,'' {\em IEEE Transactions on
  Control of Network Systems}, vol.~11, no.~1, pp.~375--388, 2024.

\bibitem{KT-KTH-AK-LE-AB-LE-ZJ-GN-AS-DF-IU-LW:18}
K.~T{\'o}th, K.~T. Hofer, A.~Kandr{\'a}cs, L.~Entz, A.~Bag{\'o},
  L.~Er{\H{o}}ss, Z.~Jord{\'a}n, G.~Nagy, A.~S{\'o}lyom, D.~Fab{\'o},
  I.~Ulbert, and L.~Wittner, ``Hyperexcitability of the network contributes to
  synchronization processes in the human epileptic neocortex,'' {\em The
  Journal of Physiology}, vol.~596, no.~2, pp.~317--342, 2018.

\bibitem{HGEM-TLE-BK-JFN-CAS-RGE-RRG-GMM-CJM-AKT-JDC-SAvG-WvD:15}
H.~G.~E. Meijer, T.~L. Eissa, B.~Kiewiet, J.~F. Neuman, C.~A. Schevon, R.~G.
  Emerson, R.~R. Goodman, G.~M.~M. Jr., C.~J. Marcuccilli, A.~K. Tryba, J.~D.
  Cowan, S.~A. van {G}ils, and W.~van {D}rongelen, ``Modeling focal epileptic
  activity in the {W}ilson-{C}owan model with depolarization block,'' {\em
  Journal of Mathematical Neuroscience}, vol.~5, pp.~1--17, 2015.

\bibitem{AC:22}
A.~Coletti, ``On {J}ansen-{R}it system modeling epilepsy phenomena,'' in {\em
  International Conference on New Trends in the Applications of Differential
  Equations in Sciences}, pp.~281--292, Springer, 2022.

\bibitem{VKJ-WCS-PPQ-AII-CB:14}
V.~K. Jirsa, W.~C. Stacey, P.~P. Quilichini, A.~I. Ivanov, and C.~Bernard, ``On
  the nature of seizure dynamics,'' {\em Brain}, vol.~137, no.~8,
  pp.~2210--2230, 2014.

\bibitem{JT-FW-PC-OF:11}
J.~Touboul, F.~Wendling, P.~Chauvel, and O.~Faugeras, ``Neural mass activity,
  bifurcations, and epilepsy,'' {\em Neural Computation}, vol.~23, no.~12,
  pp.~3232--3286, 2011.

\bibitem{DS-SVS:14}
D.~Sritharan and S.~V. Sarma, ``Fragility in dynamic networks: application to
  neural networks in the epileptic cortex,'' {\em Neural Computation}, vol.~26,
  no.~10, pp.~2294--2327, 2014.

\bibitem{DE-DS-SVS:15}
D.~Ehrens, D.~Sritharan, and S.~V. Sarma, ``Closed-loop control of a fragile
  network: application to seizure-like dynamics of an epilepsy model,'' {\em
  Frontiers in Neuroscience}, vol.~9, p.~58, 2015.

\bibitem{ETW-MV-ZH-RM-VRR-SC:22}
E.~T. Wang, M.~Vannucci, Z.~Haneef, R.~Moss, V.~R. Rao, and S.~Chiang, ``A
  bayesian switching linear dynamical system for estimating seizure
  chronotypes,'' {\em Proceedings of the National Academy of Sciences},
  vol.~119, no.~46, p.~e2200822119, 2022.

\bibitem{FLdS-WB-SNK-JP-PS-DNV:03}
F.~L.~D. Silva, W.~Blanes, S.~N. Kalitzin, J.~Parra, P.~Suffczynski, and D.~N.
  Velis, ``Epilepsies as dynamical diseases of brain systems: basic models of
  the transition between normal and epileptic activity,'' {\em Epilepsia},
  vol.~44, pp.~72--83, 2003.

\bibitem{FC-AA-FP-JC:21-csl}
F.~Celi, A.~Allibhoy, F.~Pasqualetti, and J.~Cort\'es, ``Linear-threshold
  dynamics for the study of epileptic events,'' {\em IEEE Control Systems
  Letters}, vol.~5, no.~4, pp.~1405--1410, 2021.

\bibitem{AA-FC-FP-JC:22-ojcsys}
A.~Allibhoy, F.~Celi, F.~Pasqualetti, and J.~Cort\'es, ``Optimal network
  interventions to control the spreading of oscillations,'' {\em IEEE Open
  Journal of Control Systems}, vol.~1, pp.~141--151, 2022.

\bibitem{WJ:90}
W.~James, {\em The Principles of Psychology}, vol.~1.
\newblock Cosimo, Inc., 1890.

\bibitem{GM:56}
G.~Miller, ``Human memory and the storage of information,'' {\em IRE
  Transactions on Information Theory}, vol.~2, no.~3, pp.~129--137, 1956.

\bibitem{RCA-RMS:68}
R.~C. Atkinson and R.~M. Shiffrin, ``Human memory: A proposed system and its
  control processes,'' in {\em Psychology of Learning and Motivation}, vol.~2,
  pp.~89--195, Elsevier, 1968.

\bibitem{GRL-EFL:76}
G.~R. Loftus and E.~F. Loftus, {\em Human Memory: The Processing of
  Information}.
\newblock Psychology Press, 1st~ed., 1976.

\bibitem{DLS-ET:94}
D.~L. Schacter and E.~Tulving, ``What are the memory systems of 1994?,'' {\em
  Memory Systems}, vol.~1994, p.~424, 1994.

\bibitem{AB:20}
A.~Baddeley, ``Working memory,'' in {\em Memory}, pp.~71--111, Routledge, 2020.

\bibitem{JJH:82}
J.~J. Hopfield, ``Neural networks and physical systems with emergent collective
  computational abilities,'' {\em Proceedings of the National Academy of
  Sciences}, vol.~79, no.~8, pp.~2554--2558, 1982.

\bibitem{GEH-TJS:86}
G.~E. Hinton and T.~J. Sejnowski, ``Learning and relearning in {B}oltzmann
  machines,'' in {\em Parallel Distributed Processing: Explorations in the
  Microstructure of Cognition} (D.~E. Rumelhart, J.~L. McClelland, and the PDP
  Research~Group, eds.), vol.~1, pp.~282--317, 1986.

\bibitem{SG-DH-HS:08}
S.~Ganguli, D.~Huh, and H.~Sompolinsky, ``Memory traces in dynamical systems,''
  {\em Proceedings of the National Academy of Sciences}, vol.~105, no.~48,
  pp.~18970--18975, 2008.

\bibitem{RP-HJ:11}
R.~Pascanu and H.~Jaeger, ``A neurodynamical model for working memory,'' {\em
  Neural Networks}, vol.~24, no.~2, pp.~199--207, 2011.

\bibitem{SA-AK-PEK:99}
S.~Albeverio, A.~Khrennikov, and P.~E. Kloeden, ``Memory retrieval as a p-adic
  dynamical system,'' {\em Biosystems}, vol.~49, no.~2, pp.~105--115, 1999.

\bibitem{RHRH-HSS-JJS:03}
R.~H.~R. Hahnloser, H.~S. Seung, and J.~J. Slotine, ``Permitted and forbidden
  sets in symmetric threshold-linear networks,'' {\em Neural Computation},
  vol.~15, no.~3, pp.~621--638, 2003.

\bibitem{IT-EK-HS:87}
I.~Tsuda, E.~Koerner, and H.~Shimizu, ``Memory dynamics in asynchronous neural
  networks,'' {\em Progress of Theoretical Physics}, vol.~78, no.~1,
  pp.~51--71, 1987.

\bibitem{ALH-AFH:52}
A.~L. Hodgkin and A.~F. Huxley, ``A quantitative description of membrane
  current and its application to conduction and excitation in nerve,'' {\em The
  Journal of Physiology}, vol.~117, no.~4, pp.~500--544, 1952.

\bibitem{GBE-DHT:10}
G.~B. Ermentrout and D.~H. Terman, {\em Mathematical Foundations of
  Neuroscience}, vol.~35.
\newblock New York: Springer, 2010.

\bibitem{EMI:07}
E.~M. Izhikevich, {\em Dynamical Systems in Neuroscience}.
\newblock Cambridge, MA: MIT Press, 2007.

\bibitem{PD-LFA:01}
P.~Dayan and L.~F. Abbott, {\em Theoretical Neuroscience: Computational and
  Mathematical Modeling of Neural Systems}.
\newblock Computational Neuroscience, Cambridge, MA: MIT Press, 2001.

\bibitem{EN-MAB-JS-LC-EJC-XH-ASM-GJP-DSB:24}
E.~Nozari, M.~A. Bertolero, J.~Stiso, L.~Caciagli, E.~J. Cornblath, X.~He,
  A.~S. Mahadevan, G.~J. Pappas, and D.~S. Bassett, ``Macroscopic resting-state
  brain dynamics are best described by linear models,'' {\em Nature Biomedical
  Engineering}, vol.~8, no.~1, pp.~68--84, 2024.

\bibitem{MB:17}
M.~Breakspear, ``Dynamic models of large-scale brain activity,'' {\em Nature
  Neuroscience}, vol.~20, no.~3, pp.~340--352, 2017.

\bibitem{XL-DC-LM-TMM-HB:11}
X.~Li, D.~Coyle, L.~Maguire, T.~M. McGinnity, and H.~Benali, ``A model
  selection method for nonlinear system identification based {fMRI} effective
  connectivity analysis,'' {\em IEEE Transactions on Medical Imaging}, vol.~30,
  no.~7, pp.~1365--1380, 2011.

\bibitem{KES-LK-LMH-JD-HEMdO-MB-KJF:08}
K.~E. Stephan, L.~Kasper, L.~M. Harrison, J.~Daunizeau, H.~E.~M. den Ouden,
  M.~Breakspear, and K.~J. Friston, ``Nonlinear dynamic causal models for
  {fMRI},'' {\em Neuroimage}, vol.~42, no.~2, pp.~649--662, 2008.

\bibitem{GTE-CK-NKL-SP:13}
G.~T. Einevoll, C.~Kayser, N.~K. Logothetis, and S.~Panzeri, ``Modelling and
  analysis of local field potentials for studying the function of cortical
  circuits,'' {\em Nature Reviews Neuroscience}, vol.~14, no.~11, pp.~770--785,
  2013.

\bibitem{SK-AB-AF-JR:19}
S.~Keeley, {\'A}.~Byrne, A.~Fenton, and J.~Rinzel, ``Firing rate models for
  gamma oscillations,'' {\em Journal of Neurophysiology}, vol.~121, no.~6,
  pp.~2181--2190, 2019.

\bibitem{TS-AVC:19}
T.~Schwalger and A.~V. Chizhov, ``Mind the last spike — firing rate models
  for mesoscopic populations of spiking neurons,'' {\em Current Opinion in
  Neurobiology}, vol.~58, pp.~155--166, 2019.

\bibitem{JF-KPH:96}
J.~Feng and K.~P. Hadeler, ``Qualitative behaviour of some simple networks,''
  {\em Journal of Physics A: Mathematical and General}, vol.~29, no.~16,
  pp.~5019--5033, 1996.

\bibitem{DAH-ZB-JC-MAM-KDH-GB:00}
D.~A. Henze, Z.~Borhegyi, J.~Csicsvari, M.~A. Mamiya, K.~D. Harris, and
  G.~Buzsaki, ``Intracellular features predicted by extracellular recordings in
  the hippocampus in vivo,'' {\em Journal of Neurophysiology}, vol.~84, no.~1,
  pp.~390--400, 2000.

\bibitem{DAH-KDH-ZB-JC-AM-HH-AS-GB:09-crcns}
D.~A. Henze, K.~D. Harris, Z.~Borhegyi, J.~Csicsvari, A.~Mamiya, H.~Hirase,
  A.~Sirota, and G.~Buzsaki, ``Simultaneous intracellular and extracellular
  recordings from hippocampus region ca1 of anesthetized rats.'' CRCNS.org,
  2009.

\bibitem{QL-AU-BH:17}
Q.~Liu, A.~Ulloa, and B.~Horwitz, ``Using a large-scale neural model of
  cortical object processing to investigate the neural substrate for managing
  multiple items in short-term memory,'' {\em Journal of Cognitive
  Neuroscience}, vol.~29, no.~11, pp.~1860--1876, 2017.

\bibitem{MK-RN-FG-MK-YI-TA-MS:21}
M.~Kajiwara, R.~Nomura, F.~Goetze, M.~Kawabata, Y.~Isomura, T.~Akutsu, and
  M.~Shimono, ``Inhibitory neurons exhibit high controlling ability in the
  cortical microconnectome,'' {\em PLOS Computational Biology}, vol.~17, no.~4,
  p.~e1008846, 2021.

\bibitem{JC-UH-CJH:15}
J.~Chen, U.~Hasson, and C.~Honey, ``Processing timescales as an organizing
  principle for primate cortex,'' {\em Neuron}, vol.~88, no.~2, pp.~244--246,
  2015.

\bibitem{MB-BC-MAP:20}
M.~Bear, B.~Connors, and M.~A. Paradiso, {\em Neuroscience: Exploring the
  Brain}.
\newblock Jones \& Bartlett Learning, 2020.

\bibitem{SMS-RWG:06}
S.~M. Sherman and R.~W. Guillery, {\em Exploring the Thalamus and Its Role in
  Cortical Function}.
\newblock Cambridge, MA: MIT Press, 2006.

\bibitem{MW-SDV:19}
M.~Wolff and S.~D. Vann, ``The cognitive thalamus as a gateway to mental
  representations,'' {\em Journal of Neuroscience}, vol.~39, no.~1, pp.~3--14,
  2019.

\bibitem{HG:96}
H.~Gudden, ``Klinische und anatomische beitr{\"a}ge zur kenntniss der multiplen
  alkoholneuritis nebst bemerkungen {\"u}ber die regenerationsvorg{\"a}nge im
  peripheren nervensystem,'' {\em Archiv f{\"u}r Psychiatrie und
  Nervenkrankheiten}, vol.~28, no.~3, pp.~643--741, 1896.

\bibitem{EA-TO:15}
E.~Ahissar and T.~Oram, ``Thalamic relay or cortico-thalamic processing? {O}ld
  question, new answers,'' {\em Cerebral Cortex}, vol.~25, no.~4, pp.~845--848,
  2015.

\bibitem{YBS-SK:11}
Y.~B. Saalmann and S.~Kastner, ``Cognitive and perceptual functions of the
  visual thalamus,'' {\em Neuron}, vol.~71, no.~2, pp.~209--223, 2011.

\bibitem{SMS-RWG:96}
S.~M. Sherman and R.~W. Guillery, ``Functional organization of thalamocortical
  relays,'' {\em Journal of Neurophysiology}, vol.~76, no.~3, pp.~1367--1395,
  1996.

\bibitem{ASM:15}
A.~S. Mitchell, ``The mediodorsal thalamus as a higher order thalamic relay
  nucleus important for learning and decision-making,'' {\em Neuroscience \&
  Biobehavioral Reviews}, vol.~54, pp.~76--88, 2015.

\bibitem{ASM-SMS-MAS-RGM-RPV-YC:14}
A.~S. Mitchell, S.~M. Sherman, M.~A. Sommer, R.~G. Mair, R.~P. Vertes, and
  Y.~Chudasama, ``Advances in understanding mechanisms of thalamic relays in
  cognition and behavior,'' {\em Journal of Neuroscience}, vol.~34, no.~46,
  pp.~15340--15346, 2014.

\bibitem{FA-VF-ARM-EJK-EC-WM:18}
F.~Alcaraz, V.~Fresno, A.~R. Marchand, E.~J. Kremer, E.~Coutureau, and
  M.~Wolff, ``Thalamocortical and corticothalamic pathways differentially
  contribute to goal-directed behaviors in the rat,'' {\em Elife}, vol.~7,
  p.~e32517, 2018.

\bibitem{JMA-HAS:15}
J.~M. Alonso and H.~A. Swadlow, ``Thalamus controls recurrent cortical
  dynamics,'' {\em Nature Neuroscience}, vol.~18, no.~12, pp.~1703--1704, 2015.

\bibitem{LG-SPJ-DEF-MC-MS:05}
L.~Gabernet, S.~P. Jadhav, D.~E. Feldman, M.~Carandini, and M.~Scanziani,
  ``Somatosensory integration controlled by dynamic thalamocortical
  feed-forward inhibition,'' {\em Neuron}, vol.~48, no.~2, pp.~315--327, 2005.

\bibitem{MMH-LA:16}
M.~M. Halassa and L.~Acs{\'a}dy, ``Thalamic inhibition: diverse sources,
  diverse scales,'' {\em Trends in Neurosciences}, vol.~39, no.~10,
  pp.~680--693, 2016.

\bibitem{SMS:12}
S.~M. Sherman, ``Thalamocortical interactions,'' {\em Current Opinion in
  Neurobiology}, vol.~22, no.~4, pp.~575--579, 2012.

\bibitem{SJC-TJL-BWC:07}
S.~Cruikshank, T.~J. Lewis, and B.~Connors, ``Synaptic basis for intense
  thalamocortical activation of feedforward inhibitory cells in neocortex,''
  {\em Nature Neuroscience}, vol.~10, no.~4, pp.~462--468, 2007.

\bibitem{SJK-JD-KJF:08}
S.~J. Kiebel, J.~Daunizeau, and K.~J. Friston, ``A hierarchy of time-scales and
  the brain,'' {\em PLOS Computational Biology}, vol.~4, no.~11, p.~e1000209,
  2008.

\bibitem{KG-JMD:20}
K.~George and J.~M. Das, ``Neuroanatomy, thalamocortical radiations,'' {\em
  StatPearls [Internet]}, 2020.

\bibitem{KD-JT-ZJH-BL:15}
K.~Delevich, J.~Tucciarone, Z.~J. Huang, and B.~Li, ``The mediodorsal thalamus
  drives feedforward inhibition in the anterior cingulate cortex via
  parvalbumin interneurons,'' {\em Journal of Neuroscience}, vol.~35, no.~14,
  pp.~5743--5753, 2015.

\bibitem{JTP-CKJ-AA:01}
J.~T. Porter, C.~K. Johnson, and A.~Agmon, ``Diverse types of interneurons
  generate thalamus-evoked feedforward inhibition in the mouse barrel cortex,''
  {\em Journal of Neuroscience}, vol.~21, no.~8, pp.~2699--2710, 2001.

\bibitem{HAS:02}
H.~A. Swadlow, ``Thalamocortical control of feed-forward inhibition in awake
  somatosensory ‘barrel’cortex,'' {\em Philosophical Transactions of the
  Royal Society of London. Series B: Biological Sciences}, vol.~357, no.~1428,
  pp.~1717--1727, 2002.

\bibitem{EKK-JV:02}
E.~K. Kosmidis and J.~Vibert, ``Feed-forward inhibition in the visual
  thalamus,'' {\em Neurocomputing}, vol.~44, pp.~479--487, 2002.

\bibitem{JSI-MS:11}
J.~S. Isaacson and M.~Scanziani, ``How inhibition shapes cortical activity,''
  {\em Neuron}, vol.~72, no.~2, pp.~231--243, 2011.

\bibitem{JM-RD:85}
J.~Moran and R.~Desimone, ``Selective attention gates visual processing in the
  extrastriate cortex,'' {\em Science}, vol.~229, no.~4715, pp.~782--784, 1985.

\bibitem{NL:05}
N.~Lavie, ``Distracted and confused?: Selective attention under load,'' {\em
  Trends in {C}ognitive {S}ciences}, vol.~9, no.~2, pp.~75--82, 2005.

\bibitem{FP:03}
P.~Foldiak, ``Sparse coding in the primate cortex,'' in {\em The Handbook of
  Brain Theory and Neural Networks} (M.~A. Arbib, ed.), pp.~1064--1068,
  Cambridge, MA: MIT Press, 2nd~ed., 2003.

\bibitem{MI-AR-DK-TR-JMR:17}
M.~Imani, A.~Rahimi, D.~Kong, T.~Rosing, and J.~M. Rabaey, ``Exploring
  hyperdimensional associative memory,'' in {\em IEEE International Symposium
  on High Performance Computer Architecture}, (Austin, TX), pp.~445--456, 2017.

\bibitem{RAS-RGS-SST:12}
R.~A. Stefanescu, R.~G. Shivakeshavan, and S.~S. Talathi, ``Computational
  models of epilepsy,'' {\em Seizure}, vol.~21, no.~10, pp.~748--759, 2012.

\bibitem{CJS:05}
C.~J. Stam, ``Nonlinear dynamical analysis of {EEG} and {MEG}: review of an
  emerging field,'' {\em Clinical Neurophysiology}, vol.~116, no.~10,
  pp.~2266--2301, 2005.

\bibitem{EN-FHLdS:05}
E.~Niedermeyer and F.~H.~L. da~Silva, {\em Electroencephalography: Basic
  Principles, Clinical Applications, and Related Fields}.
\newblock Lippincott Williams \& Wilkins, 2005.

\bibitem{PC:96}
P.~Chauvel, ``Stereo-electroencephalography,'' {\em Multimethodological
  Assessment of the Epileptic Forms}, pp.~135--163, 1996.

\bibitem{EMI:00}
E.~M. Izhikevich, ``Neural excitability, spiking and bursting,'' {\em
  International Journal of Bifurcation and Chaos}, vol.~10, no.~06,
  pp.~1171--1266, 2000.

\bibitem{WTB:10}
W.~T. Blume, ``Focal and generalized: both here and there,'' {\em Epilepsy
  Currents}, vol.~10, no.~5, pp.~115--117, 2010.

\bibitem{JWB-JC-KRP:13}
J.~W. Bang, J.~Choi, and K.~R. Park, ``Noise reduction in brainwaves by using
  both {EEG} signals and frontal viewing camera images,'' {\em Sensors},
  vol.~13, no.~5, pp.~6272--6294, 2013.

\bibitem{CC-AD-VI:12}
C.~Curto, A.~Degeratu, and V.~Itskov, ``Flexible memory networks,'' {\em
  Bulletin of Mathematical Biology}, vol.~74, no.~3, pp.~590--614, 2012.

\bibitem{EJR-JWG-SMS:05}
E.~J. Ramcharan, J.~W. Gnadt, and S.~M. Sherman, ``Higher-order thalamic relays
  burst more than first-order relays,'' {\em Proceedings of the National
  Academy of Sciences}, vol.~102, no.~34, pp.~12236--12241, 2005.

\bibitem{MCS-SWM-JT-RCS-MW-AJP-FQY-DAL:10}
M.~C. Schmid, S.~W. Mrowka, J.~Turchi, R.~C. Saunders, M.~Wilke, A.~J. Peters,
  F.~Q. Ye, and D.~A. Leopold, ``Blindsight depends on the lateral geniculate
  nucleus,'' {\em Nature}, vol.~466, no.~7304, pp.~373--377, 2010.

\bibitem{NT:50}
N.~Tinbergen, ``The hierarchical organization of nervous mechanisms underlying
  instinctive behaviour,'' in {\em Symposium for the Society for Experimental
  Biology}, vol.~4, pp.~305--312, 1950.

\bibitem{ARL:70}
A.~R. Luria, ``The functional organization of the brain,'' {\em Scientific
  American}, vol.~222, no.~3, pp.~66--79, 1970.

\bibitem{DM-RL-AF-KE-ETB:09}
D.~Meunier, R.~Lambiotte, A.~Fornito, K.~Ersche, and E.~T. Bullmore,
  ``Hierarchical modularity in human brain functional networks,'' {\em
  Frontiers in Neuroinformatics}, vol.~3, p.~37, 2009.

\bibitem{BG-EE-GH-AG-AK:12}
B.~Gauthier, E.~Eger, G.~Hesselmann, A.~Giraud, and A.~Kleinschmidt, ``Temporal
  tuning properties along the human ventral visual stream,'' {\em Journal of
  Neuroscience}, vol.~32, no.~41, pp.~14433--14441, 2012.

\bibitem{UH-JC-CJH:15}
U.~Hasson, J.~Chen, and C.~J. Honey, ``Hierarchical process memory: memory as
  an integral component of information processing,'' {\em Trends in Cognitive
  Sciences}, vol.~19, no.~6, pp.~304--313, 2015.

\bibitem{JAH-SM-KEH-JDW-HC-AB-PB-SC-LC-AC-AF-DF-NG-CRG-NG-PAG-AMH-AH-RH-JEK-LK-XK-JL-PL-YL-JL-SM-MTM-MN-LN-SWO-BO-ES-SAS-WW-QW-YW-AW-JWP-ARJ-CK-HZ:19}
J.~A. Harris, S.~Mihalas, K.~E. Hirokawa, J.~D. Whitesell, H.~Choi, A.~Bernard,
  P.~Bohn, S.~Caldejon, L.~Casal, A.~Cho, A.~Feiner, D.~Feng, N.~Gaudreault,
  C.~R. Gerfen, N.~Graddis, P.~A. Groblewski, A.~M. Henry, A.~Ho, R.~Howard,
  J.~E. Knox, L.~Kuan, X.~Kuang, J.~Lecoq, P.~Lesnar, Y.~Li, J.~Luviano,
  S.~McConoughey, M.~T. Mortrud, M.~Naeemi, L.~Ng, S.~W. Oh, B.~Oullette,
  E.~Shen, S.~A. Sorenson, W.~Wakeman, Q.~Wang, Y.~Wang, A.~Williford, J.~W.
  Phillips, A.~R. Jones, C.~Koch, and H.~Zeng, ``Hierarchical organization of
  cortical and thalamic connectivity,'' {\em Nature}, vol.~575, no.~7781,
  pp.~195--202, 2019.

\bibitem{MM-JC:24-ojcsys}
M.~McCreesh and J.~Cort\'es, ``Control of linear-threshold brain networks via
  reservoir computing,'' {\em IEEE Open Journal of Control Systems}, vol.~3,
  pp.~325--341, 2024.

\bibitem{VV:97}
V.~Veliov, ``A generalization of the {T}ikhonov theorem for singularly
  perturbed differential inclusions,'' {\em Journal of Dynamical \& Control
  Systems}, vol.~3, no.~3, pp.~291--319, 1997.

\bibitem{AAM-RDM-BR-DHH-HHH:17}
A.~A. Moustafa, R.~D. McMullan, B.~Rostron, D.~H. Hewedi, and H.~H. Haladjian,
  ``The thalamus as a relay station and gatekeeper: relevance to brain
  disorders,'' {\em Reviews in the Neurosciences}, vol.~28, no.~2,
  pp.~203--218, 2017.

\bibitem{RWG-SMS:02}
R.~W. Guillery and S.~M. Sherman, ``Thalamic relay functions and their role in
  corticocortical communication: generalizations from the visual system,'' {\em
  Neuron}, vol.~33, no.~2, pp.~163--175, 2002.

\bibitem{KEC:86}
K.~E. Chu, ``Generalization of the {B}auer-{F}ike theorem,'' {\em Numerische
  Mathematik}, vol.~49, no.~6, pp.~685--691, 1986.

\bibitem{GKK:01}
G.~K. Kostopoulos, ``Involvement of the thalamocortical system in epileptic
  loss of consciousness,'' {\em Epilepsia}, vol.~42, pp.~13--19, 2001.

\bibitem{SRC-BV:17}
S.~R. Cole and B.~Voytek, ``Brain oscillations and the importance of waveform
  shape,'' {\em Trends in Cognitive Sciences}, vol.~21, no.~2, pp.~137--149,
  2017.

\bibitem{GB-AD:04}
G.~Buzs{\'a}ki and A.~Draguhn, ``Neuronal oscillations in cortical networks,''
  {\em Science}, vol.~304, no.~5679, pp.~1926--1929, 2004.

\bibitem{LVG-AC-AF-LF-JG-MF:13}
L.~V. Gambuzza, A.~Cardillo, A.~Fiasconaro, L.~Fortuna, J.~G{\'o}mez-Gardenes,
  and M.~Frasca, ``Analysis of remote synchronization in complex networks,''
  {\em Chaos: An Interdisciplinary Journal of Nonlinear Science}, vol.~23,
  no.~4, p.~043103, 2013.

\bibitem{HS-YK:86}
H.~Sakaguchi and Y.~Kuramoto, ``A soluble active rotater model showing phase
  transitions via mutual entertainment,'' {\em Progress of Theoretical
  Physics}, vol.~76, no.~3, pp.~576--581, 1986.

\bibitem{YQ:19}
Y.~Qin, {\em Distributed coordination and partial synchronization in complex
  networks}.
\newblock PhD thesis, University of Groningen, 2019.

\bibitem{VV-PH:14}
V.~Vuksanovi{\'c} and P.~H{\"o}vel, ``Functional connectivity of distant
  cortical regions: role of remote synchronization and symmetry in
  interactions,'' {\em NeuroImage}, vol.~97, pp.~1--8, 2014.

\bibitem{LG-CM-AV:10}
L.~Gollo, C.~Mirasso, and A.~Villa, ``Dynamic control for synchronization of
  separated cortical areas through thalamic relay,'' {\em NeuroImage}, vol.~52,
  no.~3, pp.~947--955, 2010.

\bibitem{VN-MV-MC-ADG-VL:13}
V.~Nicosia, M.~Valencia, M.~Chavez, A.~D{\'\i}az-Guilera, and V.~Latora,
  ``Remote synchronization reveals network symmetries and functional modules,''
  {\em Physical Review Letters}, vol.~110, no.~17, p.~174102, 2013.

\bibitem{RDT:03}
R.~D. Traub, ``Fast oscillations and epilepsy,'' {\em Epilepsy Currents},
  vol.~3, no.~3, pp.~77--79, 2003.

\bibitem{OD-AV-TJO-NJ-IES-MdC-PP:18}
O.~Devinsky, A.~Vezzani, T.~J. O'Brien, N.~Jette, I.~E. Scheffer, M.~de~Curtis,
  and P.~Perucca, ``Epilepsy (primer),'' {\em Nature Reviews: Disease Primers},
  vol.~4, no.~1, 2018.

\bibitem{SSS:02}
S.~Spencer, ``Neural networks in human epilepsy: evidence of and implications
  for treatment,'' {\em Epilepsia}, vol.~43, no.~3, pp.~219--227, 2002.

\bibitem{MS-MB-BHB-KL-WRM-FBM-BL-JvG-GAW:10}
M.~Stead, M.~Bower, B.~Brinkmann, K.~Lee, W.~Marsh, F.~Meyer, B.~Litt, J.~V.
  Gompel, and G.~Worrell, ``Microseizures and the spatiotemporal scales of
  human partial epilepsy,'' {\em Brain}, vol.~133, no.~9, pp.~2789--2797, 2010.

\bibitem{EAC-NL:55}
E.~A. Coddington and N.~Levinson, {\em Theory of Ordinary Differential
  Equations}.
\newblock New York: McGraw-Hill, 1955.

\bibitem{EN-RP-JC:22-auto}
E.~Nozari, R.~Planas, and J.~Cort\'es, ``Structural characterization of
  oscillations in brain networks with rate dynamics,'' {\em Automatica},
  vol.~146, p.~110653, 2022.

\bibitem{LGD-RAW-WG-DC-OCS-JLPV:05}
L.~G. Dominguez, R.~A. Wennberg, W.~Gaetz, D.~Cheyne, O.~C. Snead, and J.~L.~P.
  Velazquez, ``Enhanced synchrony in epileptiform activity? local versus
  distant phase synchronization in generalized seizures,'' {\em Journal of
  Neuroscience}, vol.~25, no.~35, pp.~8077--8084, 2005.

\bibitem{GB:06}
G.~Buzsaki, {\em Rhythms of the Brain}.
\newblock Oxford, UK: Oxford University Press, 2006.

\bibitem{CB-AM-CF:23}
C.~Beste, A.~M{\"u}nchau, and C.~Frings, ``Towards a systematization of brain
  oscillatory activity in actions,'' {\em Communications Biology}, vol.~6,
  no.~1, p.~137, 2023.

\bibitem{JBC-JRM-SR-MJK:01}
J.~B. Caplan, J.~R. Madsen, S.~Raghavachari, and M.~J. Kahana, ``Distinct
  patterns of brain oscillations underlie two basic parameters of human maze
  learning,'' {\em Journal of Neurophysiology}, vol.~86, no.~1, pp.~368--380,
  2001.

\bibitem{MM-TM-JC:23-csl}
M.~McCreesh, T.~Menara, and J.~Cort\'es, ``Sufficient conditions for
  oscillations in competitive linear-threshold brain networks,'' {\em IEEE
  Control Systems Letters}, vol.~7, pp.~2886--2891, 2023.

\bibitem{KM-AD-VI-CC:24}
K.~Morrison, A.~Degeratu, V.~Itskov, and C.~Curto, ``Diversity of emergent
  dynamics in competitive threshold-linear networks,'' {\em SIAM Journal on
  Applied Dynamical Systems}, vol.~23, no.~1, pp.~855--884, 2024.

\bibitem{CP-SM-KM-CC:22}
C.~Parmelee, S.~Moore, K.~Morrison, and C.~Curto, ``Core motifs predict dynamic
  attractors in combinatorial threshold-linear networks,'' {\em PloS One},
  vol.~17, no.~3, p.~e0264456, 2022.

\bibitem{EN-JC:19-acc}
E.~Nozari and J.~Cort\'es, ``Oscillations and coupling in interconnections of
  two-dimensional brain networks,'' in {\em {A}merican {C}ontrol {C}onference},
  (Philadelphia, PA), pp.~193--198, July 2019.

\bibitem{PF:15}
P.~Fries, ``Rhythms for cognition: Communication through coherence,'' {\em
  Neuron}, vol.~88, pp.~220--235, 2015.

\bibitem{DM-RV:16}
D.~McLelland and R.~VanRullen, ``Theta-gamma coding meets
  communication-through-coherence: neuronal oscillatory multiplexing theories
  reconciled,'' {\em PLOS Computational Biology}, vol.~12, no.~10, p.~e1005162,
  2016.

\bibitem{EN-JC:19-sfn}
E.~Nozari and J.~Cort\'es, ``Communication through coherence: Analysis of the
  optimal phase shift and the importance of waveform shape,'' in {\em
  Neuroscience}, (Chicago, IL), Oct. 2019.
\newblock Poster.

\bibitem{DR-GH:22}
D.~Reyner-Parra and G.~Huguet, ``Phase-locking patterns underlying effective
  communication in exact firing rate models of neural networks,'' {\em PLOS
  Computational Biology}, vol.~18, no.~5, p.~e1009342, 2022.

\bibitem{AHVL-JH-SL-TN-BM-AV-FB:17}
A.~H.~V. Lautz, J.~Herding, S.~Ludwig, T.~Nierhaus, B.~Maess, A.~Villringer,
  and F.~Blankenburg, ``Gamma and beta oscillations in human {MEG} encode the
  contents of vibrotactile working memory,'' {\em Frontiers in Human
  Neuroscience}, vol.~11, p.~576, 2017.

\bibitem{EKM-ML-AMB:18}
E.~K. Miller, M.~Lundqvist, and A.~M. Bastos, ``Working memory 2.0,'' {\em
  Neuron}, vol.~100, no.~2, pp.~463--475, 2018.

\bibitem{ML-PH-MRW-SLB-EKM:18}
M.~Lundqvist, P.~Herman, M.~R. Warden, S.~L. Brincat, and E.~K. Miller, ``Gamma
  and beta bursts during working memory readout suggest roles in its volitional
  control,'' {\em Nature Communications}, vol.~9, no.~1, p.~394, 2018.

\bibitem{ICF-SK:19}
I.~C. Fiebelkorn and S.~Kastner, ``A rhythmic theory of attention,'' {\em
  Trends in Cognitive Sciences}, vol.~23, no.~2, pp.~87--101, 2019.

\bibitem{EA-JJ:01}
E.~Awh and J.~Jonides, ``Overlapping mechanisms of attention and spatial
  working memory,'' {\em Trends in Cognitive Sciences}, vol.~5, no.~3,
  pp.~119--126, 2001.

\bibitem{AK-TE:13}
A.~Kiyonaga and T.~Egner, ``Working memory as internal attention: Toward an
  integrative account of internal and external selection processes,'' {\em
  Psychonomic Bulletin \& Review}, vol.~20, pp.~228--242, 2013.

\bibitem{NEM-MGS-ACN:17}
N.~E. Myers, M.~G. Stokes, and A.~C. Nobre, ``Prioritizing information during
  working memory: beyond sustained internal attention,'' {\em Trends in
  cognitive sciences}, vol.~21, no.~6, pp.~449--461, 2017.

\bibitem{ML-SLB-JR-MWW-TJB-EKM-PH:23}
M.~Lundqvist, S.~L. Brincat, J.~Rose, M.~W. Warden, T.~J. Buschman, E.~K.
  Miller, and P.~Herman, ``Working memory control dynamics follow principles of
  spatial computing,'' {\em Nature Communications}, vol.~14, no.~1, p.~1429,
  2023.

\bibitem{BK-RG:11}
B.~Kolb and R.~Gibb, ``Brain plasticity and behaviour in the developing
  brain,'' {\em Journal of the Canadian Academy of Child and Adolescent
  Psychiatry}, vol.~20, no.~4, p.~265, 2011.

\bibitem{PM-AR:19}
P.~Mateos-Aparicio and A.~Rodr{\'\i}guez-Moreno, ``The impact of studying brain
  plasticity,'' {\em Frontiers in Cellular Neuroscience}, vol.~13, p.~66, 2019.

\bibitem{JL-NS:05}
J.~Lisman and N.~Spruston, ``Postsynaptic depolarization requirements for {LTP}
  and {LTD}: a critique of spike timing-dependent plasticity,'' {\em Nature
  Neuroscience}, vol.~8, no.~7, pp.~839--841, 2005.

\bibitem{AK-VP-MAF-SKG-CK:19}
A.~Kumar, V.~Pareek, M.~A. Faiq, S.~K. Ghosh, and C.~Kumari, ``Adult
  neurogenesis in humans: a review of basic concepts, history, current
  research, and clinical implications,'' {\em Innovations in Clinical
  Neuroscience}, vol.~16, no.~5-6, p.~30, 2019.

\bibitem{DNA-JMW:15}
D.~N. Abrous and J.~M. Wojtowicz, ``Interaction between neurogenesis and
  hippocampal memory system: new vistas,'' {\em Cold Spring Harbor Perspectives
  in Biology}, vol.~7, no.~6, p.~a018952, 2015.

\bibitem{WD-JBA-FHG:10}
W.~Deng, J.~B. Aimone, and F.~H. Gage, ``New neurons and new memories: how does
  adult hippocampal neurogenesis affect learning and memory?,'' {\em Nature
  Reviews Neuroscience}, vol.~11, no.~5, pp.~339--350, 2010.

\bibitem{RCS-ST-MTdM:16}
R.~C. Cassilhas, S.~Tufik, and M.~T. de~Mello, ``Physical exercise,
  neuroplasticity, spatial learning and memory,'' {\em Cellular and Molecular
  Life Sciences}, vol.~73, pp.~975--983, 2016.

\bibitem{MF-CH-MMM-SV:09}
M.~Fisher, C.~Holland, M.~M. Merzenich, and S.~Vinogradov, ``Using
  neuroplasticity-based auditory training to improve verbal memory in
  schizophrenia,'' {\em American Journal of Psychiatry}, vol.~166, no.~7,
  pp.~805--811, 2009.

\bibitem{JZK-DSB:23}
J.~Z. Kim and D.~S. Bassett, ``A neural machine code and programming framework
  for the reservoir computer,'' {\em Nature Machine Intelligence}, vol.~5,
  no.~6, pp.~622--630, 2023.

\bibitem{FD-CCH-AG:22}
F.~Damicelli, C.~C. Hilgetag, and A.~Goulas, ``Brain connectivity meets
  reservoir computing,'' {\em PLoS Computational Biology}, vol.~18, no.~11,
  p.~e1010639, 2022.

\end{thebibliography}

\end{document}